\documentclass[a4paper,twocolumn,pra,superscriptaddress,amsmath,amssymb,showpacs,floatfix]{revtex4-1}

\usepackage{scrextend}
\usepackage{graphicx}
\usepackage{epstopdf}
\usepackage{tabularx}
\usepackage{array}
\usepackage[table]{xcolor}     
\usepackage{booktabs} 
\usepackage{booktabs,dcolumn}
\usepackage{amsmath}
\usepackage[colorlinks]{hyperref} 

\definecolor{lightgray}{gray}{0.9}
\definecolor{tablerow1}{RGB}{225,217,205}
\definecolor{tablerow2}{RGB}{236,229,221}

\usepackage[normalem]{ulem}
\usepackage{cancel}

\DeclareMathOperator{\sign}{sign}
\usepackage[version=3]{mhchem}
\hypersetup{%
linkcolor=blue,
citecolor=blue,
filecolor=black,
menucolor=black,
urlcolor=black
}

\begin{document} 

\title{Theoretical description of circular dichroism in photoelectron 
  angular distributions of randomly oriented chiral molecules after 
  multi-photon photoionization}
\author{R. E. Goetz}
\affiliation{Theoretische Physik, Universit\"{a}t Kassel,
  Heinrich-Plett-Str. 40, D-34132 Kassel, Germany}
\author{T. A. Isaev}
\affiliation{Fachbereich Chemie, Philipps-Universit\"{a}t Marburg,
  Hans-Meerwein-Strasse 4, 35032 Marburg, Germany}
\author{B. Nikoobakht}
\affiliation{Fachbereich Chemie, Philipps-Universit\"{a}t Marburg,
  Hans-Meerwein-Strasse 4, 35032 Marburg, Germany}
\author{R. Berger}
\affiliation{Fachbereich Chemie, Philipps-Universit\"{a}t Marburg,
  Hans-Meerwein-Strasse 4, 35032 Marburg, Germany}
\author{C. P. Koch}
\affiliation{Theoretische Physik, Universit\"{a}t Kassel,
 Heinrich-Plett-Str. 40, D-34132 Kassel, Germany}
\email{christiane.koch@uni-kassel.de}
\date{\today}            
\begin{abstract}
Photoelectron circular dichroism refers to the forward/backward asymmetry in
the photoelectron angular distribution with respect to the propagation axis
of circularly polarized light. It has recently been demonstrated in
femtosecond multi-photon photoionization experiments with randomly oriented
camphor and fenchone molecules [C. Lux et al., Angew. Chem. Int. Ed. 51, 5001
(2012); C. S.  Lehmann et al., J. Chem. Phys. 139, 234307 (2013)]. A
theoretical framework describing this process as (2+1) resonantly enhanced
multi-photon ionization is constructed, which consists of two-photon
photoselection from randomly oriented molecules and successive one-photon
ionisation of the photoselected molecules. It combines perturbation theory
for the light-matter interaction with \textit{ab initio} calculations for the
two-photon absorption and a single-center expansion of the photoelectron
wavefunction  in terms of hydrogenic continuum functions. It is verified that
the model correctly reproduces the basic symmetry behavior expected under
exchange of handedness and light helicity. When applied it to fenchone and
camphor, semi-quantitative agreement with the experimental data is found, for
which a sufficient $d$ wave character of the electronically excited
intermediate state is crucial. 
\end{abstract}
 \maketitle
\section{Introduction}
\label{sec:introduction}

Photoelectron spectroscopy is a powerful tool for studying
photoionization dynamics. Intense short laser pulses for the 
ionization, which easily drive multi-photon transitions, allow to 
observe effects in table-top experiments that otherwise would require
synchrotron radiation.  
A recent example is the photoelectron circular dichroism (PECD) of chiral 
molecules~\cite{LuxACIE12,LehmannJCP13,Janssen2014,LuxCPC15,FanoodPCCP15}.  
It refers to the forward/backward asymmetry with respect to the light
propagation axis in the photoelectron angular distribution (PAD) obtained 
after excitation with circularly polarized
light~\cite{Ritchie1976,PowisAdvCP08,Nahon2010}. 
When the PAD is expanded in Legendre polynomials, a PECD is characterized 
by the expansion coefficients of the odd-order polynomials with the highest 
order polynomial being determined by the order of the process, i.e.,  
the number of absorbed photons~\cite{Ritchie1976,DixitPRA1983}. 

A theoretical description of such experiments with intense femtosecond laser
pulses requires proper account of the multi-photon excitation pathways. In
the pioneering work of McClain and
co-workers~\cite{pmw1970,Monson1970,McClain1972,wm1974}, a model for the
simultaneous absorption of two photons including the corresponding modified
molecular selection rules was formulated. Two-photon circular dichroism was
developed in Ref.~\cite{Itjr1974}, attributing the effect to a difference in
the absorption coefficient for the two left and two right polarized photons.
These approaches are based on a perturbation expansion of the light-matter
interaction. The strong-field approximation provides an alternative
description which is particularly suited for very intense
fields~\cite{Keldysh1965,Faisal1973}. 

Multi-photon transitions driven by strong femtosecond laser pulses may or may
not involve intermediate states.

In recent experiments with bicyclic
ketones~\cite{LuxACIE12,LehmannJCP13,Janssen2014,LuxCPC15,FanoodPCCP15}, a
2+1-REMPI process was employed. The nature of the intermediate state remains
yet  to be clarified. A first theoretical study used the strong-field
approximation~\cite{LeinPRA2014}. While the standard strong-field
approximation using a plane wave basis for the photoelectron was found to
fail in describing PECD, accounting for the Coulomb interaction between
photoelectron and photoion in the Born approximation allowed for observation
of PECD. However, the PAD did not agree with the epxerimental ones. This may
be explained by the role fo the intermediate state in the REMPI process which
necessarily is ignored in the strong-field approximation~\cite{LeinPRA2014}. 

Here, we take the opposite approach, starting with a 
perturbation theory treatment of the multi-photon process. 
Thus, ionization is viewed as a (weak) one-photon 
transition into the continuum,  the 'initial' state of which is 
prepared by non-resonant two-photon absorption. 
Such an approach is motivated by the moderate intensities,
of the order of $10^{12}\,$W/cm$^2$,  used in the
experiments~\cite{LuxACIE12,LehmannJCP13,Janssen2014,LuxCPC15,FanoodPCCP15}.
Although clearly in the multi-photon regime, such 
intensities can be described comparatively well by low order
perturbation theory~\cite{AmitayPRL08,RybakPRL11,LevinPRL15}.
 
The non-resonant two-photon preparation step yields an important difference
compared to pure one-photon excitation~\cite{RitchiePRA1976}. In the latter
case, the first order Legendre polynomial alone accounts for the
PECD~\cite{Reid2003,cooper,ChandraJPhysB87}.  This results from the random
orientation of the molecules, or, in more technical terms, from integrating
the differential cross section over  the Euler  angles.  In contrast,
non-resonant two-photon excitation may lead to an orientation-dependent
probability distribution of the molecules in the resonant intermediate
state~\cite{LehmannJCP13}.  In this case, the maximum order of Legendre
polynomials contributing to the PAD is not limited to 2, but 6 for a 2+1
process. Whether the two-photon absorption is orientation-dependent is
determined by the two-photon transition matrix elements. Here, we calculate
the two-photon transition matrix elements using state of the art \textit{ab
initio} methods.  
However, for molecules as complex as camphor and fenchone, it is extremely
challenging to model the complete photoionization process from first
principles, even when using the most advanced \textit{ab initio}
methods. We therefore split the theoretical description into two
parts. 

As long as all electrons remain bound, state of the art quantum
chemical approaches, for example the coupled cluster methods,
can be used to accurately determine the electronic
wave functions. However, once an electron starts to leave the ionic
core, the standard basis sets of electronic structure theory are not
well adapted. An alternative is offered by a single-center expansion
into  eigenfunctions of a hydrogen-like atom for which both
bound and
continuum functions are known analytically. The hydrogenic
continuum functions properly account for the
long-range Coulomb interaction between ionic core and ejected electron but
neglect the effect of short-range correlations in the ionization
step. 
The basis functions for the single center expansion are chosen such as
to yield the simplest possible model that is able to reproduce the 
laboratory-frame photoelectron angular 
distributions (LF-PADs) resulting from a 2+1-REMPI process in
randomly oriented chiral molecules. The two descriptions are
matched at the resonant, electronically excited intermediate state
by projecting the numerically calculated wavefunction onto the basis
functions of the single center expansion. 

Our approach of calculating the PAD as a one-photon absorption cross section
for an effective ``initial''  state in a single center expansion, while
neglecting dynamical effects, allows us to generalize our findings to chiral
molecules other than fenchone or camphore. In particular, we analyze the role
of the laser polarization for each step in the 2+1 ionization process and
determine the conditions on the two-photon absorption matrix elements for
yielding PECD. 

The remainder of the paper is organized as follows: Our theoretical framework
is introduced in Sec.~\ref{sec:model}. In detail, Sec.~\ref{subsec:ansatz}
defines the PAD as one-photon photoionization cross section and summarizes
the single center expansion.  To make connection with experiment, the cross
sections need to be transformed from the molecule-fixed frame into the
laboratory frame and averaged over the random orientations of the molecules.
The corresponding expressions for a 2+1 REMPI process are presented in
Sec.~\ref{subsec:PAD} with the details of the derivation given in the
appendix. The symmetry properties required for observing PECD are analyzed in
Sec.~\ref{subsec:parity}.  Section~\ref{sec:abinitio} is dedicated to
\textit{ab initio} calculations for the  intermediate, electronically
excited states and the two-photon absorption matrix elements.
Section~\ref{subsec:compdetails}  presents the computational details and
Sec.~\ref{subsec:abinitioresults} the results. The one-center reexpansion
required for matching the numerical results to the single-center
description derived in Sec.~\ref{sec:model} is described in
Sec.~\ref{subsec:reexpansion}. Our numerical results for the PAD of camphor
and fenchone and the corresponding PECD are presented in
Sec.~\ref{sec:pad_results} with Sec.~\ref{subsec:fenchone} dedicated to
fenchone and Sec.~\ref{subsec:camphor} to camphore. Our findings are
summarized and discussed in Sec.~\ref{subsec:discussion}.
Section~\ref{sec:conclusions} concludes.

\section{Model}
\label{sec:model}
We model the resonantly enhanced multi-photon photoionization as a 2+1
process, assuming the last photon to constitute a weak probe of the molecular
state that is prepared by  non-resonant two-photon absorption. 
For simplicity, we
employ the strict electric dipole approximation. That is, contributions from
magnetic dipole terms, which are important for circular polarization
dependent differences in absorption cross sections, and higher order electric
and magnetic multipole terms are neglected.

Defining two
coordinates systems, the molecular frame of reference $\mathcal{R}$ and the
laboratory frame $\mathcal{R}^\prime$, $\epsilon_{\varrho_2}^\prime$ denotes
the polarization of the laser field with respect to the laboratory frame
(where we distinguish the polarization of the ionizing photon,
$\epsilon_{\varrho_2}^\prime$ from that of the first two photons,
$\epsilon_{\varrho_1}^\prime$). 
For convenience,
  we work in the spherical basis. Thus, $\epsilon_{\varrho_2}^\prime$ and  $\epsilon_{\varrho_1}^\prime$ correspond to the spherical unit vectors in the laboratory frame, with $\varrho_{1,2}=\pm 1,0$ denoting left/right circular and linear polarization of the laser
beam which propagates in the positive $z^\prime$ direction (the relation between the spherical and Cartesian unit vectors is found in 
Eq.~\eqref{eq:cartesian_spherical}).  
Primed (unprimed) coordinates refer the laboratory
(molecular) frame of reference throughout.
Both frames, $\mathcal{R}^\prime$ and $\mathcal{R}$, are related by
an arbitrary  coordinate rotation  $D(\alpha\beta\gamma)$,
where $\omega=(\alpha,\beta,\gamma)$ denote  the Euler angles
defining the orientation of $\mathcal{R}$ with respect to
$\mathcal{R}^\prime$.

Consider a one-photon ($1\mathrm{P}$) transition in a molecule whose
orientation  
with respect to $\mathcal{R}^\prime$ is given by the Euler angles
$\omega$. The corresponding differential photoionization cross
section, when measured in the molecular frame $\mathcal{R}$, 
reads, within perturbation theory and the electric dipole
approximation and in SI units~\cite{Hans1957}, 
\begin{eqnarray}
  \label{eq:diffX} 
  \frac{d^2\sigma_{1\mathrm{P}}}{d\omega\; d\Omega_{\mathbf{k}}} &=& c_0\,
  \left|\langle\Psi_{\mathbf{k}}|
    \mathbf{\epsilon}^\prime_{\varrho_2}\cdot\mathbf{r}       
    |\Psi_o\rangle\right|^2\,, 
\end{eqnarray}
where $c_0 = 4\pi^2\alpha\hbar\omega_{{\mathrm{ph}}}$ with  $\alpha$ being
the fine-structure constant, $\hbar\omega_{\mathrm{ph}}$ the energy of the
ionizing photon, $\hbar$ the reduced Planck constant and $\mathbf{r}$ the
position operator of the electron (or a sum of the various position operators
in the multi-electron case). 
The
polarization of the electric field in the laboratory frame of reference is
specified by $\epsilon^\prime_{\varrho_2}$, where $\varrho_2$  takes the
value 0 for linear and $+1 (-1)$ for left (right) circular polarization,
respectively.  $|\Psi_{\mathbf{k}}\rangle$ denotes an energy
normalized molecular state with one electron transfered to the ionisation
continuum with asymptotic electron linear momentum $\mathbf{k}$.
$|\Psi_o\rangle$ is the (bound, unity normalized) molecular state prepared by
the non-resonant two-photon absorption, which is defined in the molecular
frame of reference.  In Eq.~\eqref{eq:diffX}, we employ the standard notation
for doubly differential cross sections in the molecular frame of
reference~\cite{ChandraJPhysB87,ChengPRA2010,LuchessePRA1982} that depend not
only on the solid angle $\Omega_{\mathbf{k}}$ but also on the orientation of
the molecule via the Euler angles $\omega$. 
We utilize a single-center approximation~\cite{Bishop67} which
allows us to calculate the matrix elements in
Eq.~\eqref{eq:diffX} explicitly. That is, we project the
multi-electron wave function obtained from \textit{ab
  initio} calculations, $|\Psi_o\rangle$, on one-electron basis
functions and neglect 
electron correlations in the continuum description. 
We first discuss in Sec.~\ref{subsec:ansatz}
our choice of $|\Psi_o\rangle$ and then 
explain below in Sec.~\ref{subsec:PAD} how to
connect the differential ionization cross section to the
experimentally measured photoelectron angular distributions.

\subsection{Single center expansion}
\label{subsec:ansatz}
The ``initial'' state for the one-photon ionization is a
multi-electron wavefunction which is usually expanded in specially
adapted basis functions developed in quantum chemistry. In contrast, 
the single center expansion 
is based on the fact that any molecular wavefunction can be written as 
a linear combination of functions about a single arbitrary
point~\cite{Bishop67}. Of course, such an ansatz will converge very
slowly, if the multi-center character of the wavefunction is
important. Writing the wavefunction of  the
electronically excited state of the neutral molecule, that is prepared
by the two-photon absorption process, as 
$\langle\mathbf{r}| \Psi_o\rangle=\Psi_o(\mathbf{r})$, we expand it 
into eigenfunctions of a hydrogen-like atom, 
\begin{eqnarray}
  \label{eq:exited_state}
  \Psi_o(\mathbf{r}) = \sum_{n_o=0}^\infty\sum_{\ell_o=0}^{n_o-1}
  \sum_{m_o=-\ell_o}^{\ell_o} a^{\ell_o}_{m_o}(n_o)\,                   
  R^{n_o}_{\ell_o}(r)\, Y^{\ell_o}_{m_o}(\Omega_{\mathbf{r}})\,.
\end{eqnarray}
Here, $a^{\ell_o}_{m_o}(n_o)$                   
stands for the unknown expansion coefficients, 
$R^{n_o}_{\ell_o}(r)$  denotes the radial eigenfunctions of the
hydrogen-like atom, and $Y^{\ell_o}_{m_o}(\Omega_{\mathbf{r}})$ are the spherical
harmonics. $\Omega_{\mathbf{r}}=(\vartheta_{\mathbf{r}},\phi_{\mathbf{r}})$
refers  to the polar and azimuthal angles of the position 
vector $\mathbf{r}$ in the molecular frame of reference.
Note that  all information about the geometry and the symmetry 
properties of the ``initial'' electronically excited state is
contained in the expansion coefficients $a^{\ell_o}_{m_o}(n_o)$.         
The number of basis
functions must be truncated in any actual calculation, i.e.,
\begin{eqnarray}
  \label{eq:exc_state}
  \Psi_o(\mathbf{r}) \approx \sum_{n_o=n_o^{\mathrm{min}}}^{n_o^{\mathrm{max}}}\sum_{\ell_o=0}^{n_o-1}
  \sum_{m_o=-\ell_o}^{\ell_o} a^{\ell_o}_{m_o}(n_o)\,  
  R^{n_o}_{\ell_o}(r)\, Y^{\ell_o}_{m_o}(\Omega_{\mathbf{r}})\,.\quad
\end{eqnarray}
     
Strictly speaking, all molecular orbitals that are involved
in Slater determinants describing the excited state should be subject
to the single center expansion. In the present model, we 
employ an effective one-electron picture by expanding only one
representative virtual orbital around the single center, namely the
one that is additionally  occupied in the supposedly leading configuration for the respective excited state.

We will also ask what  the simplest possible model is that gives rise
to PECD. In this case, we assume a single quantum
number $n$,  $n=n_o$, to contribute to Eq.~\eqref{eq:exited_state},
i.e.,  
\begin{eqnarray}
\label{eq:exited_state2}
\Psi^{s}_o(\mathbf{r}) \approx \sum_{\ell_o=0}^{L_{o,\rm max}}
\sum_{m_o=-\ell_o}^{\ell_o} a^{\ell_o}_{m_o}\, 
R^{n_o}_{\ell_o}(r)\, Y^{\ell_o}_{m_o}(\Omega_{\mathbf{r}})\,,
\end{eqnarray}
where $L_{o,\mathrm {max}}$ refers to the highest angular momentum state
appearing in the ``initial'' wavefunction. It follows from basic symmetry
arguments that the minimal value of $L_{o,\rm max}$ for which a PECD
can be expected is $L_{o,\mathrm{max}}=2$, that is, at least $d$-orbitals are
required.

We model the photoionization as a one-electron process arising from a
hydrogenic-like system exclusively, which allows for neglecting
the bound molecular part (the remaining molecular parent ion) in $|\Psi_{\mathbf{k}}\rangle$. Thus, 
the resulting continuum wave functions, $\Psi_{\mathbf{k}}(\mathbf{r})$, are
expanded into  partial waves in a way that allows for  an explicit
expression of the photoionization cross section in terms of the
scattering solid angle
$\Omega_{\mathbf{k}}$~\cite{ChengPRA10,LuchessePRA1982,DillChemphys87,cooper},  
\begin{eqnarray}
\label{eq:solution}
\Psi_{\mathbf{k}}(\mathbf{r}) = 4\pi\sum_{l=0}^{\infty}\sum_{m=-l}^{l}
\mathrm{i}^{\ell}
\phi_{k,\ell,m}(r)\, Y^{*\,\ell}_{m}(\Omega_{\mathbf{k}})\, 
Y^{\ell}_{m}(\Omega_{\mathbf{r}})\,.
\end{eqnarray}
Here, $Y^{\ell}_{m}(\Omega_{\mathbf{r}})$ and $Y^{\ell}_{m}(\Omega_{\mathbf{k}})$
correspond to the spherical harmonics describing the orientation of
the photoelectron position and momentum,
respectively, and $\phi_{k,\ell,m}(r)$ is the radial part of the
photoelectron wavefunction. For simplicity, 
we use here and in the following $Y^{*\, \ell}_m(\Omega_{\mathbf{k}})$ 
as an abbreviation for $(Y^{\ell}_m(\Omega_{\mathbf{k}}))^*$. 
Modeling photoionization as a one-electron process, we can approximate  
\begin{eqnarray}
\label{eq:approximation}
\phi_{k,\ell,m}(r)\approx e^{-\mathrm{i}\delta_{\ell}} G_{k,\ell}(r)\,,
\end{eqnarray}
where $G_{k,\ell}(r)$ are the well-known radial continuum wavefunctions
of the hydrogen atom, recalled in  Appendix~\ref{subsec:H-cont}, 
and $\delta_{\ell}$ stands for the Coulomb phase
shift of the $\ell-$th scattered partial wave, 
with $\delta_{\ell} =
\Gamma(\ell+1-\mathrm{i}/k)$~\cite{LuchessePRA1982,ChandraJPhysB87,DillChemphys87}.  
Note that we expect the phase shift for molecules to depend 
on $\ell_o$ and  $m_o$ since the molecular potential of chiral
molecules is  not spherically symmetric. 
Neglecting the
$m_o$-dependence of the phase shift involves no approximation when
using Eq.~\eqref{eq:exited_state} since 
the hydrogen eigenfunctions form a complete
orthonormal basis. 
However, this is not true anymore when truncating
the basis, cf. Eq.~\eqref{eq:exc_state}. Our ansatz thus involves an
additional approximation, namely Eq.~\eqref{eq:approximation}. 

By construction, Eq.~\eqref{eq:approximation} yields 
orthogonality between bound and unbound wavefunctions which is
required to avoid spurious singularities~\cite{ChengPRA10} and
reproduce the correct threshold behavior of the  
photoioization cross-sections~\cite{OanaJCP09}. 
With the approximation of Eq.~\eqref{eq:approximation}, we
account for the long-range Coulomb interaction between photoelectron
and a point charge representing the  ionic 
core but neglect the short-range static exchange. Also, 
dynamic changes in the electron distribution, such as 
adjustments of the electronic cloud due to nuclear
motion, as well as the interaction of the outgoing photoelectron with
the driving electric field  upon photoionization are neglected.

Inserting Eq.~\eqref{eq:approximation} into Eq.~\eqref{eq:solution}
yields 
\begin{equation}
\label{eq:photoelectron}
\Psi_{\mathbf{k}}(\mathbf{r}) = 4\pi\sum_{l=0}^{\infty}\sum_{m=-l}^{l}
\mathrm{i}^{\ell}
e^{-\mathrm{i}\delta_{\ell}} G_{k,\ell}(r)\,
Y^{*\, \ell}_{m}(\Omega_{\mathbf{k}})\, 
Y^{\ell}_{m}(\Omega_{\mathbf{r}})\,,
\end{equation}
and we can evaluate the matrix element in
Eq.~\eqref{eq:diffX}. Because  
the wavefunctions are given in the molecular frame of reference, we
need to rotate the spherical unit vector 
$\mathbf{\epsilon}^\prime_{\varrho_2}$  in
Eq.~\eqref{eq:diffX} into that frame~\cite{ChandraJPhysB87}.
Expanding the rotation operator  $D(\alpha\beta\gamma)$ 
connecting $\mathbf{r}$ and $\mathbf{r}^\prime$ into
irreducible rank 1 tensor representations, cf. Appendix~\ref{subsec:rotmat}, 
Eq.~\eqref{eq:diffX} becomes
\begin{eqnarray}\label{eq:diffXa}
  \frac{d^2\sigma_{1\mathrm{P}}}{d\omega d\Omega_{\mathbf{k}}} &=& c_0 
   \sum^1_{q=-1}\sum^1_{q^\prime=-1}
  \mathcal{D}^{(1)}_{q, \varrho_2}(\omega) 
  \mathcal{D}^{(1)}_{-q^\prime,-\varrho_2}(\omega)\\\nonumber
  && \times (-1)^{q^\prime-\varrho_2}
  \langle\Psi_{\mathbf{k}}|\mathbf{r}_{q}|\Psi_o\rangle
  \langle\Psi_{\mathbf{k}}|\mathbf{r}_{q^\prime}|\Psi_o\rangle^{*}\,.
\end{eqnarray}
Inserting Eqs.~\eqref{eq:exited_state2} and~\eqref{eq:photoelectron}
to evaluate the overlap integrals yields 
\begin{eqnarray}
  \label{eq:diffX2}
  \frac{d^2\sigma_{1\mathrm{P}}}{d\omega\; d\Omega_{\mathbf{k}}} &=&c_0\,
\sum_{\substack{\ell,m \\ n_o\ell_o,m_o}}\, \sum_{\substack{\ell^\prime,m^\prime\\  
n^\prime_o,\ell^\prime_o,m^\prime_o}}\, \sum_{q=-1}^{1}\sum_{q^\prime=-1}^{1}
(-\mathrm{i})^{\ell-\ell^\prime}e^{\mathrm{i}(\delta_{\ell}-\delta_{\ell^\prime})}\nonumber\\ 
&&\times a^{\ell_o}_{m_o}(n_o)\,a^{*\ell^\prime_o}_{m^\prime_o}(n^\prime_o) I^{n_o}_{_{k}}(\ell,\ell_o)
I^{n^\prime_o}_{_{k}}(\ell^\prime,\ell^\prime_o) \nonumber\\ 
&&\times Y^{\ell}_{m}(\Omega_{\mathbf{k}})Y^{*\ell^\prime}_{m^\prime}(\Omega_{\mathbf{k}})\,  
\mathcal{D}^{(1)}_{q,\varrho_2}(\omega)\mathcal{D}^{*{(1)}}_{q^\prime,\varrho_2}({\omega})\nonumber\\
&&
\times\mathcal{S}^{\ell,m}_{\ell_o,m_o}(q) 
\mathcal{S}^{*\ell^\prime,m^\prime}_{\ell^\prime_o,m^\prime_o}(q^\prime)\,.
\end{eqnarray}
In Eq.~\eqref{eq:diffX2}, we have introduced  radial and angular 
integrals $I_{k}(\ell,\ell_o)$ and
$\mathcal{S}^{\ell,m}_{\ell_o,m_o}(q)$, given by 
\begin{subequations}\label{eq:Ik_and_S}
\begin{eqnarray}
  \label{eq:Ik}
  I^{n_o}_{k}(\ell,\ell_o) = I_o\,
  \int^{+\infty}_{0} r^3\,G_{k,\ell}(r)\,R^{n_o}_{\ell_o}(r)\,dr
\end{eqnarray}
for a fixed $n_o$ in Eq.~\eqref{eq:exited_state} 
with $I_o = 4\pi/3$, and 
\begin{eqnarray}
  \label{eq:S}
  \mathcal{S}^{\ell,m}_{\ell_o,m_o}(q) &=& \int Y^{\ell\,*}_{m}(\Omega_r)\, 
  Y^{1}_{q}(\Omega_r)Y^{\ell_o}_{m_o}(\Omega_r)\,d\Omega_r \\\nonumber
  &=& (-1)^{-m}\, b_{\ell,\rm \ell_o}
\begin{pmatrix} 
\ell & 1 & \ell_o \vspace*{0.33cm} \\ 0 & 0 & 0 
\end{pmatrix}
\begin{pmatrix} 
\ell & 1 & \ell_o \vspace*{0.33cm} \\ -m & q & m_o
\end{pmatrix}
\end{eqnarray}
\end{subequations}
with 
\[
b_{\ell,\rm \ell_o}=\sqrt{3\, (2\ell+1)(2\ell_o+1)/4\pi}
\]
and using Wigner $3j$
symbols~\cite{edmonds,irreducible,Rose1967,Varshalovich1988}. 
The angular integral $\mathcal{S}^{\ell,m}_{\ell_o,m_o}(q)$ 
determines, for each spherical unit vector $q=0,\pm 1$, 
the selection rules between the angular components of the bound exited 
electronic state with quantum numbers $\ell_o$, $m_o$
and the partial wave components of the 
continuum wavefunction with quantum numbers $\ell$, $m$. 
Equation~\eqref{eq:S} implies that transitions are
allowed if and only if $\ell+1+\ell_o$ is
even and $m_o+q-m=0$ for all $|\ell_o-1|\le\ell\le\ell_o+1$.
This is a special case of the more general rule for multipole
transitions derived in Ref.~\cite{DixitPRA1983}.
The angular integrals can be evaluated analytically using the standard 
angular momentum algebra, whereas the radial integrals in
Eq.~\eqref{eq:Ik} are computed numerically. 

The choice of basis to describe the radial part of the continuum 
wavefunction determines the weight with which each 
excited state expansion coefficient $a^{\ell_o}_{m_o}(n_o)$ 
contributes to the PAD, cf. Eqs.~\eqref{eq:diffX2}
and~\eqref{eq:Ik}. Thus, choosing for
example planes waves, i.e., the eigenfunctions of the ``free''
photoelectron,  
which is described in terms of the Bessel functions~\cite{edmonds,irreducible,Varshalovich1988}, and does not take into 
account the Coulomb interaction between the outgoing photoelectron and the
remaining ion,
would translate
into a PAD different from  the one  obtained with the 
hydrogenic continuum wavefunctions of Eq.~\eqref{eq:photoelectron}.
Whether or not the model is able to reproduce the measured Legendre
coefficients will to some extent
depend on the choice of basis for the radial part in 
Eq.~\eqref{eq:solution}. 

The missing ingredient to determine the differential
photoionization cross section, Eq.~\eqref{eq:diffX},
are the expansion coefficients, $a^{\ell_o}_{m_o}(n_o)$, 
of the intermediate excited state wavefunction. 
They can either be used as fitting parameters or 
determined from \textit{ab initio} calculations, see
Sec.~\ref{sec:abinitio} . 

Two more steps are then required to connect the differential ionization
cross section to the experimentally measured
PAD. First, the PAD is measured in the laboratory frame and the
differential ionization cross section thus needs to be rotated
from the molecular into the laboratory frame. Second, the orientation
of the molecule with respect to the laboratory frame, defined by the
polarization axis of the laser electric field, is arbitrary. We
therefore need 
to average over all possible orientations, i.e., integrate over the
Euler angles $\omega = (\alpha$, $\beta$, $\gamma)$, as we consider
a randomly oriented initial ensemble of molecules.

\subsection{Photoelectron Angular Distributions}
\label{subsec:PAD}
Rotating the differential cross section from the molecular into the
laboratory frame requires rotation of the continuum state 
$|\Psi_{\mathbf{k}}\rangle$ into $|\Psi_{\mathbf{k}^\prime}\rangle$
using the inverse of Eq.~\eqref{eq:eq1}. This leads to  
\begin{widetext}
\begin{eqnarray} 
\label{eq:differential2}
  \frac{d^2\sigma_{1\mathrm{P}}}{d\omega \;d\Omega_{{\mathbf{k}}^\prime}} &=& 
  c_0\, \sum_{\substack{\ell,m \\ n_o,\ell_o,m_o}}\, 
  \sum_{\substack{\ell^\prime,m^\prime \\ n^\prime_o,\ell^\prime_o,m^\prime_o}}\, \sum_{q,q^\prime}
  (-\mathrm{i})^{\ell-\ell^\prime}e^{\mathrm{i}(\delta_{\ell}-\delta_{\ell^\prime})}\, 
  a^{\ell_o}_{m_o}(n_o)\,a^{*\ell^\prime_o}_{m^\prime_o}(n^\prime_o)
  I^{n_o}_{_{k}}(\ell,\ell_o)\, I^{n^\prime_o}_{_{k}}(\ell^\prime,\ell^\prime_o) \mathcal{S}^{\ell,m}_{\ell_o,m_o}(q) 
  \mathcal{S}^{*\ell^\prime,m^\prime}_{\ell^\prime_o,m^\prime_o}(q^\prime)\nonumber\\          
  && \times\sum^{\ell+\ell^\prime}_{\mathcal{L}=\vert\ell-\ell^\prime\vert}
\begin{pmatrix} \ell & \ell^\prime & \mathcal{L}\vspace*{0.33cm}\\ 0 & 0 & 0 \end{pmatrix}
\begin{pmatrix} \ell & \ell^\prime & \mathcal{L} 
\vspace*{0.33cm} \\ m & -m^\prime & -(m-m^\prime) \end{pmatrix} 
\sum_{\mu=-\mathcal{L}}^{\mathcal{L}}
\mathcal{D}^{(1)}_{q,\varrho_2}(\omega)\mathcal{D}^{(1)}_{-q^\prime,-\varrho_2}(\omega)
    \mathcal{D}^{(\mathcal{L})}_{m^\prime-m,-\mu}(\omega)
     P^{\mu}_{\mathcal{L}}(\cos\vartheta^\prime_{k})\,e^{\mathrm{i}\mu\varphi^\prime_k}\nonumber\\
     &&\times (2\mathcal{L}+1)\,\varsigma^\mu_{\mathcal{L}}(\ell,\ell^\prime)\, (-1)^{m^\prime + q^\prime- \varrho_2}\,, 
\end{eqnarray}
\end{widetext}
where $\varsigma^\mu_{\mathcal{L}}(\ell,\ell^\prime)$ is  defined in
Eq.~\eqref{eq:varsigma:def}  in Appendix~\ref{subsec:X1P}. 
$P^{\mu}_{\mathcal{L}}(\cos\vartheta^\prime_{k})$ denotes the associate 
Legendre polynomials. A detailed derivation of
Eq.~\eqref{eq:differential2} is found in Appendix~\ref{subsec:X1P}.

When averaging over all orientations in the second step, we need to
account for the fact that the probability for non-resonant two-photon
absorption from the ground state to the intermediate electronically
excited state is,  
depending on the properties of the two-photon absorption tensor, 
not isotropic.  The differential ionization cross section in the
laboratory frame therefore needs to be weighted by the probability of
the electronically excited state to be occupied after absorption  
of the first two (identical) photons. Thus, 
the cross section for  photoemission into a solid angle
$d\Omega_{\mathbf{k}^\prime}$ around the axis $\mathbf{k}^\prime$  in 
the laboratory frame, after one-photon transition from 
the electronically excited intermediate state, is given by 
\begin{eqnarray}
\label{eq:weighted0}
\frac{d^2\sigma_{2+1}}{d\omega \;d\Omega_{\mathbf{k}^\prime}}
= \rho_{2\mathrm{P}}(\omega)
~\frac{d^2\sigma_{1\mathrm{P}}}{d\omega \;d\Omega_{\mathbf{k}^\prime}},
\end{eqnarray}
where $\rho_{2\mathrm{P}}(\omega)$ denotes the orientation-dependent
probability to reach the intermediate excited state
by absorption of two identical photons from the ground state.
Equation~\eqref{eq:weighted0} assumes  a molecule to have, in its 
electronic ground state, an initial orientation  of
$\omega=(\alpha,\beta,\gamma)$ with respect to the laboratory 
frame of reference. Note that Eq.~\eqref{eq:weighted0} makes an
additional assumption, namely the relative phase between the
two-photon and one-photon steps to be irrelevant for the photoelectron
spectrum and angular distribution. For a discussion of similar approximations 
in related multiphoton transitions between bound states, see for instance
Refs.~\cite{McClain1972,Monson1970}.

The experimentally measured PAD contains
contributions from all molecules in the sample, each of them with a specific
orientation $\omega$. The total photoelectron signal 
is therefore obtained by an incoherent summation
over the contributions from all molecules. This is equivalent to
integrating  Eq.~\eqref{eq:weighted0} over the Euler angles weighted by the
probability of two-photon absorption. The
``averaged'' photoionization cross section in the laboratory frame
therefore reads, 
\begin{eqnarray}
\label{eq:weighted}
\frac{d\sigma_{2+1}}{d\Omega_{\mathbf{k}^\prime}}
= \int\rho_{2\mathrm{P}}(\omega)
~\frac{d^2\sigma_{1\mathrm{P}}}{d\omega\; d\Omega_{\mathbf{k}^\prime}}
d\omega\,,
\end{eqnarray}
where the integration is carried over the Euler angles $\alpha,\beta,\gamma$. 

The orientation-dependent probability to reach the intermediate excited state, $\rho_{2P}(\omega)$, is obtained from 
the transition probability for two-photon absorption from the ground state $|\Psi_g\rangle$ to the intermediate electronically excited state $|\Psi_o\rangle$. The latter in general is defined as~\cite{Peticolas1967}
\begin{subequations}
  \label{eq:peticolas1b}
  \begin{equation}
    \label{eq:probatlitypeticolas}
    A^{(2)}_{o,g} = \tilde{\mathcal{N}}_0(\omega_{\mathrm{ph}})\, |\mathcal{M}|^2\,,    
  \end{equation}
where $\mathcal{M}$,
in the strict electric dipole approximation, 
$\exp(i\mathbf{k}\cdot\mathbf{r})\approx1$, reads
\begin{eqnarray}
  \label{eq:peticolas2b}
  \mathcal{M} &=& \sum_{n} \left\{\dfrac{(\mathbf{e}_1\cdot \langle\Psi_o|\mathbf{r}|\Psi_n\rangle)(\langle\Psi_n|\mathbf{r}|\Psi_g\rangle\cdot \mathbf{e}_2)}{\hbar\omega_g-\hbar\omega_n +\hbar\omega_{\mathrm{ph},2}}\right.\nonumber\\ 
              &&\quad\quad\left.+\dfrac{(\mathbf{e}_1\cdot \langle\Psi_o|\mathbf{r}|\Psi_n\rangle)(\langle\Psi_n|\mathbf{r}|\Psi_g\rangle\cdot \mathbf{e}_2)}{\hbar\omega_g-\hbar\omega_n +\hbar\omega_{\mathrm{ph},1}}\right\}\quad\quad 
\end{eqnarray}
In Eq.~\eqref{eq:peticolas2b}, 
$\mathbf{e}_j$ denotes the polarization direction (without specifying a certain frame of reference) of photon $j$ ($j=1,2$) with energy $\hbar\omega_{\mathrm{ph},j}$.
\end{subequations}
To shorten notation, the polarization independent quantity $\tilde{\mathcal{N}}_0(\omega_{\mathrm{ph}})$ 
in Eq.~\eqref{eq:probatlitypeticolas} contains all prefactors, 
\[ 
  \tilde{\mathcal{N}}_0(\omega_{\mathrm{ph}}) = \dfrac{2\pi e^4_0}{\hbar^3 c^2} (F_1\, \hbar\omega_{\mathrm{ph},1})\,I(\omega_{\mathrm{ph},2})\,,
\]
with $e_0$ being the elementary charge, and where $F_1$ and $I(\omega_{\mathrm{ph},2})$ refer to the incident laser-photon-flux (of type $1$) and the energy flux per unity frequency  (of type $2$), respectively~\cite{Peticolas1967}.
Evaluation of Eq.~\eqref{eq:peticolas2b} requires a frame transformation, since 
the wavefunctions involved in the two-photon transition matrices are known in the molecular frame whereas the polarization directions of the photons  are given in the laboratory frame of reference. 
As before, transformation of the polarization directions from the laboratory frame to the molecular frame is  
carried out  by means of the Wigner rotation matrices around the Euler angles $\omega = (\alpha,\beta,\gamma)$. Consequently, the orientation dependent two-photon absorption probability is obtained as 
\begin{subequations}
  \label{eq:newtensor}
  \begin{eqnarray}
\rho_{2\mathrm{P}}(\omega)
&=&
\left(\dfrac{8\pi^2\hbar}{3}\right)^2\tilde{\mathcal{N}}_0(\omega_{\mathrm{ph}})\left| \sum_{q_1,q_2}\mathcal{D}^{(1)}_{q_1,\rm\varrho_1}(\omega)
\mathcal{D}^{(1)}_{q_2,\rm\varrho_1}(\omega)\,
T_{q_1,q_2}\right|^2\,,\nonumber\\
\label{eq:density2}
\end{eqnarray}
where we have applied the properties of the rotation matrices between both
frames, detailed in Appendix~\ref{subsec:rotmat}, to Eq.~\eqref{eq:peticolas2b}. In Eq.~\eqref{eq:density2}, $T_{q_1,q_2}$ denotes 
the two-photon absorption tensor in the
molecular frame of reference, whose tensor elements reads,
\begin{eqnarray}
T_{q_1,q_2} &=& \sum_{n}\dfrac{\langle\Psi_o| r_{q_1}|n\rangle\langle n| r_{q_2}|\Psi_g\rangle}{\hbar\omega_g - \hbar\omega_n + \hbar\omega_{\mathrm{ph},2}}
     +  \dfrac{\langle\Psi_o| r_{q_2}|n\rangle\langle n| r_{q_1}|\Psi_g\rangle}{\hbar\omega_g - \hbar\omega_n + \hbar\omega_{\mathrm{ph},1}}\,.\nonumber\\
\label{eq:tensordef}
\end{eqnarray}
\end{subequations}
In Eq.~\eqref{eq:density2}, $\varrho_1$ denotes
the  polarization direction in the laboratory frame of reference, i.e. $\varrho_1 = \pm 1,0$, driving the two-photon absorption process, both photons
having the same polarization direction.
Additionally, the indexes $q_1$ and $q_2$ take the values $\pm 1,0$. Finally, $r_{q_k}$ denotes the spherical component of the
position operator $\hat{\mathbf{r}}$, with $q_k = \pm 1,0$.
The correspondence between the spherical and Cartesian components of $r_k$ are
detailed in Eq.~\eqref{eq:cartesian_spherical}. Hence, it is straightforward
to write $T_{q_1,q_2}$ in terms of the tensor elements written in the
Cartesian basis, $T_{og}^{\alpha\beta}(\omega_{\mathrm{ph}})$, for $\alpha,\beta=x,y,z$, cf. Eq.~\eqref{tanmoments}.
The correspondences are detailed in Eq.~\eqref{eq:tensor_def2}, in
Appendix~\ref{subsec:rotmat}.

 A further step consist of normalizing the
 probability density, such that the normalization condition,
\begin{eqnarray}
  \label{eq:normalization_condition}
  \int \rho_{2P}(\omega)\,d\omega = 1
\end{eqnarray}
is fulfilled. Using the properties of addition of angular momenta, it is
straightforward to find that the normalization factor reads, upon integration
of Eq.~\eqref{eq:density2} over the Euler angles,
\begin{subequations}
\begin{eqnarray}
  \tilde{\mathcal{N}}_0(\varrho_1) &=&\tilde{\gamma}(\omega_{\mathrm{ph}})\mathcal{B}(\varrho_1)
\end{eqnarray}
where we have defined,
\begin{widetext}
 \begin{eqnarray}
 \label{eq:normalization_factor}
 \mathcal{B}(\varrho_1) &=&\sum_{\substack{q_1,q_2\\q^\prime_1,q^\prime_2}} T_{q_1,q_2}\,T^{*}_{q^\prime_1,q^\prime_2}\sum_{\substack{\mathcal{Q}=0}}^2
 \left(2\mathcal{Q}+1 \right)
 \begin{pmatrix} 1 & 1 & \mathcal{Q}\vspace*{0.31cm}\\ q^\prime_1 & q^\prime_2 & -q^\prime_1-q^\prime_2 \end{pmatrix}
 \begin{pmatrix} 1 & 1 & \mathcal{Q}\vspace*{0.31cm}\\ \varrho_1 & \varrho_1 & -2\varrho_1 \end{pmatrix}
  \begin{pmatrix} 1 & 1 & \mathcal{Q}\vspace*{0.31cm}\\ q_1 & q_2 & -q^\prime_1-q^\prime_2 \end{pmatrix}
  \begin{pmatrix} 1 & 1 & \mathcal{Q}\vspace*{0.31cm}\\ \varrho_1 & \varrho_1 & -2\varrho_1 \end{pmatrix}\,,\nonumber\\
 \end{eqnarray}
 \end{widetext}
 \label{eq:prob_density1}
\end{subequations}
with  $\tilde{\gamma}(\omega_{\mathrm{ph}})\equiv (8\pi^2\hbar/3)^2\,\tilde{\mathcal{N}}_0(\omega_{\mathrm{ph}})$.
To retrieve Eqs.~\eqref{eq:normalization_factor}, we have made use of the properties
involving the product of two Wigner rotations matrices, as well as the
integration involving a product of three Wigner rotations matrices, and apply
them to Eq.~\eqref{eq:density2}. These properties are outlined in 
Eq.~\eqref{subeq:properties1:2} and Eq.~\eqref{eq:three_integral}, in  
in
Appendix~\ref{subsec:rotmat} and Appendix~\ref{subsec:X2+1}, respectively.

Finally, the orientation dependent probability density reads,
\begin{eqnarray}
  \label{eq:final_proba2p}
  \rho_{2P}(\omega) &=& \mathcal{N}_0(\varrho_1)
\left| \sum_{q_1,q_2}\mathcal{D}^{(1)}_{q_1,\rm\varrho_1}(\omega)
\mathcal{D}^{(1)}_{q_2,\rm\varrho_1}(\omega)\,
T_{q_1,q_2}\right|^2\,,\quad\quad
\end{eqnarray}
with $\mathcal{N}_0(\varrho_1) = \mathcal{B}^{-1}(\varrho_1)$.
In order to alleviate notations, and unless otherwise stated, we write $\mathcal{N}_0=\mathcal{N}_0(\varrho_1)$.
It is important to note, however, that in practice, computation of $\mathcal{N}_0$ is
not required, since this factor is common to all Legendre coefficients, and all
of them are given, as in the experiment~\cite{LuxACIE12,LuxCPC15}, normalized with respect to $c_0$.

Each component of the second-rank tensor $T_{q_1, q_2}$ determines
a property of the system, namely, the average transition rate. As a result of that 
the tensor $T_{q_1, q_2}$ has two types of symmetry properties. The first one is due to 
an intrinsic symmetry originated from the property itself.                          
For instance, $T_{q_1, q_2}$ defines the probability of a absorption of two identical 
photons. Since two photons of the same energy and polarization are not the same, 
$T_{q_1, q_2}$ has to be symmetric. The second type of symmetry comes from the geometric
symmetry of the molecule, and that specifies which of tensor components have to be 
zero~\cite{McCLAIN19771,Marco1983}.

In the isotropic case, $\rho_{2\mathrm{P}}(\alpha,\beta,\gamma)=1$, and 
evaluation of
Eq.~\eqref{eq:weighted} is analogous to integrating over 
Eq.~\eqref{eq:differential2}, resulting in the standard
expressions for the differential photoionization cross
section~\cite{Reid2003,ChandraJPhysB87,ChengPRA10,cooper,Yang1948,Janssen2014}:
If the weak probe photon is linearly polarized
($\epsilon^\prime_{\varrho_2}=\epsilon^\prime_0$),  
only $P_0$ and $P_2$ can become non-zero, whereas for circularly polarized
light, $P_0$, $P_1$ and $P_2$ can have non-vanishing
values. Moreover, the laboratory frame PAD preserves the cylindrical symmetry with
respect to the propagation direction of the light $z^\prime$, i.e.,
$\mu=\varrho_2-\varrho_2=0$ in Eq.~\eqref{eq:differential2}.

The situation changes if the probability to populate the
intermediate electronically excited state becomes anisotropic. 
If this probability depends 
on the initial orientation of the molecule, given in terms of the
Euler angles $\omega$ with respect to the laboratory frame
$\mathcal{R}^\prime$,  
the Wigner rotation matrices in Eq.~\eqref{eq:density2} 
couple to those in Eq.~\eqref{eq:differential2}. Upon
integration over the Euler angles in Eq.~\eqref{eq:weighted}, this 
gives rise to higher order Legendre polynomials in the PAD, as we show
now. 
To evaluate the angular momentum coupling in Eq.~\eqref{eq:weighted},
we  expand the norm squared in
Eq.~\eqref{eq:density2}. Making use of the product rule for 
Wigner rotation matrices, Eq.~\eqref{eq:density2} then becomes 
\begin{subequations}\label{eq:rho}
\begin{widetext}
\begin{eqnarray}
\label{eq:density3}
\rho_{2\mathrm{P}}(\omega) &=&\mathcal{N}_0\sum_{\substack{q_1,q_2 \\ q_3,q_4}}  
(-1)^{q_3+q_4}\,T_{q_1,q_2}T^{*}_{q_3,q_4} 
\sum^4_{K=0}\,g^{(K)}_{q_1,q_2,q_3,q_4}
\mathcal{D}^{(K)}_{s,0}(\omega)\,,
\end{eqnarray}
with $s=q_1+q_2-q_3-q_4$, and where we have defined
\begin{eqnarray}
\label{eq:gK}
g^{(K)}_{q_1,q_2,q_3,q_4}(\varrho_1) &=&
\sum^2_{Q=0}
\sum^2_{Q^\prime=0}
\sum^{Q+Q^\prime}_{K=|Q-Q^\prime|}
\gamma^{(K)}_{Q,\rm Q^\prime}    
\begin{pmatrix} 1 & 1 & Q \vspace*{0.33cm} \\ q_1 & q_2 & -q_1-q_2\end{pmatrix}
\begin{pmatrix} 1 & 1 & Q \vspace*{0.33cm} \\ \varrho_1 & \varrho_1 & -2\varrho_1 \end{pmatrix}\\\nonumber
&&\times
\begin{pmatrix} 1 & 1 & Q^\prime \vspace*{0.33cm} \\ q_3 & q_4 & -q_3-q_4\end{pmatrix}
\begin{pmatrix} 1 & 1 & Q^\prime \vspace*{0.33cm} \\ \varrho_1 & \varrho_1 & -2\varrho_1 \end{pmatrix}
\begin{pmatrix} Q & Q^\prime & K \vspace*{0.33cm} \\ q_1+q_2 & -q_3-q_4 & -s\end{pmatrix}
\begin{pmatrix} Q & Q^\prime & K \vspace*{0.33cm} \\ 2\varrho_1 & -2\varrho_1 & 0\end{pmatrix}
\end{eqnarray}
\end{widetext}
\end{subequations}
with $\gamma^{(K)}_{Q,\rm Q^\prime}=(2Q+1)(2Q^\prime+1)(2K+1)$.
In Eq.~\eqref{eq:density3}, the orientation dependence is contained
in $\mathcal{D}$, the polarisation dependence in $g$ and the dependence on
molecular parameters in $T$.
The derivation of Eqs.~\eqref{eq:rho}, 
employing the standard angular momentum  
algebra,  is presented in
Appendix~\ref{subsec:2Ptensor}. We make once more use of the product
rule for two rotation matrices, namely those 
involving the laser polarization in Eq.~\eqref{eq:differential2}, 
cf. Eq.~\eqref{subeq:property1}                                        
in Appendix~\ref{subsec:X2+1}.
Thus, a product of three rotation matrices is obtained 
when inserting Eqs.~\eqref{eq:rho} and~\eqref{eq:differential3_appendix}, 
into Eq.~\eqref{eq:weighted0}.
Evaluating the products of the Wigner $3j$ symbols, the
differential cross section, Eq.~\eqref{eq:weighted0},  
for a specific orientation $\omega$ of the molecule  becomes
\begin{subequations}\label{eq:diffXLF}
\begin{eqnarray}
\label{eq:diff1}
\frac{d^2\sigma_{2+1}}{d\omega d\Omega_{{\mathbf{k}}^\prime}} &=&
c_o\,
\sum^{\infty}_{\mathcal{L}=0}
\sum^{+\mathcal{L}}_{\mu=-\mathcal{L}}
b^{\mu}_\mathcal{L}(\omega)
P^{\mu}_{\mathcal{L}}(\cos{\vartheta^\prime_{k}})\,e^{i\mu\phi^{\prime}_{k}}\,,\quad\quad
\end{eqnarray}
where the only orientation-dependent quantity, 
$b^{\mu}_\mathcal{L}(\omega)$,  is given by 
\begin{eqnarray}
\label{eq:a_coef}
b^{\mu}_\mathcal{L}(\omega) &=&
\sum_{\lambda}
\kappa(\lambda)\,\,
\mathcal{D}^{K}_{s,\rm 0}(\omega)
\mathcal{D}^{\nu}_{q-q^\prime,\rm 0}(\omega)
\mathcal{D}^{\mathcal{L}}_{m^\prime-m,\rm -\mu}(\omega)\,.\quad\quad\quad
\end{eqnarray}
\end{subequations}
Note that the summation in Eq.~\eqref{eq:a_coef}
runs over all indices, except $\mathcal{L}$ and $\mu$, i.e., 
$\lambda=\{K,\nu,Q,Q^\prime,q,q^\prime,q_k,n_o,n^\prime_o,\ell,\ell^\prime,\ell_o,\ell^\prime_o\}$,
with $K=1,2,3,4$ and $\nu=0,1,2$ appearing from the coupling of the
first and second  
Wigner rotation matrices in Eq.~\eqref{eq:differential2},
c.f. Eq.~\eqref{eq:condense_1}. The specific form of 
$\kappa^\mu_{\mathcal{L}}(\lambda)$ is detailed in
Eq.~\eqref{eq:kappa_def}, in  Appendix~\ref{subsec:X2+1}.

We can now use the integral properties of a product
of three Wigner rotation
matrices~\cite{Varshalovich1988,irreducible,edmonds},  
c.f. Eq.~\eqref{eq:three_integral} in
Appendix~\ref{subsec:X2+1}. Integration of $b^{\mathcal{L}}_{\mu,\rm\nu}(\omega)$ 
over the Euler angles then yields 
\begin{eqnarray}
  \label{a1_integrated}
  c^{\mu}_{\mathcal{L},\rm\lambda}&=&
  \int b^{\mu}_{\mathcal{L},\rm\lambda}(\omega)\; d^3\omega\\
  &=&\sum_{\lambda}\kappa^\mu_{\mathcal{L}}(\lambda)
  \begin{pmatrix} K & \nu & \mathcal{L}\\s & q-q^\prime & m^\prime-m\end{pmatrix}
  \begin{pmatrix} K & \nu & \mathcal{L}\\0 & 0 & -\mu\end{pmatrix}\nonumber\\
  &=& \nonumber
  \sum_{\lambda}\kappa^\mu_{\mathcal{L}}(\lambda)
  \begin{pmatrix} K & \nu & \mathcal{L}\\s & q-q^\prime & m^\prime-m\end{pmatrix}
  \begin{pmatrix} K & \nu & \mathcal{L}\\0 & 0 & 0\end{pmatrix}\,\delta_{\mu,0}\,.
\end{eqnarray}
Note that the second Wigner symbol in the right-hand side of
Eq.~\eqref{a1_integrated} is non-zero only if $\mu=0$ and
$K+\nu+\mathcal{L}$ is even with $|K-\nu|\le\mathcal{L}\le K+\nu$. 
Because $\mu=0$, the terms depending on the azimuthal angle
in Eq.~\eqref{eq:differential2} do not contribute and
we retrieve cylindrical symmetry for the PAD of
Eq.~\eqref{eq:weighted} which can thus be expressed in terms of
Legendre polynomials. Furthermore, according to the fifth and sixth
Wigner symbols in 
Eq.~\eqref{eq:gK}$, K=0,\ldots,4$, because $|Q-Q^\prime|\le K\le Q+Q^\prime$, and
$0\le Q\le2$ according to the first and second Wigner symbols in
Eq.~\eqref{eq:gK}. The same applies to $Q^\prime$, reflecting the
addition of angular momentum in a two-photon absorption process.

Making use, in Eq.~\eqref{a1_integrated}, of  the fact that 
the non-zero contributions for $\nu$ are given by $\nu=0,1,2$,
c.f. Eq.~\eqref{eq:condense_1}, one obtains that $\mathcal{L}$
runs from $0$ to $6$, and higher orders give only vanishing  
contributions. Therefore, the highest order
Legendre polynomial that contributes to the PAD is 
$\mathcal{L}_{\mathrm{max}}=6$, as expected for a 2+1 process from the
$2(m+n)-1$ rule~\cite{Reid2003}.

Finally, evaluating Eq.~\eqref{eq:weighted} with the help of 
Eq.~\eqref{a1_integrated} yields the experimentally measured PAD that
is obtained for an initial ensemble of  randomly oriented
molecules, 
\begin{subequations}\label{eq:PADfinal}
\begin{eqnarray}
\label{eq:final_LFPAD}
\frac{d\sigma_{2+1}}{d{\Omega_{\mathbf{k}^\prime}}} &=&
\sum^{6}_{\mathcal{L}=0}\, c_{\mathcal{L}}\,
P_{\mathcal{L}}\left(\cos{\vartheta^\prime_{\mathbf{k}}}\right)\,, 
\end{eqnarray}
with coefficients 
\begin{widetext}
\begin{eqnarray}
\label{eq:final_coeff}
c_{\mathcal{L}}({\varrho_1},{\varrho_2}) &=& \tilde{c}_o\,\mathcal{N}_0 \sum_{\substack{\ell,m \\ n_o,\ell_o,m_o}}\, 
\sum_{\substack{\ell^\prime,m^\prime \\ n^\prime_o\ell^\prime_o,m^\prime_o}}\, \sum_{q,q^\prime}
\sum_{\substack{q_1,q_2 \\ q_3,q_4}}
\sum^2_{\nu=0} \sum^4_{K=0} 
(-1)^{q3+q4}\,(2\nu+1)(2\mathcal{L}+1)
a^{\ell_o}_{m_o}(n_o)\, a^{*\ell^\prime_o}_{m^\prime_o}(n^\prime_o)\, T_{q_1,q_2} T^{*}_{q_3,q_4}
\\\nonumber
&&\times (-\mathrm{i})^{\ell-\ell^\prime}\,
(-1)^{m^\prime-q-\varrho_2}\,e^{\mathrm{i}(\delta_{\ell}-\delta_{\ell^\prime})}\,\,g^{(K)}_{q_1,q_2,q_3,q_4}(\varrho_1)
\,\,I^{n_o}_{_{k}}(\ell,\ell_o)\,\, I^{n^\prime_o}_{_{k}}(\ell^\prime,\ell^\prime_o)\,\,\mathcal{S}^{\ell,m}_{\ell_o,m_o}(q) 
\,\,\mathcal{S}^{\ell^\prime,m^\prime}_{\ell^\prime_o,m^\prime_o}(q^\prime)\,\hat{\varsigma}(\ell,\ell^\prime)\,\\\nonumber 
&&
\times
\begin{pmatrix} \ell & \ell^\prime & \mathcal{L}\vspace*{0.33cm}\\ m & -m^\prime & m^\prime-m\end{pmatrix}
\begin{pmatrix} \ell & \ell^\prime & \mathcal{L}\vspace*{0.33cm}\\ 0 & 0 & 0 \end{pmatrix}
\begin{pmatrix} 1 & 1 & \nu\vspace*{0.33cm}\\ q & -q^\prime & q^\prime-q \end{pmatrix}
\begin{pmatrix} 1 & 1 & \nu\vspace*{0.33cm}\\ \varrho_2 & -\varrho_2 & 0 \end{pmatrix}
\begin{pmatrix} K & \nu & \mathcal{L}\vspace*{0.33cm}\\s & q-q^\prime & m^\prime-m\end{pmatrix}
\begin{pmatrix} K & \nu & \mathcal{L}\vspace*{0.33cm}\\0 & 0 & 0\end{pmatrix}\,.
\end{eqnarray}
\end{widetext}
\end{subequations}
with $\tilde{c}_o=4\pi c_o$, and $\hat{\varsigma}(\ell,\ell^\prime)= \sqrt{(2\ell+1)(2\ell^\prime+1)}$. Derivation
of Eq.~\eqref{eq:PADfinal} is explicitly detailed in Appendix~\ref{subsec:X2+1}. Note that the coefficients
$c_{\mathcal{L}}(\varrho_1,\varrho_2)$ depend on
the expansion coefficients $a_{m_o}^{\ell_o}(n_o)$ describing the 
intermediate electronically exited state, the two-photon absorption
tensor elements, $T_{q_1,q_2}$, and the laser polarization directions of the
two-photon absorption step, $\varrho_1$, and of the one-photon
ionization, $\varrho_2$.

We would like to emphasize
that the contribution of Legendre polynomials with order higher
than 2  in Eq.~\eqref{eq:PADfinal} is due to the 
orientation dependence of populating the intermediate electronically
excited state by two-photon absorption from the electronic ground
state. 
That is, the density $\rho(\omega)$ expresses the fact that molecules
with a certain orientation $\omega=\omega_1$ have a larger probability to
undergo  non-resonant two-photon absorption than molecules with
some other orientation $\omega=\omega_2$. So although  the
molecules are assumed to be completely randomly oriented with respect
to the laser beam axis when they are in their electronic ground state,
an effective alignment 
results for those molecules that absorb two photons. This effective
alignment results from selection of certain orientations rather than
rotational dynamics which would occur on a much slower timescale. 
The contribution of higher order Legendre polynomials to the PAD is
then entirely determined by the properties of the two-photon absorption
tensor and the electronically excited state. In order to interpret the 
experimentally observed PADs for fenchone and camphor
in terms of their expansion in Legendre
polynomials, at least qualitatively, we 
estimate $a_{m_o}^{\ell_o}(n_o)$ and $T_{q_1,q_2}$ using \textit{ab initio}
calculations or via fitting. 
Before presenting the corresponding details  in Sec.~\ref{sec:abinitio}, we
discuss below the basic symmetry properties of these parameters of our
model as well as  the dependence on the laser
polarization directions $\varrho_1$, $\varrho_2$.

\subsection{PECD and symmetry}
\label{subsec:parity}
By definition, PECD is obtained if the sign of the odd Legendre
coefficients change when the helicity of the electric field changes. 
Analogously, for fixed electric field helicity, the odd Legendre
coefficients change sign when enantiomers are interchanged. 
We therefore first inspect sign changes in the Legendre
coefficients for molecules of opposite handedness within our 
one-center expansion framework.
The relation between a given enantiomer and its mirror image is given
by the parity operator, which changes the
coordinates $\mathbf{r}$ to $-\mathbf{r}$.
We therefore check, in the following, that our model transforms properly
under parity. 

Moreover, we determine the role that the excited state
coefficients $a^{\ell_o}_{m_o}(n_o)$ and two-photon absorption tensor
elements play for each Legendre 
coefficient that contributes to the PAD. To this end, we rewrite 
Eq.~\eqref{eq:final_coeff}, expressing each
$c_{\mathcal{L}}({\varrho_1},{\varrho_2})$ 
explicitly in terms of the $a^{\ell_o}_{m_o}(n_o)$ and $T_{q,q^\prime}$,
\begin{eqnarray}
\label{eq:final_cj}
c_{\mathcal{L}}({\varrho_1},{\varrho_2})&=&
\sum_{\substack{n_o,\ell_o,m_o\\ n^\prime_o,\ell^\prime_o,
m^\prime_o}}\sum_{\substack{q_1,q_2 \\ q_3,q_4}}\, 
\gamma^{n_o,\ell_o,m,n^\prime_o,\ell^\prime_o,m^\prime_o}_{q_1,q_2,q_3,q_4}(\mathcal{L},\epsilon^\prime_{\varrho_1},\epsilon^\prime_{\varrho_2})\nonumber\\         
&&\times
a^{\ell_o}_{m_o}(n_o)\,a^{*\ell^\prime_o}_{m^\prime_o}(n^\prime_o)\,
T_{q_1,q_2}\, T^{*}_{q_3,q_4}\nonumber\\
\end{eqnarray}
Equation~\eqref{eq:final_cj}
allows for determining  each Legendre coefficient 
as a function of the 
intermediate electronically excited state via $a^{\ell_o}_{m_o}(n_o)$  
and  $T_{q,q^\prime}$, i.e., it connects the measured Legendre
coefficients to the electronic structure properties. 
We can thus compare the contribution of different
$a^{\ell}_{m_o}(n_o)$ to different Legendre coefficients $c_{\mathcal{L}}$, and explain 
differences, observed e.g. for different molecules, in
terms of the electronic structure. This is important because 
investigation of camphor and fenchone revealed, for example, 
the same order of magnitude for the first and third Legendre
coefficient in camphor, in contrast to fenchone where 
$c_3$ is about one order of magnitude smaller than
$c_1$~\cite{LuxACIE12,LuxCPC15}. This 
observation suggests a significantly different electronic structure
despite the fact that  the two bicyclic
monoketones are constitutional isomers
which differ only in the position of the geminal methyl
groups~\cite{PollmannSpectro97}.   

In the following, 
we discuss the behavior under parity and the contribution of the 
$a^{\ell_o}_{m_o}(n_o)$ and $T_{q,q^\prime}$ to the 
$c_{\mathcal{L}}({\varrho_1},{\varrho_2})$ separately for the excited state coefficients, the 
two-photon absorption tensor and the laser polarization. 

\subsubsection{Role of the excited state expansion coefficients}
\label{subsec:role_expansion_coefficients}
In this section, we  explicitly show that our single-center expansion for the
$(2+1)$ REMPI process properly transforms under parity. Note that 
the two-photon absorption process conserves parity, which implies 
that exchanging enantiomers results in a parity change  of the
expansion coefficients of the intermediate electronically excited state,
from $a^{\ell_o}_{m_o}(n_o)$ to
$(-1)^{\ell_o}\,a^{\ell_o}_{m_o}(n_o)$. For practical convenience, we
define the following quantity present in 
Eq.~\eqref{eq:final_coeff} depending on $\ell_o$ and $m_o$,
\begin{eqnarray}
  \label{eq:S_parity1}
  \mathcal{P}_{\mathcal{L}}&=& a^{\ell_o}_{m_o}(n_o) a^{\ell^\prime_o}_{m^\prime_o}(n^\prime_o)\mathcal{S}^{\ell,m}_{\ell_o,m_o}(q) 
  \,\,\mathcal{S}^{\ell^\prime,m^\prime}_{\ell^\prime_o,m^\prime_o}(q^\prime)
  \begin{pmatrix} \ell & \ell^\prime & \mathcal{L}\vspace*{0.33cm}\\ 
    0 & 0 & 0\end{pmatrix}\,.\nonumber\\
\end{eqnarray}
Upon application of the parity operator, 
Eq.~\eqref{eq:S_parity1} becomes 
\begin{eqnarray}
  \label{eq:S_parity11}
  \tilde{\mathcal{P}}_{\mathcal{L}}&=& (-1)^{\ell_o+\ell^\prime_o}\,a^{\ell_o}_{m_o}(n_o) a^{\ell^\prime_o}_{m^\prime_o}(n^\prime_o)\nonumber\\
 &&
  \times\,
  \mathcal{S}^{\ell,m}_{\ell_o,m_o}(q) 
  \,\mathcal{S}^{\ell^\prime,m^\prime}_{\ell^\prime_o,m^\prime_o}(q^\prime)
  \begin{pmatrix} \ell & \ell^\prime & \mathcal{L}\vspace*{0.33cm}\\ 
    0 & 0 & 0\end{pmatrix}\,.\nonumber\\
\end{eqnarray}
Furthermore, we make use of the following property of the Wigner $3j$
symbols~\cite{irreducible,edmonds,Hans1957,Varshalovich1988}, 
\begin{eqnarray}
\label{eq:wigner_prop1}
\begin{pmatrix} j & j^\prime & J \vspace*{0.33cm} \\ m & m^\prime & M \end{pmatrix}
=(-1)^{j+j^\prime+J}
\begin{pmatrix} j & j^\prime & J \vspace*{0.33cm} \\ -m & -m^\prime &
  -M \end{pmatrix}\,,\quad
\end{eqnarray}
and apply it to the first Wigner $3j$ symbol in the expressions for 
$\mathcal{S}^{\ell,m}_{\ell_o,m_o}(q)$ and $\mathcal{S}^{\ell^\prime,m^\prime}_{\ell^\prime_o,m^\prime_o}(q^\prime)$, i.e.
Eq.~\eqref{eq:S}, containing triple zeros in the second row.   
The parity-transformed $\mathcal P_{\mathcal{L}}$ thus becomes
\begin{eqnarray}
  \label{eq:S_tilde2}
  \tilde{\mathcal{P}}_{\mathcal{L}}&=&
  (-1)^{\ell_o+\ell^\prime_o}\, (-1)^{\ell+\ell_o+\ell^\prime+\ell^\prime_o}\,\nonumber\\
  &&\times\,\mathcal{S}^{\ell,m}_{\ell_o,m_o}(q) 
  \,\,\mathcal{S}^{\ell^\prime,m^\prime}_{\ell^\prime_o,m^\prime_o}(q^\prime)
  \begin{pmatrix} \ell & \ell^\prime & \mathcal{L}\vspace*{0.33cm}\\ 0 & 0 & 0\end{pmatrix}\,.
\end{eqnarray}
Applying Eq.~\eqref{eq:wigner_prop1} once more to the Wigner $3j$
symbol in Eq.~\eqref{eq:S_tilde2} allows for eliminating the
explicit dependence of $\tilde{\mathcal{P}}_{\mathcal{L}}$
on the partial waves $\ell$ and $\ell^\prime$, 
\begin{eqnarray}
  \label{eq:parity_coeff}
  \tilde{\mathcal{P}}_{\mathcal{L}}&=&
  (-1)^{\ell_o+\ell^\prime_o}\,(-1)^{\ell+\ell_o+\ell^\prime+\ell^\prime_o} 
  \mathcal{S}^{\ell,m}_{\ell_o,m_o}(q) 
\,\mathcal{S}^{\ell^\prime,m^\prime}_{\ell^\prime_o,m^\prime_o}(q^\prime)\nonumber\\
&&\times
(-1)^{\ell+\ell^\prime+\mathcal{L}}
\begin{pmatrix} \ell & \ell^\prime & \mathcal{L}\vspace*{0.33cm}\\ 0 & 0 & 0\end{pmatrix}\nonumber\\
                     &=&
(-1)^{\mathcal{L}}
\mathcal{S}^{\ell,m}_{\ell_o,m_o}(q) 
\,\mathcal{S}^{\ell^\prime,m^\prime}_{\ell^\prime_o,m^\prime_o}(q^\prime)
\begin{pmatrix} \ell & \ell^\prime & \mathcal{L}\vspace*{0.33cm}\\ 0 & 0 & 0\end{pmatrix}\nonumber\\
                     &=&
(-1)^{\mathcal{L}}\mathcal{P}_{\mathcal{L}}\,.
\end{eqnarray}
Because $\mathcal P_{\mathcal{L}}$ and $\tilde{\mathcal{P}}_{\mathcal{L}}$
refer, by construction, to enantiomers of opposite
handedness, Eq.~\eqref{eq:parity_coeff} implies a change of sign for 
$\mathcal{L}$ odd, cf. Eq.~\eqref{eq:PADfinal}, when interchanging
enantiomers, and no sign change for $\mathcal{L}$ even. 
Our model properly reproduces this basic symmetry behavior.
The corresponding behavior under change of the light helicity, keeping
the same enantiomer, is checked below in
Sec.~\ref{subsec:role_polarization}.

Next we check the dependence of the non-zero Legendre coefficients contributing
to the PAD on the maximum order $L_{o,\mathrm{max}}$ of the excited 
state coefficients, $a^{\ell_o}_{m_o}(n_o)$, cf. Eq.~\eqref{eq:exited_state2}.
According to Equation~\eqref{eq:final_coeff}, a non-zero projection of
the electronically excited state onto 
$d$-orbitals ($\ell_o=2$) is required 
to ensure that higher orders $c_{\mathcal{L}}$  are non-zero.
In fact, an additional requirement to reach
$\mathcal{L}_{\mathrm{max}}=6$ 
is that $L_{o,\mathrm{max}}\ge 2$. This is straightforward to see by
inspecting the term 
\begin{eqnarray*}
\begin{pmatrix} \ell & \ell^\prime & \mathcal{L}\vspace*{0.33cm}\\ 0 & 0 & 0\end{pmatrix}
\end{eqnarray*}
in Eq.~\eqref{eq:final_coeff}, 
defining the PAD for a $(2+1)$ REMPI process. This term vanishes 
unless $\ell+\ell^\prime+\mathcal{L}$ is even and
$|\ell-\ell^\prime|\le\mathcal{L}\le\ell+\ell^\prime$. In order to 
reach $\mathcal{L}_{\mathrm{max}}=6$, the minimal requirement in terms
of the angular momentum for the continuum 
wavepacket is $\ell_{\mathrm{max}}=3$. Together with the selection rule
$\ell_{\mathrm{max}}=L_{o,\mathrm{max}}+1$, cf. Eq.~\eqref{eq:S}, this implies 
$L_{o,\mathrm{max}}=2$, i.e., presence of $d$-waves in the 
resonantly excited state. Note that a contribution from 
higher partial waves only modifies the algebraic value of the Legendre
coefficients, but does not lead to higher orders because, as we have
already pointed out, the maximal order of the Legendre coefficients 
is also limited by the term 
\begin{eqnarray*}
\begin{pmatrix} K & \nu & \mathcal{L}\vspace*{0.33cm}\\
0 & 0 & 0\end{pmatrix}
\end{eqnarray*}
in Eq.~\eqref{eq:final_coeff}. 

Perhaps even more interestingly, for circular polarization direction
($\varrho_1=\varrho_2=\pm 1$),  $c_5$ vanishes if the 
projection of the electronically excited state onto $\ell_o=3$
is zero. In other words, expansion of the electronically  
excited state in terms of $s$, $p$ and $d$ orbitals results in
non-zero Legendre coefficients $c_\mathcal{L}$ for $\mathcal{L}$ up to
6, except for $c_5$. In fact, we found $c_5$ to appear only in
presence of a non-vanishing
contribution of $f$ orbitals. This does not result from selection
rules as discussed before, but rather from an accidental compensation
of terms in the summations in Eq.~\eqref{eq:final_coeff} which arises
from the central symmetry of our single center basis functions. 

Given the experimental observation of Ref.~\cite{LuxACIE12,LuxCPC15},  
we expect the electronically excited state for fenchone and
camphor to have non-vanishing projections onto $s$-, $p$-, $d$- and
possibly $f$-orbitals.  Also, the eventual expansion coefficients of the 
electronically excited state will most likely be different for
fenchone and camphor to account for the different ratios of $c_3$ and
$c_1$ observed for the two molecules~\cite{LuxACIE12,LuxCPC15}.

\subsubsection{Role of Polarizations $\varrho_1$ and $\varrho_2$}
\label{subsec:role_polarization}
Having shown sign inversion for the odd Legendre coefficients for
enantiomers of opposite handedness and a fixed circular polarization
direction, we outline,  in the following, an analogous symmetry
property that is relevant when considering the same enantiomer but
inverting the 
polarization direction. By definition, PECD  requires all odd
Legendre  expansion coefficients for a given enantiomer to change sign
when changing circular  
polarization from left to right, and vice versa. 
In order to show that our approach also properly reproduces this 
behavior, we employ again the symmetry
properties of the Wigner $3j$ symbols in
Eq.~\eqref{eq:final_coeff}, similarly to 
Sec.~\ref{subsec:role_expansion_coefficients}.
For the sake of completeness, we consider the general case of
independent polarizations for the two-photon absorption and the 
one-photon ionization processes.

First, we consider all terms in Eq.~\eqref{eq:final_coeff} 
depending on $\epsilon^\prime_{\varrho_2}$. We apply Eq.~\eqref{eq:wigner_prop1}
to the fourth and sixth Wigner $3j$ symbol in Eq.~\eqref{eq:final_coeff} for
$c_{\mathcal{L}}(-\varrho_1,-\varrho_2)$. This yields 
\begin{subequations}\label{eq:factors}
\begin{eqnarray}
\label{eq:factor2}
\begin{pmatrix} 1 & 1 & \nu \vspace*{0.33cm} \\ -\varrho_2 &
  +\varrho_2 & 0 \end{pmatrix}  
  =   (-1)^{2+\nu}
\begin{pmatrix} 1 & 1 & \nu \vspace*{0.33cm} \\ \varrho_2 & -\varrho_2
  & 0 \end{pmatrix} 
\end{eqnarray}
for the fourth  Wigner $3j$ symbol, and 
\begin{eqnarray}
\label{eq:factor3}
\begin{pmatrix} K & \nu & \mathcal{L} \vspace*{0.33cm} \\ 0 & 0 &
  0 \end{pmatrix}  
  =   (-1)^{K+\nu+\mathcal{L}}
\begin{pmatrix} K & \nu & \mathcal{L} \vspace*{0.33cm} \\ 0 & 0 &
  0 \end{pmatrix} 
\end{eqnarray}
for the sixth Wigner $3j$ symbol in Eq.~\eqref{eq:final_coeff} when the
polarization direction driving the ionization proceess is
$-\varrho_2$. Next, we evaluate the expression containing the
information about the polarization direction 
driving the two-photon absorption process. For
$\epsilon_{-\varrho_1}$, the term $g^K_{\varrho_1}(q_1,q_2,q_3,q_4)$,
defined in Eq.~\eqref{eq:gK}, reads 
  \begin{eqnarray}
  \label{eq:gK_tilde}
    g^K_{-\varrho_1}(q_1,q_2,q_3,q_4) = (-1)^{K}\,g^K_{+\varrho_1}(q_1,q_2,q_3,q_4)\,,\quad
\end{eqnarray}
\end{subequations}
when changing $\varrho_1$ to $-\varrho_1$. In Eq.~\eqref{eq:gK_tilde},
we have applied Eq.~\eqref{eq:wigner_prop1} to the second,
fourth and sixth  Wigner $3j$ symbols in Eq.~\eqref{eq:gK}. 
The Legendre coefficient $c_{\mathcal{L}}(-\varrho_1,-\varrho_2)$
involves, according to Eq.~\eqref{eq:final_coeff}, 
the triple product of Eqs.~\eqref{eq:factors}, that is,
\begin{widetext}
\begin{eqnarray}
  \label{eq:gK_sym}
g^K_{-\varrho_1}(q_1,q_2,q_3,q_4)  
\begin{pmatrix} 1 & 1 & \nu \vspace*{0.33cm} \\ -\varrho_2 & +\varrho_2 & 0 \end{pmatrix} 
\begin{pmatrix} K & \nu & \mathcal{L} \vspace*{0.33cm} \\ 0 & 0 & 0 \end{pmatrix} 
                  &=&
(-1)^{\mathcal{L}}
g^K_{+\varrho_1}(q_1,q_2,q_3,q_4)  
\begin{pmatrix} 1 & 1 & \nu \vspace*{0.33cm} \\ +\varrho_2 & -\varrho_2 & 0 \end{pmatrix} 
\begin{pmatrix} K & \nu & \mathcal{L} \vspace*{0.33cm} \\ 0 & 0 & 0 \end{pmatrix} \,.
\end{eqnarray}
\end{widetext}
This implies, according to Eq.~\eqref{eq:final_coeff},
\begin{eqnarray}
\label{eq:c_sym}
c_{\mathcal{L}}(-\varrho_1,-\varrho_2) = (-1)^{\mathcal{L}}\, c_{\mathcal{L}}(+\varrho_1,+\varrho_2)\,,
\end{eqnarray}
i.e., indeed,  only  odd Legendre coefficients change
sign  when changing simultaneously the
polarization directions $\varrho_1$ and $\varrho_2$, whereas all even
coefficients remain unchanged.

Next, we evaluate all non-vanishing Legendre coefficients as a function
of the polarization directions $\varrho_1$ and $\varrho_2$ without
making any assumptions on the two-photon absorption tensor $T$.
To this end, we first consider the case where the two-photon
absorption process is 
driven by  linearly polarized light, $\varrho_1=0$. 
The second  Wigner $3j$ symbol in Eq.~\eqref{eq:gK} then becomes 
\begin{eqnarray*}
\label{eq:sigma0}
\begin{pmatrix} 1 & 1 & Q \vspace*{0.33cm} \\ \varrho_1 & \varrho_1 & -2\varrho_1 \end{pmatrix}
                  &=&
\begin{pmatrix} 1 & 1 & Q \vspace*{0.33cm} \\ 0 & 0 & 0 \end{pmatrix}\,.
\end{eqnarray*}

It does not vanish if and only if  $Q=0,2$; 
and analogously for the fourth Wigner symbol in Eq.~\eqref{eq:gK}
involving $Q^\prime$. Furthermore, the sixth Wigner $3j$
symbol in Eq.~\eqref{eq:gK} becomes
\begin{eqnarray*}
\begin{pmatrix} Q & Q^\prime & K \vspace*{0.33cm} \\ 0 & 0 & 0 \end{pmatrix}\,,
\end{eqnarray*}
which is non-zero only if $K$ is even, because $Q$ and $Q^\prime$ are
even, and $|Q-Q^\prime| \le K\le Q+Q^\prime$.  As a consequence,
because both $Q$ and $Q^\prime$ are 
restricted to $0$ and $2$, $K$ must be equal to 
0, 2 or 4.
Now, we consider the fourth Wigner $3j$ symbol in
Eq.~\eqref{eq:final_coeff}, namely 
\begin{eqnarray}
\label{eq:sigma02}
\begin{pmatrix} 1 & 1 & \nu\vspace*{0.33cm}\\ \varrho_2 & -\varrho_2 & 0 \end{pmatrix}\,,
\end{eqnarray}
which contains the information about the photoionization transition.
If the photoionization process is driven by linearly polarized light
($\varrho_2=0$), 
the allowed values for $\nu$ in Eq.~\eqref{eq:sigma02} are
$\nu=0,2$. Therefore, the last Wigner symbol in Eq.~\eqref{eq:final_coeff},
\begin{eqnarray}
\label{eq:sigma03}
\begin{pmatrix} K & \nu & \mathcal{L}\vspace*{0.33cm}\\0 & 0 &
  0\end{pmatrix}\,, 
\end{eqnarray}
has non-vanishing values only for $|K-\nu|\le\mathcal{L}\le K+\nu$ and
$K+v+\mathcal{L}$ must be even due to the triple zeros in the second
row. Because $K=[0,2,4]$ for $\varrho_1=0$  and $\nu=0,2$ for
$\varrho_2=0$, the maximal order of Legendre coefficients is
$\mathcal{L}_{\mathrm{max}}=6$ and the non-vanishing Legendre
coefficients are those 
for $\mathcal{L}=0,2,4,6$, i.e., there are no odd Legendre polynomials
in the PAD for $\varrho_1=\varrho_2=0$.

On the other hand, if we keep $\varrho_1=0$ but the photoionization
transition is driven by circularly polarized light
($\varrho_2=\pm 1$), the non-vanishing
values in Eq.~\eqref{eq:sigma02} are not anymore restricted to even 
$\nu$, but instead to  $\nu=0,1,2$. Using these values for $\nu$
together with the requirement
$|K-\nu|\le\mathcal{L}\le K+\nu$ in Eq.~\eqref{eq:sigma03}, we obtain, 
for $K=0,2,4$ (due to $\varrho_1=0$), even as well as odd Legendre
polynomials in the PAD, i.e., $\mathcal{L}=0,1,\ldots,6$. 
Next we check whether PECD can arise, i.e., whether the non-zero odd
coefficients change sign under changing the light helicity, for
$\varrho_1=0$ and $\varrho_2=\pm 1$. To 
this end, we  explicitly write out the dependence of 
Eq.~\eqref{eq:final_coeff} on the polarization direction $\varrho_2$
driving the ionization step and define
 \begin{subequations}
 \begin{eqnarray}
 \label{eq:final_coeff_reduced1}
 \zeta^{K,\nu}_{\mathcal{L}}(\varrho_2) &=& 
 \begin{pmatrix} 1 & 1 & \nu\vspace*{0.33cm}\\ \varrho_2 & -\varrho_2 & 0 \end{pmatrix}
 \begin{pmatrix} K & \nu & \mathcal{L}\vspace*{0.33cm}\\0 & 0 &
   0\end{pmatrix}\,, 
 \end{eqnarray}
 corresponding to the fourth and sixth Wigner $3j$ symbol in
 Eq.~\eqref{eq:final_coeff}. For the opposite polarization direction
 $-\varrho_2$, this quantity becomes
 \begin{eqnarray}
 \label{eq:final_coeff_reduced2}
 \zeta^{K,\nu}_{\mathcal{L}}(-\varrho_2) &=& 
 \begin{pmatrix} 1 & 1 & \nu\vspace*{0.33cm}\\ -\varrho_2 & \varrho_2 & 0 \end{pmatrix}
 \begin{pmatrix} K & \nu & \mathcal{L}\vspace*{0.33cm}\\0
                   & 0 & 0\end{pmatrix}\nonumber\\
 &=&
 (-1)^{2\nu+K+\mathcal{L}}
 \begin{pmatrix} 1 & 1 & \nu\vspace*{0.33cm}\\ \varrho_2 & -\varrho_2 & 0 \end{pmatrix}
 \begin{pmatrix} K & \nu & \mathcal{L}\vspace*{0.33cm}\\0
                  & 0 & 0\end{pmatrix}\nonumber\\
 &=&
 (-1)^{\mathcal{L}}\, \zeta^{K,\nu}_{\mathcal{L}}(\varrho_2)\,,
 \label{eq:c_linear_circular}
 \end{eqnarray}
 \end{subequations}
where we have applied Eq.~\eqref{eq:wigner_prop1} to both Wigner $3j$ symbols
in Eq.~\eqref{eq:final_coeff_reduced2},
together with the fact that $K$ is even for $\varrho_1=0$, as previously 
discussed. Finally, inserting Eq.~\eqref{eq:c_linear_circular} 
into Eq.~\eqref{eq:final_coeff} yields
\begin{eqnarray}
c_{\mathcal{L}}(\varrho_1=0,-\varrho_2) = (-1)^{\mathcal{L}}
c_{\mathcal{L}}(\varrho_1=0,+\varrho_2)\,.
\end{eqnarray}
As a consequence, also for linearly polarized light driving the two-photon
absorption process, odd Legendre
coefficients change sign when the polarization direction of the
ionizing field is changed from right to left, and vice versa.
Whereas  $K$ must be even for $\varrho_1=0$, 
$\nu$ is $\nu=0,1,2$ for $\varrho_2=\pm1$,
allowing $\mathcal{L}$ to take odd and even values in
Eq.~\eqref{eq:c_linear_circular}. This implies that there is no need for
circular polarization to drive the two-photon absorption
process: Two-photon absorption driven by linearly polarized light
followed by photoionization with circularly polarized light is
sufficient for observing PECD in chiral molecules.
In Section~\ref{subsec:role_tensor} we investigate the specific role of
the two-photon aborption tensor for all the cases discussed above.
Conversely, the two-photon transition may be driven by circularly
polarized light followed by photoionization with linearly polarized
light, i.e.,  $\varrho_1=\pm 1$ and $\varrho_2=0$. As shown in
Eq.~\eqref{eq:demo_finished:0} in 
Appendix~\ref{subsec:ci:0}, such a configuration leads to a PAD consisting 
exclusively of even Legendre contributions.

In Eq.~\eqref{eq:c_sym} we have shown that only odd Legendre
coefficients change sign when changing simultaneously the polarization
direction driving the two-photon absorption and the one-photon
ionization. In
Appendix~\ref{subsec:ci},  we show that 
\begin{eqnarray}
  \label{eq:ci_rho2}
  c_{\mathcal{L}}(\varrho_1,\varrho_2) 
  = (-1)^{\mathcal{L}}c_{\mathcal{L}}(\varrho_1,-\varrho_2)\,,  
\end{eqnarray}
i.e., odd Legendre coefficients change sign when the polarization
direction of the photoionization transition is changed, whereas 
the polarization of the field driving the two-photon absorption is
kept fixed. This suggests the polarization direction of the ionizing
field alone to impose the sign for all odd Legendre 
coefficients; the polarization direction in the two-photon
absorption process plays no role. To verify this statement, we calculate
 $c_{\mathcal{L}}(-\varrho_1,\varrho_2)$ in Appendix~\ref{subsec:ci2}
and find indeed
\begin{eqnarray}
\label{eq:main_sym_result2}
c_{\mathcal{L}}(-\varrho_1,\varrho_2) = c_{\mathcal{L}}(+\varrho_1,\varrho_2)\,.
\end{eqnarray}
That is, the two-photon process determines only the degree of
anisotropy prior to ionization. 

To summarize, using linearly polarized light for both two-photon
absorption and one-photon ionization results in a PAD consisting only
of even Legendre polynomials, i.e., vanishing PECD. In contrast, when
the $(2+1)$ REMI process is driven by circularly polarized light,
higher order odd Legendre polynomials may contribute, depending on the
geometric properties of the resonantly excited state. The occurrence
of non-zero Legendre coefficients for all  polarization combinations
is summarized in  Table~\ref{table:all_in_one} below. 

\subsubsection{Role of two-photon absorption tensor}
\label{subsec:role_tensor}
The number of Legendre coefficients that contribute to  PECD in our
model of the 2+1 REMPI process is determined by how anisotropic the
ensemble of electronically excited molecules is. This, in turn, follows
from the properties of the two-photon absorption tensor. Here, we
check the conditions that $T_{q_1,q_2}$,
in order to give rise to this anisotropy. To this end, 
we introduce the two-photon absorption amplitude
$\mathcal{A}_{2\mathrm{P}}(\omega)$, 
where for convenience the multiplying factor in Eq.~\eqref{eq:final_proba2p} has been dropped,
\begin{eqnarray}
  \mathcal{A}_{2\mathrm{P}}(\omega) &=& \sum_{q_1}\sum_{q_2}
\mathcal{D}^{(1)}_{q_1,\varrho_1}(\omega)\,
\mathcal{D}^{(1)}_{q_2,\varrho_1}(\omega)\,
T_{q_1,q_2} \,,
\end{eqnarray}
i.e.,
$\rho_{2\mathrm{P}}(\omega)\propto|\mathcal{A}_{2\mathrm{P}}(\omega)|^2$,
cf. Eq.~\eqref{eq:final_proba2p}. For simplicity, we define
$\tilde{\mathcal{A}}_{2\mathrm{P}}(\omega)$
such that $\mathcal{A}_{2\mathrm{P}}(\omega) =
\frac{4\pi}{3} \tilde{\mathcal{A}}_{2\mathrm{P}}(\omega)$.
We first check the 'trivial' case of an isotropic two-photon
absorption tensor, i.e., a two-photon tensor that is diagonal in  the
Cartesian basis  with equal elements. 
In this case, $\tilde{\mathcal{A}}_{2\mathrm{P}}(\omega)$ becomes 
\begin{eqnarray*}
  \tilde{\mathcal{A}}_{2\mathrm{P}}(\omega) &=& 
+\mathcal{D}^{(1)}_{0,\varrho_1}(\omega)\,\mathcal{D}^{(0)}_{0,\varrho_1}(\omega)\,
T_{zz}\\\nonumber
&&
-\frac{1}{2}\mathcal{D}^{(1)}_{-1,\varrho_1}(\omega)\,\mathcal{D}^{(1)}_{+1,\varrho_1}(\omega)\,
\left(T_{xx}+T_{yy}\right)\\\nonumber
&&
-\frac{1}{2}\mathcal{D}^{(1)}_{+1,\varrho_1}(\omega)\,\mathcal{D}^{(1)}_{-1,\varrho_1}(\omega)\,
\left(T_{xx}+T_{yy}\right)\,,
\end{eqnarray*}
where we have employed the transformation between 
spherical and Cartesian basis,
cf. Eq.~\eqref{eq:cartesian_spherical}. Taking the elements to be 
equal, $T_{xx}=T_{yy}=T_{zz}=1$ without loss
of generality, $\tilde{\mathcal{A}}_{2\mathrm{P}}(\omega)$ can be written as 
\begin{eqnarray}
\tilde{\mathcal{A}}_{2\mathrm{P}}(\omega) &=& 
\mathcal{D}^{(1)}_{0,\varrho_1}(\omega)\,
\mathcal{D}^{(1)}_{0,\varrho_1}(\omega)\,
-
2\mathcal{D}^{(1)}_{-1,\varrho_1}(\omega)\,
\mathcal{D}^{(1)}_{+1,\varrho_1}(\omega)\nonumber\\
&=&
\sum_{\mu=0,\pm 1}\,(-1)^{\mu}
\mathcal{D}^{(1)}_{\mu,\varrho_1}(\omega)\,
\mathcal{D}^{(1)}_{-\mu,\varrho_1}(\omega)\nonumber\\
&=&
\sum_{\mu=0,\pm 1}\,(-1)^{-\varrho_1}
\mathcal{D}^{(1)}_{\mu,\varrho_1}(\omega)\,
\mathcal{D}^{{*}(1)}_{\mu,-\varrho_1}(\omega)\nonumber\\
&=&(-1)^{-\varrho_1}\,
\delta_{\varrho_1,-\varrho_1}\,,
\label{eq:first_law}
\end{eqnarray}
where we have used Eq.~\eqref{eq:wigner_unitary}.
That is, for an isotropic two-photon tensor, 
it is not possible to reach an anisotropic distribution by
absorption of two identical photons.
The PAD for the $(2+1)$ REMPI process then reduces to the well-known one
for one-photon ionization of randomly oriented molecules, i.e., only $P_0$
and $P_2$ contribute if  $\varrho_2=0$, and $P_0$, $P_1$ and $P_2$
are non-zero for $\varrho_2=\pm 1$. 

\begin{table*}[tb]
\newcommand{\ra}[1]{\renewcommand{\arraystretch}{#1}}
  \centering
  \ra{1.0}
\caption{\label{table:all_in_one} Contribution of Legendre
  coefficients to the PAD as a function of the partial wave cut-off
  in Eq.~\eqref{eq:exited_state2}  and 
  the polarizations $\epsilon_{\varrho_1}'$ and $\epsilon_{\varrho_2}'$
  of two-photon absorption and photoionization, respectively, for an
  isotropic and anisotropic  
  two-photon absorption tensor $\mathrm{T}$ 
  within the strict electric
  dipole approximation.  }
\begin{tabular}{cccccccccccccccccccccccccc} 
 \toprule[1.0pt] 
 \addlinespace[0.1cm]
 & &\multicolumn{4}{c}{$\epsilon_{0}'/\epsilon_{\pm 1}'$} & \phantom{abcd}& \multicolumn{4}{c}{$\epsilon_{\pm 1}'/\epsilon_{0}'$} & \phantom{abcd} & \multicolumn{4}{c}{$\epsilon_{0}'/\epsilon_{0}'$} & \phantom{abcd} & \multicolumn{4}{c}{$\epsilon_{\pm 1}'/\epsilon_{\pm 1}'$}&
 \phantom{abcd}&\multicolumn{3}{c}{$\epsilon_{\pm 1}'/\epsilon_{\mp 1}'$}\\
 \addlinespace[0.2cm]
 isotropic& &$s$       & $p$      & $d$ &$f$ &\phantom{abc}& $s$ & $p$ & $d$ & $f$ &\phantom{abc}& $s$ & $p$ & $d$ &$f$ &\phantom{abc}& $s$ & $p$ & $d$ &$f$ &\phantom{abc}&  $s$ & $p$ &$d$ &$f$\\
\hline
 \addlinespace[0.1cm]
$c_0$&& $\bullet$ &$\bullet$&  $\bullet$ & $\bullet$ &&$-$ &$-$ & $-$ & $-$  &&$\bullet$ &$\bullet$ & $\bullet$ & $\bullet$  &&$-$ &$-$ & $-$ & $-$   &&$-$ &$-$ & $-$ & $-$ \\
$c_1$&& $-      $ &$-      $&  $\bullet$ & $\bullet$ &&$-$ &$-$ & $-$ & $-$ &&$-$       &$-$       & $-$       & $-$        &&$-$ &$-$ & $-$ & $-$   &&$-$ &$-$ & $-$ & $-$ \\
$c_2$&& $\bullet$ &$\bullet$&  $\bullet$ & $\bullet$ &&$-$ &$-$ & $-$ & $-$ &&$\bullet$ &$\bullet$ & $\bullet$ & $\bullet$  &&$-$ &$-$ & $-$ & $-$   &&$-$ &$-$ & $-$ & $-$ \\
$c_3$&& $-$       &$-$      &  $-      $ & $-      $ &&$-$ &$-$ & $-$ & $-$ &&$-$ &$-$       & $-$       & $-$        &&$-$ &$-$ & $-$ & $-$   &&$-$ &$-$ & $-$ & $-$ \\
$c_4$&& $-$       &$-$      &  $-$       & $-$       &&$-$ &$-$ & $-$ & $-$ &&$-$ &$-$       & $-$       & $-$        &&$-$ &$-$ & $-$ & $-$   &&$-$ &$-$ & $-$ & $-$ \\
$c_5$&& $-$       &$-$      &  $-$       & $-$       &&$-$ &$-$ & $-$ & $-$ &&$-$ &$-$       & $-$       & $-$        &&$-$ &$-$ & $-$ & $-$   &&$-$ &$-$ & $-$ & $-$ \\
$c_6$&& $-$       &$-$      &  $-$       & $-$       &&$-$ &$-$ & $-$ & $-$ &&$-$ &$-$       & $-$       & $-$        &&$-$ &$-$ & $-$ & $-$   &&$-$ &$-$ & $-$ & $-$ \\
\addlinespace[0.05cm]
\toprule[1.0pt] 
\addlinespace[0.05cm]
 & &\multicolumn{4}{c}{$\epsilon_{0}'/\epsilon_{\pm 1}'$} & \phantom{abcd}& \multicolumn{4}{c}{$\epsilon_{\pm 1}'/\epsilon_{0}'$} & \phantom{abcd} & \multicolumn{4}{c}{$\epsilon_{0}'/\epsilon_{0}'$} & \phantom{abcd} & \multicolumn{4}{c}{$\epsilon_{\pm 1}'/\epsilon_{\pm 1}'$}&
 \phantom{abcd}&\multicolumn{3}{c}{$\epsilon_{\pm 1}'/\epsilon_{\mp 1}'$}\\
 anisotropic& &$s$       & $p$      & $d$ &$f$ &\phantom{abc}& $s$ & $p$ & $d$ & $f$ &\phantom{abc}& $s$ & $p$ & $d$ &$f$ &\phantom{abc}& $s$ & $p$ & $d$ &$f$ &\phantom{abc}&  $s$ & $p$ &$d$ &$f$\\
\hline
 $c_0$&&$\bullet$ &$\bullet$ & $\bullet$ & $\bullet$ &&$\bullet$ &$\bullet$ & $\bullet$&  $\bullet$   &&$\bullet$ &$\bullet$ & $\bullet$ & $\bullet$  &&$\bullet$ &$\bullet$ & $\bullet$ & $\bullet$ &&$\bullet$ &$\bullet$ & $\bullet$ & $\bullet$\\
 $c_1$&&$-      $ &$-      $ & $\bullet$ & $\bullet$ &&$-$ &$-      $ & $-$      &  $-$               &&$-$       &$-      $ & $-$       & $-$        &&$-$       &$-      $ & $\bullet$ & $\bullet$ &&$-$       &$-$       & $\bullet$ & $\bullet$\\
 $c_2$&&$\bullet$ &$\bullet$ & $\bullet$ & $\bullet$ &&$\bullet $ &$\bullet$ & $\bullet$&  $\bullet$  &&$\bullet$ &$\bullet$ & $\bullet$ & $\bullet$  &&$\bullet$ &$\bullet$ & $\bullet$ & $\bullet$ &&$\bullet$ &$\bullet$ & $\bullet$ & $\bullet$\\
 $c_3$&&$-      $ &$-      $ & $\bullet$ & $\bullet$ &&$-$ &$-      $ & $-$      &  $-$               &&$-$       &$-      $ & $-$       & $-$        &&$-$       &$-      $ & $\bullet$ & $\bullet$ &&$-$       &$-$       & $\bullet$ & $\bullet$\\
 $c_4$&&$-      $ &$\bullet$ & $\bullet$ & $\bullet$ &&$-$ &$\bullet$ & $\bullet$&  $\bullet$         &&$-$       &$\bullet$ & $\bullet$ & $\bullet$  &&$-$       &$\bullet$ & $\bullet$ & $\bullet$ &&$-$       &$\bullet$ & $\bullet$ & $\bullet$\\
 $c_5$&&$-$       &$-$       & $-$       & $\bullet$ &&$-$ &$-$       & $-$      &  $-$               &&$-$       &$-$       & $-$       & $-$        &&$-$       &$-$       & $-$       & $\bullet$ &&$-$       &$-$       & $-$       & $\bullet$\\
 $c_6$&&$-$       &$-$       & $\bullet$ & $\bullet$ &&$-$ &$-$       & $\bullet$&  $\bullet$         &&$-$ &$-$       & $\bullet$ & $\bullet$  &&$-$       &$-$       & $\bullet$ & $\bullet$ &&$-$       &$-$       & $\bullet$ & $\bullet$\\
 \hline
\bottomrule[0.8pt]
\addlinespace[0.1cm]
\multicolumn{16}{l}{$\bullet$\, contributing to the PAD}\\
\multicolumn{16}{l}{$-$\, not contributing to the PAD}\\
\end{tabular}
\end{table*}
In what follows, we discuss a general two-photon absorption tensor,
decomposing it as 
\begin{eqnarray}
  \label{eq:tensor_gral}
  \mathrm{T} &=& \alpha_o\openone_{3\times 3}+  
  \begin{pmatrix} \beta_{xx} & 0 & 0 \\ 
  0 & \beta_{yy} & 0 \\
0 & 0 & \beta_{zz} \end{pmatrix}
+ 
\begin{pmatrix} 0 & \mathrm{T}_{xy} & \mathrm{T}_{xz} \\ 
  \mathrm{T}_{xy} & 0 & \mathrm{T}_{yz} \\
  \mathrm{T}_{xz} & \mathrm{T}_{yz} & 0 \end{pmatrix}\nonumber\\
         &\equiv& \mathrm{T}_{\mathrm{Id}} +
\mathrm{T}_{\mathrm{d}} +\mathrm{T}_{\mathrm{nd}}\,,
\end{eqnarray}
where we have split the diagonal elements into
$\mathrm{T}_{\mathrm{Id}}$
and $\mathrm{T}_{\mathrm{d}}$ in order to differentiate between isotropic and
anisotropic two-photon tensors.
The contributions of odd and even Legendre polynomials to the PAD 
as a function of $L_{o,\rm max}$, the number of partial waves in the
electronically  excited state, the polarizations
$\epsilon^\prime_{\varrho_1}$ and $\epsilon^\prime_{\varrho_2}$, 
and the two-photon absorption tensor are 
summarized in Table~\ref{table:all_in_one}. 
If  the complete $(2+1)$ REMPI process is driven by linearly polarized
light and only $\alpha_0\neq 0$, then $P_o$ and $P_2$ contribute to the
PAD as just discussed. If the two-photon absorption tensor is
anisotropic, even Legendre polynomials of  higher order can appear. 
For a molecule characterized by such a two-photon absorption tensor,
odd Legendre polynomials can contribute to the PAD if the
polarization  of the ionization step is circular
($\epsilon^\prime_{\varrho_2}=\epsilon^\prime_{\pm 1}$).
Analogously,  
both even and odd Legendre polynomials can appear 
if $\epsilon^\prime_{\varrho_1}=
\epsilon^\prime_{\varrho_2}=\epsilon^\prime_{\pm 1}$.  
Note that anisotropy of the two-photon tensor is sufficient, i.e., it
does not matter whether the anisotropy is due to diagonal or
non-diagonal elements of the Cartesian tensor. 
The latter case is the one discussed in
Ref.~\cite{LehmannJCP13}, where a ``nearly'' diagonal two-photon
absorption tensor was used.  In other words, an anisotropic tensor
with non-zero off-diagonal elements in the Cartesian basis also yields
the pattern in the lower part of Table~\ref{table:all_in_one}. 

As indicated, the point group symmetry of the molecule determines which tensor
components of $T_{q_1,q_2}$ must be zero. This tensor pattern
is a property of the states involved in the transition and is determined by the 
symmetry of the initial and final states. For instance, in molecular systems
with point groups T and O, the photon absorption tensor
becomes more selective. The 2+1 process between two states 
that transform like the totally symmetric representation of these point groups 
will only take place with linearly polarized laser light. In this case the 
isotropic part $\mathrm{T}_{\mathrm{Id}}$ of Eq.~(\ref{eq:tensor_gral}) can remain nonzero.
If the 2+1 process involves initial and final states that transform like non-totally symmetric 
representations of the point group, the tensor pattern changes and thus the tensor 
might have isotropic or anisotropic parts. This determines whether the 2+1 process is
allowed or not. We refer the reader to Refs.~\cite{McCLAIN19771,Marco1983}
for more detailed discussion of this issue.

\section{Ab initio calculations}
\label{sec:abinitio}
The theoretical framework to model PECD presented above involves a
number of molecular parameters. These can either be obtained by
fitting the theoretical PAD to the experimental results or from     
  \textit{ab initio} calculations.  
Below we provide \textit{ab initio} results for the two-photon
absorption tensor for non-resonant transitions from the electronic 
ground state to the lowest-lying electronically excited states of
fenchone and camphor. To assess the quality of these calculations, we
employ different basis sets and different levels of treating
electronic correlation. 

\subsection{Computational details}
\label{subsec:compdetails}
\begin{figure}[tb]
\centering
\includegraphics[width=\linewidth]{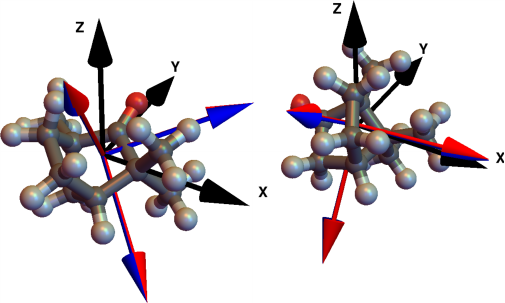} 
\caption{The oriented structures of fenchone (left) and camphor
(right). The black vectors represent the Cartesian coordinate system 
located at the center of mass of the molecular systems. The blue and
red vectors refer to the eigenvectors of the right and left two-photon
tensors corresponding to the third excited state (for more information
see Appendix~\ref{appen_twophoton}).} 
\label{fig:optimiz}       
\end{figure}

The linear response coupled cluster method with single and double
(CC-SD) cluster amplitudes is used to calculate the intermediate
electronicallyexcited state and the two-photon absorption tensor in
the electric  dipole approximation. Moreover, time-dependent density
functional theory (TD-DFT) calculations with the {\sc b3lyp}
exchange-correlation functional are performed. The molecular structure 
was energy minimized in all cases by performing DFT calculations with 
the {\sc b3lyp} exchange-correlation functional and the def2-TZVP basis 
set on all atoms, using the {\sc turbomole} program package \cite{turbomole6}. 
In Fig.~\ref{fig:optimiz}, the energy-minimized molecular structures of fenchone 
and camphor are shown, where the black vectors represent the Cartesian 
coordinate system located at the center of mass of the molecular systems.
These structures and orientations correspond to the ones used 
subsequently for the calculation of the two-photon absorption tensors.
Cartesian coordinates of the oriented structures are reported in the
Supplemental Material~\cite{SuppMat}.

Calculations for the two-photon transition strength 
tensor were performed using the {\sc dalton} program package
\cite{dalton:2005}.  
Details of the implementation of the two-photon absorption 
tensors within the linear response coupled cluster (CC) scheme 
are found in Refs.~\cite{paterson:2006,hattig:1998}. The orbital
unrelaxed methodology was employed in the linear response calculations of 
the two-photon absorption tensors on the coupled cluster level. Electrons 
occupying the $11$ energetically lowest-lying molecular orbitals that
are dominated by $1$s orbitals of the various carbon atoms or the
oxygen were excluded from 
the correlation treatment on the coupled cluster levels (so-called frozen
core approximation). The evaluation of the two-photon absorption tensor
was performed at the CC-SD/Rydberg-TZ level of theory. It is worth noting that 
two-photon transition strength tensor $T_{i,j}$, 
($i$,$j$=$x,y,z$) is calculated in the coupled cluster framework as 
a symmetric product of two-photon transition moments from initial to final state 
and from final to initial state (the left and right two-photon transition moments). 
As explained in more detail in Appendix~\ref{appen_twophoton}, 
in coupled cluster theory, the symmetrized biorthogonal structure
inhibits  identification of the left and right two-photon absorption
tensors. Thus, using the results of coupled cluster theory directly
in the calculation of PAD might be problematic, because the model
constructed in Sec.~\ref{sec:model} depends on only one two-photon
absorption tensor. We present a solution to this problem in
Appendix~\ref{appen_twophoton}. 
In Fig.~\ref{fig:optimiz}, the eigenvectors of the left and right two-photon
absorption tensors for the third exited state of fenchone and camphor
are shown (blue and red vectors).  

To benchmark the quality of the electronic structure calculations,
electronic excitation energies for transitions to the energetically lowest
lying singlet states are performed on the CCSD and approximate second order
couplet cluster (CC2) level for the $n$-aug-cc-pV$N$Z hierarchy of
basis sets
(see below). The {\sc turbomole} program package \cite{turbomole6} was used
for calculations on the CC2 level within the resolution of the identity
(RI) approximation. Select results were compared to conventional
CC2 calculations with the {\sc molpro} program package \cite{molpro},
confirming that the RI approximation has little impact on the computed
excitation energies (typically less than 10~meV). CCSD calculations for
excitation energies were performed with {\sc molpro}. Again, electrons
occupying 
the $11$ energetically lowest-lying molecular orbitals were kept frozen in
all coupled cluster calculations. 

The following basis sets were employed:
\begin{itemize}\setlength{\itemsep}{0pt}
\item Turbomole-TZVP with H:[3s,1p], C:[5s,3p,1d], O:[5s,3p,1d].
\item Rydberg-TZ with H: [2s], C: [5s,3p,1d], O:[5s,4p,3d,2f],
  'q':[1s,1p,1d], where 
  'q' is a ''dummy'' center, positioned at the center of mass of the
  molecule. Primitive diffuse $s$, $p$, $d$ Gaussian basis  
  functions with the exponent coefficients equal to 0.015~$a_0$ were placed on this center.
  With this basis we can expect quite a reliable description of the
  higher excited states (which, according to
  Ref.~\cite{pulm:1997}, are diffuse Rydberg states) but most likely
  not for the lowest lying excited state.
\item The ($n$-aug-)cc-pV$N$Z hierarchy of basis sets which are
correlation consistent polarized valence $N$-tuple zeta basis sets, with
$N$ = D, T, Q, referring to double-$\zeta$, triple-$\zeta$ and
quadruple-$\zeta$, respectively. On the oxygen nucleus, these basis
sets have been also augmented by further diffuse functions with $n$ =
s, d, t, q  implying single, double, triple and quadruple
augmentation, respectively. We used the procedure described in 
Ref.~\cite{DEW1994} for producing these aforementioned 
augmented basis sets.
\end{itemize}

The single center reexpansion is performed  in two steps.
First, the orbitals of the hydrogen atom are calculated
with a large uncontracted basis set [$13$s$11$p$9$d$8$f]. 
For manually adjusting the phases of the atomic orbitals, we have computed
numerically the radial part of the hydrogenic wavefunction
using the following procedure. The atomic wavefunction
\begin{equation}
 \vert \psi_i\rangle=\sum_j \vert \chi_j \rangle C_{ji}
 \label{wfabini}
\end{equation}
 is considered, where $\vert \chi_j \rangle$ is a gaussian basis function and reads
 \begin{align}
  \vert \chi_j \rangle=\frac{1}{\sqrt{2^{-3/2-l_j}\alpha_j^{-1/2-l_j}\Gamma[\frac{1}{2}+l_j]}} e^{-\alpha_j r^2} r^{l_j},
 \end{align}
where $\Gamma$ refers to the gamma function.
The $C_{ji}$, the atomic orbital coefficients, are calculated
by using the quantum chemical software Turbomole. The angular part 
can be chosen as the so-called real valued 
spherical harmonic and the integral over angular part is
\begin{align}
 \langle Y_{l_jm_j} (\theta,\phi) \vert Y_{l_km_k} (\theta,\phi) \rangle = \delta_{l_jl_k}\delta_{m_jm_k}.
\end{align}
In this way, one can calculate the radial part of Eq.~(\ref{wfabini}) and
compare it with Eq.~(\ref{radial:excitedstates})(which was used originally 
for reexpansion of the the electronically excited state of the neutral 
molecules under investigating (see Eq.~(\ref{eq:exited_state}))) and 
thus adjust the phases of atomic orbitals.

In the second step, the relevant molecular orbitals were 
calculated by projecting them onto hydrogen-like atom orbitals placed 
at the center-of-mass of camphor and fenchone, respectively, which is called
the blowup procedure in the Turbomole context~\cite{turbomole6}. This
calculation was carried out at the Hartree Fock(HF)/TZVP level of theory.

\subsection{Results and discussion}
\label{subsec:abinitioresults}
\begin{table}[tb]
\caption{\label{tab:exci} Experimental 
  and calculated excitation energies
  (in eV) for fenchone (top) and camphor (bottom) obtained by
  TD-DFT and CC-SD/Rydberg-TZ used for subsequent calculation of
  the two-photon transition tensor.}
\begin{tabular}{|c|c|c|c|}
\hline
 state                                                  &
 experiment~\cite{pulm:1997} &DFT-{\sc b3lyp} & CC-SD\\
\hline\hline
A / $\mathrm{n}\rightarrow \pi^*$               &  4.25                       &  4.24    & 4.44 \\
B / $\mathrm{n}\rightarrow 3\mathrm{s}$         &  6.10                       &  5.41    & 6.19 \\
$\text{C}_1$/$\mathrm{n}\rightarrow 3\mathrm{p}$&  6.58                       &  5.75    & 6.53  \\
$\text{C}_2$                                    &                             &  5.82    & 6.60  \\
$\text{C}_3$                                    &                             &  5.86    & 6.62  \\
$\text{D}_1$/$\mathrm{n}\rightarrow 3\mathrm{d}$&  7.14                       &          & 7.04  \\
$\text{D}_2$                                    &                             &          & 7.09  \\
$\text{D}_3$                                    &                             &          & 7.10  \\
$\text{D}_4$                                    &                             &          & 7.12  \\
$\text{D}_5$                                    &                             &          & 7.14  \\                                                
\hline\hline
A / $\mathrm{n}\rightarrow \pi^*$               &  4.21                       &  4.15    & 4.37 \\
B / $\mathrm{n}\rightarrow 3\mathrm{s}$         &  6.26                       &  5.53    & 6.33 \\ 
$\text{C}_1$/$\mathrm{n}\rightarrow 3\mathrm{p}$&  6.72                       &  5.87    & 6.73 \\
                                                &                             &          &       \\ 
$\text{C}_2$                                    &                             &  5.90    & 6.75  \\
$\text{C}_3$                                    &                             &  5.98    & 6.78 \\
$\text{D}_1$/$\mathrm{n}\rightarrow 3\mathrm{d}$&  7.28                       &          & 7.21  \\
$\text{D}_2$                                    &                             &          & 7.27  \\
$\text{D}_3$                                    &                             &          & 7.29  \\
$\text{D}_4$                                    &                             &          & 7.31  \\
$\text{D}_5$                                    &                             &          & 7.33  \\
\hline
\end{tabular}
\end{table}
The excitation energies of the lowest lying excited states for
fenchone and camphor are presented in Tables~\ref{tab:exci}-\ref{tab:camphortz}. The
labeling of the states follows the one for the absorption spectra of
Ref.~\cite{pulm:1997}. The states B, C and D are comparatively close in
energy. In principle, the order in which the states are obtained in
the calculations is unknown and the states may be interchanged due to
an insufficient level of the correlation treatment or the smallness of
the basis set. Nevertheless, we suppose that if the difference between
the theoretical excitation energies and the experimental ones is smaller than the
energy difference between the two states, then the order of the states
is correctly reproduced. Table~\ref{tab:exci} shows that 
the quite accurate excitation energies for the states B, C and D
are obtained in the CC-SD calculations with the Rydberg-TZ basis set
for both camphor and fenchone. The A state
is less accurately described with this basis set, while TDDFT result
for state A is very close to the corresponding experimental value.
For Rydberg states, it is well documented that the TDDFT method 
has severe limitations~\cite{Cme1998}, and we observe, indeed,
relatively large deviations between the computed excitation energies into Rydberg 
states and the corresponding experimental excitation energies as shown in Table~\ref{tab:exci}.
We this did not perform the calculation of excitation energies into even higher Rydberg states
for the present molecular systems.

Tables~\ref{tab:fenchonedz} and \ref{tab:fenchonetz} report
more detailed information on the electronic structure of fenchone,
obtained by employing both the CC$2$ and CCSD methods with
systematically improved basis sets. Enlarging the set of 
augmenting diffuse functions on the \ce{O} atom improves 
the excitation energies of the molecule under investigation. 
The energy of state A changes only mildly with increasing number of
diffuse functions and increasing the multiple zeta quality. 
Excitation energies for the state A evaluated at the  CC$2$/d-aug-ccpVQZ and
CC-SD/t-aug-pVDZ level of theory are in good agreement 
with the experimental one reported in Ref.~\cite{pulm:1997}. For 
state B, a similar dependence on  changing the augmented basis sets on
the \ce{O} atom and increasing the multiple zeta quantity can be
observed. Furthermore, we report a quite  
clear description for all members of the $n \rightarrow 3p$ Rydberg
transitions, corresponding to the C band of the experimental spectrum
reported in Ref.~\cite{pulm:1997}, whose individual components are
experimentally not resolved. The theoretical spacing among 
all components of the band C approaches to the experimental one when
increasing the augmented basis sets on the \ce{O} atom and the multiple
zeta quality. Strictly speaking, the theoretical spacing among
all components of the C band is less than $0.1$ eV which is in general 
in line with the experimental finding. 
The D state is composed of the $n \rightarrow 3d$ Rydberg transition. 
Here, we again report all individual components, which were not resolved
experimentally. The theoretical spacing among all components of the D band,
which is less than $0.1$ eV on average, approaches the experimental finding
when  increasing the augmented basis sets on the \ce{O} atom and the
multiple zeta quality. For the state A, the CC$2$ and CC-SD produce
the results close to each other, whereas for Rydberg states, deviation between
the results obtained by employing the CC$2$ and CC-SD methods is getting larger 
as was seen previously for different molecular systems~\cite{heid2009}.
Based on results of excitation energies evaluated
at CC$2$/t-aug-cc-pVDZ, d-aug-cc-pVTZ, t-aug-cc-pVTZ and d-aug-cc-pVQZ as well
as CC-SD/t-aug-cc-pVDZ, we estimate the excitation energies for fenchone
at CC-SD/t-aug-cc-pVQZ as described in the following. We add $\Delta E_1$ 
(which is the energy difference calculated using the CC$2$ method 
for basis sets d-aug-cc-pVQZ and d-aug-cc-pVTZ) as well as $\Delta E_2$ 
(which is the energy difference evaluated using CC$2$ for basis sets t-aug-cc-pVTZ 
and t-aug-cc-pVDZ) to the excitation energies calculated at the CC-SD/t-aug-ccpVDZ
level of theory. This procedure allows to estimate only few excitation energies
of the fenchone molecule at the CCSD/t-aug-cc-pVQZ level of theory. This way of estimation
does not work for all Rydberg states because the CC$2$ method is not accurate enough
for calculating excitation energies of these states. We should mention that the direct 
calculation at the CCSD/t-aug-cc-pVQZ 
level of theory was beyond our computational facilities. The corresponding results are 
shown in Table.~\ref{tab:estimate}. In order to justify this way of estimation, we employed
it for acetone, for which it is possible to calculate the excitation
energies at the CC-SD/t-aug-cc-pVQZ level of theory. This allows us to compare the excitation
energies at the CC-SD/t-aug-cc-pVQZ level of theory with the estimated ones.
The corresponding results were presented in Tables.~S$7$ and $8$ of the supporting 
information. It can be seen that the estimate values are very close to the corresponding 
ones calculated at the CC-SD/t-aug-cc-pVQZ level of theory. 
As an important remark, the excitation energies produced in Table~\ref{tab:exci} using the CC-SD/Rydberg-TZ
level of theory are closer to the experimental values than those generated using the CC-SD/t-aug-cc-pVDZ level of theory or the estimated 
values at the CC-SD/t-aug-cc-pVQZ level of theory (see Tables\ref{tab:fenchonedz} and \ref{tab:estimate}).

\begin{table*}[tb]
\caption{\label{tab:fenchonedz} 
Lowest vertical electronic singlet excitation energies (in eV) for 
fenchone as computed with the CC$2$ and CCSD method. The column heading
indicates the basis set, but augmented 
basis functions were only used on \ce{O} and deleted from  \ce{H} and
\ce{C}. Thus, for \ce{H} and \ce{C} the cc-pVDZ basis
set was used throughout.}
\begin{ruledtabular}
\begin{tabular}{lccccccccccc}
&Exp.~\cite{pulm:1997}&&\multicolumn{2}{c}{cc-pVDZ} &\multicolumn{2}{c}{aug-cc-pVDZ} &\multicolumn{2}{c}{d-aug-cc-pVDZ} &\multicolumn{2}{c}{t-aug-cc-pVDZ}\\
\cline{4-5} \cline{6-7} \cline{8-9} \cline{10-11}
State&&transition & CC2& CCSD  &CC2&CCSD  &CC2&CCSD &CC2&CCSD    \\
\hline
A&4.25&$\mathrm{n}\rightarrow \pi^*$               &4.38 &4.35  & 4.36 &4.35 &4.35&4.35  &4.34  &4.34\\
B&6.10&$\mathrm{n}\rightarrow 3\mathrm{s}$         &7.32 &7.94  & 7.23 &7.77 &5.80&6.39  &5.56  &6.15\\
$\text{C}_1$ &6.58&$\mathrm{n}\rightarrow 3\mathrm{p}$         &7.92 &8.27  & 7.72 &8.07 &6.18&6.85  &5.99  &6.71\\
$\text{C}_2$ &    &                                            &8.07 &8.52  & 7.93 &8.31 &6.28&6.97  &6.01  &6.74\\
$\text{C}_3$ &    &                                            &8.11 &8.76  & 7.99 &8.66 &6.38&7.10  &6.03  &6.79\\
D&7.14& $\mathrm{n}\rightarrow 3\mathrm{d}$        &8.22 &8.83  & 8.20 &8.78 &7.71&8.00  &6.65  &7.39\\
 &    &                                            &8.57 &8.95  & 8.28 &8.79 &7.92&8.31  &6.76  &7.57\\
 &    &                                            &8.63 &9.02  & 8.36 &8.81 &8.15&8.59  &6.84  &7.63\\
 &    &                                            &8.72 &9.25  & 8.53 &8.87 &8.25&8.74  &6.89  &7.68\\
 &    &                                            &8.74 &9.31  & 8.59 &9.10 &8.29&8.76  &7.26  &7.95\\
 &8.27&                                            &9.02 &9.35  & 8.85 &9.20 &8.33&8.79  &7.36  &8.04\\
 &    &                                            &9.19 &9.52  & 9.03 &9.35 &8.50&8.96  &7.46  &8.06 \\
\end{tabular}
\end{ruledtabular}
\end{table*}
\begin{table*}[tb]
\caption{\label{tab:fenchonetz} 
Lowest vertical electronic singlet excitation energies (in eV) for
fenchone as computed with the CC$2$ method. The column heading indicates the
basis set, but augmented basis functions were 
only used on \ce{O} and deleted from  \ce{H} and \ce{C}.}
\begin{ruledtabular}
\begin{tabular}{lcccccc}
State&Exp.~\cite{pulm:1997}  &cc-pVTZ & aug-cc-pVTZ & d-aug-cc-pVTZ& t-aug-cc-pVTZ&d-aug-cc-pVQZ\footnotemark[1]\\
\hline
A&4.25&4.32 &  4.29 & 4.29&4.27&4.28\\
B&6.10&6.83 &  6.15 & 6.01&5.68&5.96\\
$\text{C}_1$ &6.58&7.51 &  7.40 & 6.32&6.13&6.36\\
$\text{C}_2$ &    &7.53 &  7.50 & 6.39&6.14&6.41\\
$\text{C}_3$ &    &7.69 &  7.62 & 6.46&6.17&6.45\\
D&7.14&7.90 &  7.70 & 7.58&6.83&7.36\\
 &    &7.97 &  7.82 & 7.68&6.95&7.56\\
 &    &8.19 &  8.05 & 7.80&7.02&7.68\\
 &    &8.30 &  8.22 & 8.08&7.04&7.77\\
 &    &8.48 &  8.40 & 8.19&7.20&7.88\\
 &8.27&8.63 &  8.47 & 8.20&7.28&8.04\\
 &    &8.77 &  8.66 & 8.22&7.32&8.06\\
\end{tabular}
\footnotetext[1]{In this calculation, the basis set cc-pVQZ on 
\ce{C} and \ce{O} atoms is used.}
\end{ruledtabular}
\end{table*}

\begin{table}[tb]
\caption{\label{tab:estimate} 
The estimated lowest vertical electronic singlet excitation energies (in eV) for
fenchone and camphor at CC-SD/t-aug-cc-pVQZ level of theory. }
\begin{ruledtabular}
\begin{tabular}{lcc}
State&fenchone&camphor\\
\hline
A            &4.45& 4.17\\
B            &6.22& 6.52\\
$\text{C}_1$ &6.89&7.00 \\
$\text{C}_2$ &6.90&7.02 \\
$\text{C}_3$ &6.92&7.06 \\
D            &7.79&7.73 \\
             &    &7.81 \\
             &    &7.88 \\
\end{tabular}
\end{ruledtabular}
\end{table}

For camphor, the calculated excitation energies for state A, the lowest
excited state, are in reasonable agreement with experiment for all methods
and basis sets, cf. Tables~\ref{tab:exci}, \ref{tab:camphortz} and
\ref{tab:camphordz}.  
Here, we again observe that enlarging the set of augment diffuse function
on the \ce{O} atom and the multiple zeta quality improves the results for
the excitation energies.  Furthermore, increasing the augmented  
basis sets on the \ce{O} atom and the multiple zeta quality leads to a
decrease (of less than $0.1$ eV) in the theoretical spacing among 
all components of the C and D states, which again is in line with 
the experimental finding~\cite{pulm:1997}. The estimated excitation 
energies at CC-SD/t-aug-cc-pVQZ level of theory are calculated in 
the same way as done for fenchone. These results are shown in 
Table.~\ref{tab:estimate}. We should mention that the excitation energies 
produced in Table~\ref{tab:exci} using the CC-SD/Rydberg-TZ
level of theory are better than those generated using the CC-SD/t-aug-cc-pVDZ level 
of theory or the estimated values at the CC-SD/t-aug-cc-pVQZ level of theory 
(see Tables\ref{tab:camphordz} and \ref{tab:estimate}).
\begin{table*}[tb]
\caption{\label{tab:camphordz} 
Lowest vertical electronic singlet excitation energies (in eV) for 
camphor as computed with the CC$2$ and CCSD method. 
The column  heading indicates the basis set, but augmented 
basis functions were only used on \ce{O} and deleted from  \ce{H} and
\ce{C}. Thus, for \ce{H} and \ce{C} the cc-pVDZ basis
set was use throughout.}
\begin{ruledtabular}
\begin{tabular}{lccccccccccc}
&Exp.~\cite{pulm:1997}&&\multicolumn{2}{c}{cc-pVDZ} &\multicolumn{2}{c}{aug-cc-pVDZ} &\multicolumn{2}{c}{d-aug-cc-pVDZ} &\multicolumn{2}{c}{t-aug-cc-pVDZ}\\
\cline{4-5} \cline{6-7} \cline{8-9} \cline{10-11}
State&& transition & CC2& CCSD  &CC2&CCSD  &CC2&CCSD &CC2&CCSD    \\
\hline
A&4.21&$\mathrm{n}\rightarrow \pi^*$               &4.27 &4.25  & 4.23 &4.25 &4.22&4.24  &4.22  &4.24 \\
B&6.26&$\mathrm{n}\rightarrow 3\mathrm{s}$         &7.40 &8.05  & 7.32 &7.87 &5.83&6.44  &5.64  &6.34 \\
$\text{C}_1$ &6.72&$\mathrm{n}\rightarrow 3\mathrm{p}$         &7.69 &8.10  & 7.46 &7.90 &6.25&6.93  &6.07  &6.81 \\
$\text{C}_2$ &    &                                            &8.04 &8.35  & 7.81 &8.11 &6.30&7.00  &6.09  &6.84 \\
$\text{C}_3$ &    &                                            &8.23 &8.84  & 8.11 &8.63 &6.60&7.32  &6.15  &6.93 \\
D&7.28&$\mathrm{n}\rightarrow 3\mathrm{d}$         &8.38 &8.90  & 8.19 &8.69 &7.43&7.85  &6.75  &7.56 \\
 &    &                                            &8.47 &8.98  & 8.24 &8.78 &7.79&8.10  &6.84  &7.67 \\
 &    &                                            &8.56 &9.03  & 8.33 &8.28 &7.91&8.46  &6.90  &7.73 \\
 &    &                                            &8.62 &9.22  & 8.36 &8.90 &8.14&8.62  &7.05  &7.82 \\
 &    &                                            &8.79 &9.27  & 8.62 &9.02 &8.25&8.71  &7.26  &7.85 \\
 &7.94&                                            &8.91 &9.36  & 8.77 &9.16 &8.28&8.84  &7.35  &7.95 \\
 &    &                                            &9.04 &9.51  & 8.83 &9.45 &8.33&8.85  &7.39  &8.05  \\
 \end{tabular}
\end{ruledtabular}
\end{table*}
\begin{table*}[tb]
\caption{\label{tab:camphortz} 
Lowest vertical electronic singlet excitation energies (in eV) for
camphor as computed with the CC$2$ method. The column heading indicates the 
basis set, but augmented basis functions were 
only used on \ce{O} and deleted from  \ce{H} and \ce{C}.}
\begin{ruledtabular}
\begin{tabular}{lcccccc}
State&Exp.~\cite{pulm:1997} &cc-pVTZ & aug-cc-pVTZ & d-aug-cc-pVTZ& t-aug-cc-pVTZ&d-aug-cc-pVQZ\footnotemark[1]\\
\hline
A&4.21&4.20 &  4.17 & 4.17& 4.15&4.17 \\
B&6.26&6.94 &  6.85 & 5.98& 5.78&6.02 \\
$\text{C}_1$&6.72&7.41 &  7.32 & 6.39& 6.22&6.43\\
$\text{C}_2$ &    &7.66 &  7.57 & 6.43& 6.23&6.47\\
$\text{C}_3$ &    &7.75 &  7.63 & 6.67& 6.30&6.65\\
D&7.28&7.85 &  7.65 & 7.31& 6.95&7.28\\
 &    &7.97 &  7.82 & 7.66& 7.02&7.62 \\
 &    &8.13 &  8.04 & 7.93& 7.08&7.63\\
 &    &8.19 &  8.09 & 7.98& 7.19&7.72\\
 &    &8.28 &  8.19 & 8.02& 7.25&7.94\\
 &7.94&8.62 &  8.53 & 8.08& 7.27&7.96\\
 &    &8.66 &  8.63 & 8.17& 7.34&7.99\\
\end{tabular}
\footnotetext[1]{In this calculation, the basis set cc-pVQZ on 
\ce{C} and \ce{O} atoms is used.}
\end{ruledtabular}
\end{table*}

In the following, we report the two-photon absorption tensor elements
for fenchone and camphor calculated with the TD-DFT and CC-SD
methods. The computational details for the coupled cluster
calculations are presented in Appendix~\ref{appen_twophoton}.
The elements of the two-photon absorption tensor for fenchone and camphor
in the Cartesian basis are generally independent because the molecules
have the $C_1$ point group symmetry~\cite{McCLAIN19771}. 
However, as we consider absorption of two photons with same the
frequency, the two-photon tensor must be symmetric~\cite{McCLAIN19771}.
Table~\ref{tab:twophotonA} presents the results for fenchone.
The A state in terms of the
excitation energy is of no real concern for our present purposes
because the wavelength and spectral width of the  
laser pulses employed in the $2+1$ REMPI process~\cite{LuxACIE12,LuxCPC15}
practically rule out that A is the relevant intermediate
state. As inferred from Table~\ref{tab:twophotonA}, changing the method accounting
for the electron correlations {\it i.e} TD-DFT and CC-SD, alters considerably
the skeleton of the two-photon transition matrix and in particular there are 
changes in the signs of matrix elements when employing different electron
correlation methods.
As the excitation energies for the B and C states, calculated 
with the CC-SD/Rydberg-TZ level of theory, are in good agreement with experimental ones,
cf. Table~\ref{tab:exci}, we expect the corresponding  
two-photon absorption tensor elements to be more reliable for the
evaluation of PECD than  those obtained with TD-DFT.
We therefore use the two-photon absorption tensor elements
calculated at the CC-SD/Rydberg-TZ level of theory for calculating PAD in
Sec.~\ref{sec:pad_results}.

\begin{table}
\caption{\label{tab:twophotonA} Two-photon transition matrix elements
  (in units of $a^2_0~E_{\mathrm{h}}^{-1}$ with $a_0$ being the Bohr radius and
  $E_{\mathrm{h}}$ being the Hartree energy)
at the {\sc b3lyp}/Rydberg-TZ level of theory (top) and symmetric effective
two-photon transition matrix elements at the CC-SD/Rydberg-TZ level of
theory (bottom) for fenchone. The specific orientation used is shown in Fig.~\ref{fig:optimiz}. }
\begin{ruledtabular}
  \begin{tabular}{@{}l*{7}{D{.}{.}{0}}@{}}
     \toprule
    States&T^{xx}_{go}&T^{xy}_{go}&T^{xz}_{go}&T^{yy}_{go}&T_{go}^{xz}&T^{zz}_{go}\\
\hline
\hline
    \midrule
    $\text{A}$&+$0.50$&+$0.50$&$+0.50$&+$0.20$&-$0.30$&-$0.30$\\
    $\text{B}$&+$1.60$&-$0.70$&-$2.60$&+$20.80$&+$8.20$&-$0.70$\\
$\text{C}_1$&-$40.60$&-$11.50$&-$6.30$&+$1.60$&+$1.40$&-$1.60$\\
    $\text{C}_2$&+$3.20$&+$1.30$&+$2.40$&+$5.30$&-$1.20$&-$1.40$\\
$\text{C}_3$&-$8.60$&-$3.00$&-$5.00$&-$1.90$&+$8.70$&+$0.10$\\
    \bottomrule
   \hline \hline \addlinespace[0.5ex]
state  & \tilde{T}^{xx}_{go} &\tilde{T}^{xy}_{go}& \tilde{T}^{xz}_{go} & \tilde{T}^{yy}_{go}&\tilde{T}^{yz}_{go}&\tilde{T}^{zz}_{go}\\
\hline 
$\text{A}$  &-$0.11$&-$0.03$&+$0.08$&-$0.27$&+$0.20$&-$0.27$\\
$\text{B}$  &+$1.58$&+$17.10$&+$7.50$&-$1.67$&-$0.24$&-$2.48$\\
$\text{C}_1$&-$0.21$&-$7.57$&-$4.10$&+$1.13$&+$1.02$&+$0.96$\\
$\text{C}_2$&-$21.24$&+$5.45$&-$1.32$&-$6.00$&-$1.87$&-$2.02$\\
$\text{C}_3$&-$28.67$&-$1.54$&+$4.10$&-$7.88$&+$0.04$&-$6.69$\\  
  \end{tabular}
\end{ruledtabular}
\end{table}

\begin{table}
\caption{\label{tab:twophotonCam}  The same as
  Table~\ref{tab:twophotonA} but for camphor.}
\begin{ruledtabular}
  \begin{tabular}{@{}l*{7}{D{.}{.}{0}}@{}}
     \toprule
    States&T^{xx}_{go}&T^{xy}_{go}&T^{xz}_{go}&T^{yy}_{go}&T_{go}^{xz}&T^{zz}_{go}\\
\hline
\hline
    \midrule
$\text{A}$  &-$0.30$&+$0.50$&-$1.90$&-$0.40$&-$1.00$&-$0.10$\\
$\text{B}$  &+$10.90$&-$5.40$&-$8.30$&-$8.30$&-$13.40$&-$4.10$\\
$\text{C}_1$&-$3.50$&-$4.80$&-$0.70$&-$1.90$&+$1.40$&-$3.40$\\
$\text{C}_2$&-$4.20$&+$1.00$&+$2.20$&-$0.30$&$0.00$&+$1.10$\\
$\text{C}_3$&-$23.70$&-$5.50$&-$3.10$&-$3.20$&-$2.20$&-$2.90$\\

    \bottomrule
   \hline \hline \addlinespace[0.5ex]
state  & \tilde{T}^{xx}_{go} &\tilde{T}^{xy}_{go}& \tilde{T}^{xz}_{go} & \tilde{T}^{yy}_{go}&\tilde{T}^{yz}_{go}&\tilde{T}^{zz}_{go}\\
\hline 
$\text{A}$  &-$0.35$&-$0.27$&-$0.48$&+$0.41$&-$0.03$&-$1.17$\\
$\text{B}$  &+$1.29$&+$9.36$&+$12.63$&+$6.58$&+$4.67$&+$7.55$\\
$\text{C}_1$&+$7.48$&+$0.41$&+$0.82$&-$3.46$&-$3.54$&-$5.11$\\
$\text{C}_2$&+$3.07$&+$0.28$&-$4.10$&+$4.10$&+$1.92$&-$5.88$\\
$\text{C}_3$&-$21.48$&+$0.98$&+$2.83$&-$1.95$&-$1.13$&-$0.81$\\
  \end{tabular}
\end{ruledtabular}
\end{table}

Table~\ref{tab:twophotonCam} presents the two-photon absorption tensor
elements for camphor. 
Changing the method accounting of electron correlations, TD-DFT or CC-SD,
alters considerably the skeleton of the two-photon transition
matrix.   
For camphor, similar observation as mentioned for fenchone can 
be mentioned here; the A state is very unlikely  to be the intermediate 
state probed in the $2+1$ REMPI process. As inferred from Table~\ref{tab:twophotonCam}, 
changing the method accounting for the electron correlations {\it i.e} TD-DFT 
and CC-SD, alters considerably the skeleton of the two-photon transition matrix 
and in particular there are changes in the signs of matrix elements when employing 
different electron correlation methods.
\begin{figure*}[tb]
\centering
\includegraphics[width=0.05\linewidth]{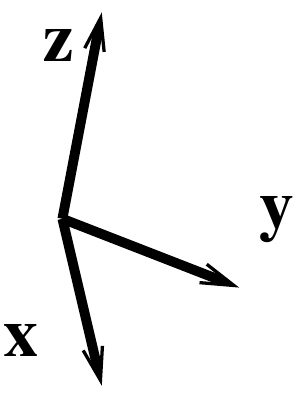} \quad \includegraphics[width=0.15\linewidth]{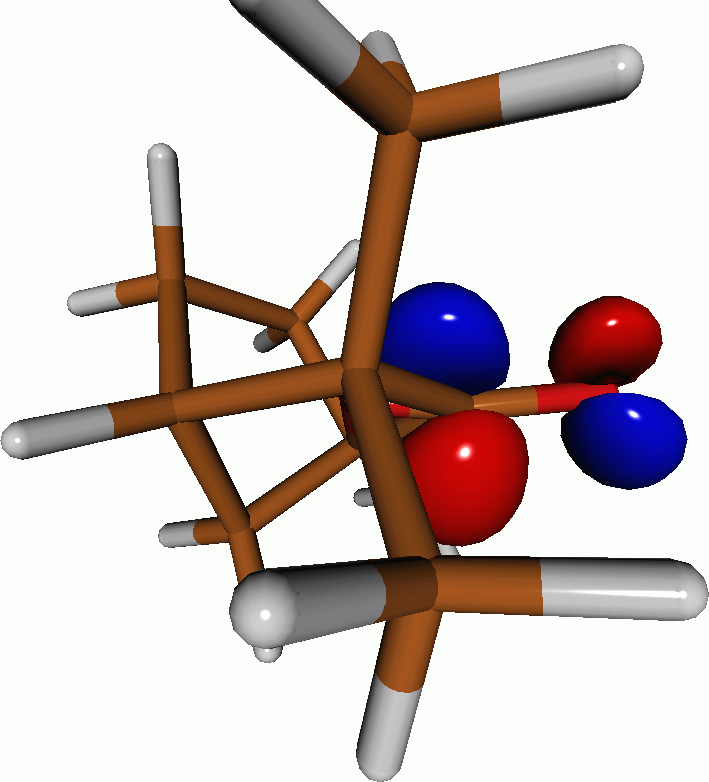} \quad \includegraphics[width=0.15\linewidth]{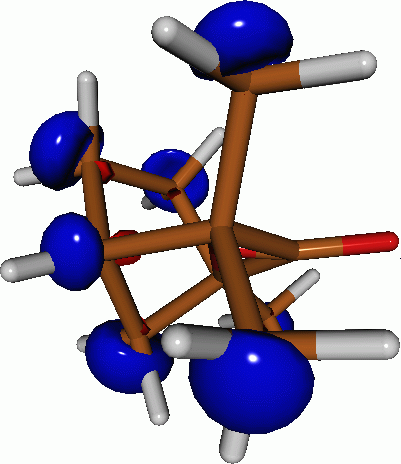} \quad \includegraphics[width=0.15\linewidth]{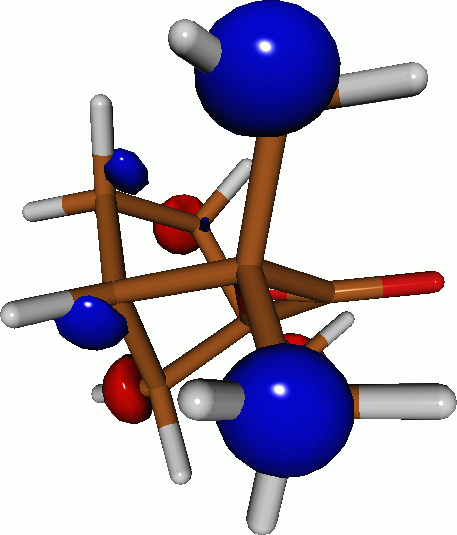}
\quad \includegraphics[width=0.15\linewidth]{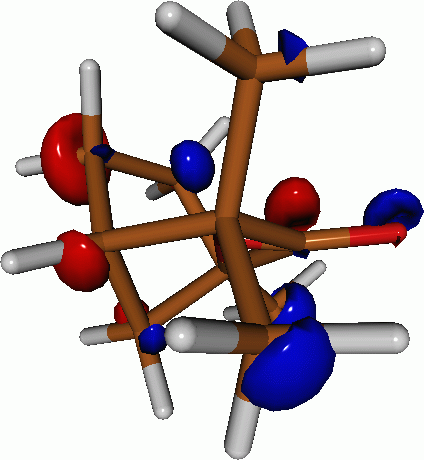} \quad \includegraphics[width=0.15\linewidth]{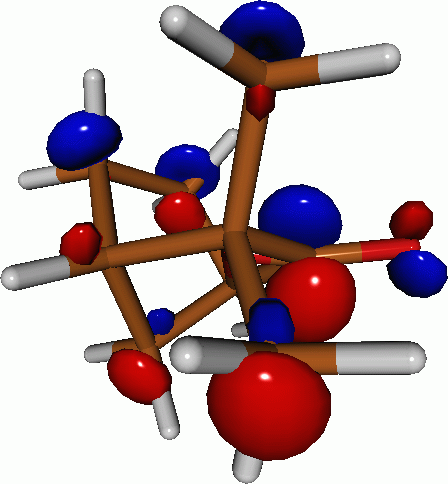}
\caption{ The molecular orbitals $43$, $44$, $45$, $46$ and $47$ of
fenchone corresponding to the excited 
states A, B, $\text{C}_1$, $\text{C}_2$ and $\text{C}_3$, respectively. These molecular orbitals are calculated  at the HF/TZVP
level of theory.}
\label{fig:orbitalfenchon}       
\end{figure*}
\begin{figure*}[tb]
\centering
\includegraphics[width=0.15\linewidth]{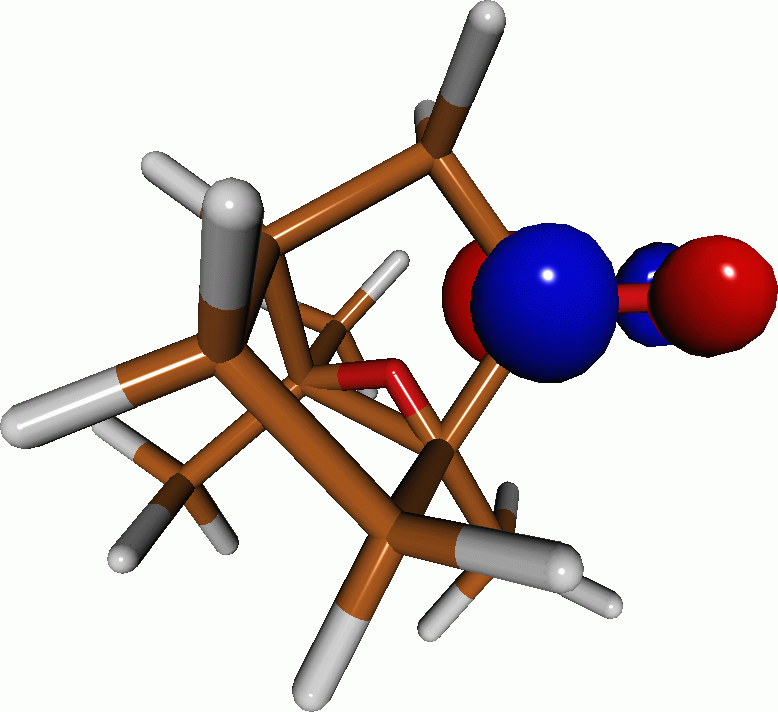} \quad \includegraphics[width=0.15\linewidth]{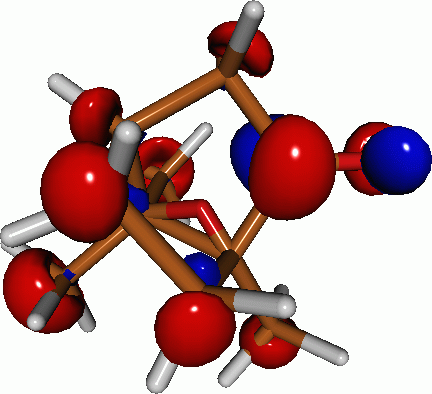} \quad \includegraphics[width=0.15\linewidth]{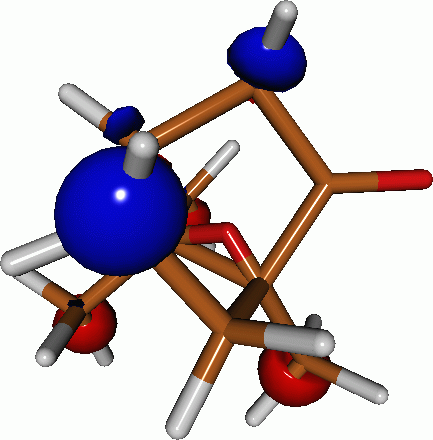}
\quad \includegraphics[width=0.15\linewidth]{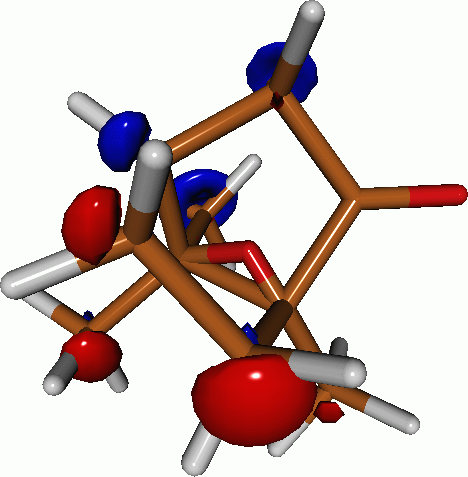} \quad \includegraphics[width=0.15\linewidth]{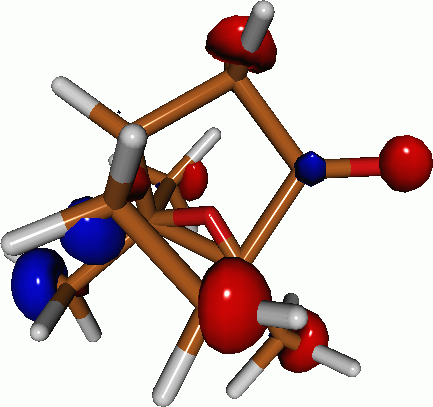}
\caption{The molecular orbitals $43$, $44$, $45$, $46$ and $47$ of camphor corresponding to the excited 
states A, B, $\text{C}_1$, $\text{C}_2$ and $\text{C}_3$, respectively. These molecular orbitals are calculated  at the HF/TZVP
level of theory.}
\label{fig:orbitalcamphor}       
\end{figure*}

\subsection{Single center reexpansion of molecular wavefunctions}
\label{subsec:reexpansion}
In order to match the \textit{ab initio} results with our model for
the 2+1 REMPI process, we perform a single center reexpansion of the 
relevant molecular orbitals (see Figs.~\ref{fig:orbitalfenchon} 
and \ref{fig:orbitalcamphor}) obtained from a HF calculation with 
the TZVP basis set, projecting them onto hydrogenic atomic orbitals 
placed at the center-of-mass of the molecule.
The hydrogenic orbitals are chosen in the form 
$\varphi=\sum_i\tilde{a}_i R_i(r)\Upsilon_i(\theta,\phi)$, where $i$ denotes a
complete set of quantum numbers, $i\equiv(n_o,\ell_o,m_o)$. $R_i(r)$
are the radial functions the hydrogen  and
$\Upsilon_i(\theta,\phi)$ the {\it real} spherical harmonics.
The transformation between the expansion coefficients $\tilde{a}_i$
and $a_i$, defined in Eq.~\eqref{eq:exited_state} with the standard
complex spherical harmonics, is given in
Appendix~\ref{subsec:real_spherical}.   

The projection quality  of the orbitals $42$ (highest occupied
molecular orbital (HOMO) for the electronic ground state)
and $43$ (one of the two singly occupied molecular orbitals (SOMOs) 
for state A) for both camphor and fenchone
is rather low. It amounts  to 28\% and 45\% for fenchone and to  24\%
and 51\%  for camphor. This is expected for the HOMO and SOMO 
which are localized orbitals. In contrast,  
for the orbitals representative of the Rydberg states B and C in all cases the projection quality is
higher than 90 \% for the corresponding SOMO. For these states, the results
of the reexpansion are presented in the Supplemental Material~\cite{SuppMat}.
We find the B state to be of $s$-type, that is, the $s$-wave
contributes more than all other waves together; whereas the C states
are of $p$-type. This is in agreement with the results of
Refs.~\cite{pulm:1997,Diedrich:2003}, where these states were also 
found to be of $s$- and $p$-type, respectively. 
The $d$ wave contributions for SOMOs orbitals corresponding to the B and 
$\text{C}_1$, $\text{C}_2$ and $\text{C}_3$ states in fenchone and camphor 
are 2\% , 3\% , 5\% and 6\%, respectively.

\section{Photoelectron angular distributions}
\label{sec:pad_results}
The experimental measurements indicate a PECD effect of 10\% for
fenchone and 6.6\% for camphor~\cite{LuxCPC15}.
We first check the range of PECD that our model allows for.
To this end, we  optimize, as a preliminary test, PECD, allowing all molecular parameters, i.e., 
two-photon absorption tensor elements and excited state expansion coefficients, to vary freely.
We expand up to $d$ and $f$ waves for a single
quantum number $n_o$, taken to be $n_o=3$ and $4$,
respectively. 
The optimization target is to maximize (or minimize,
depending on the sign) PECD in order to
determine the upper bounds. Following the definitions in 
Refs.~\cite{LuxCPC15,LeinPRA2014}, we define an optimization
functional, 
\begin{eqnarray}
  \label{eq:opt_func}
  J = \dfrac{1}{c_0}\left(2c_1 -\dfrac{1}{2}c_3
  + \dfrac{1}{4}c_5\right)\,, 
\end{eqnarray}
where the Legendre coefficients
are calculated according to Eq.~\eqref{eq:final_coeff}. 
All optimizations are carried out 
using the genetic algorithm  for constrained multivariate problems as
implemented  in Ref.~\cite{simulinkR2014a}, using  $500$  
iterations. We find numerical
bounds of about 35\% for both expansion cut-offs. The experimentally observed PECD effects are well within these bounds.

We now present calculations of the PAD for fenchone and camphor,
using two different strategies to evaluate
Eq.~\eqref{eq:PADfinal}. First, 
we aim at identifying the minimal requirement in terms of structure
and symmetry properties of the intermediate electronically excited
state for reproducing, at least qualitatively, the experimental
data. To this end, we minimize the difference between theoretically
and experimentally obtained Legendre coefficients,
$\delta_j=|(c_j-c^{\mathrm{exp}}_{j})/c^{\mathrm{exp}}_{j}|$,
taking the excited state expansion coefficients, $a^{\ell_o}_{m_o}$, cf. 
Eq.~\eqref{eq:exited_state2}, as optimization parameters, with
$n_o=3$ fixed. This 
allows for $L_{o,\text{max}}=2$, i.e., $s$, $p$ and $d$ waves.
Second, we test the agreement between theoretically
and experimentally obtained Legendre coefficients when utilizing the
expansion coefficients
and two-photon tensor elements obtained by \textit{ab initio}
calculations, cf. Section~\ref{sec:abinitio}. 
Here, our aim is  to explain the differences observed experimentally
in the PADs for fenchone and camphor 
in terms of the  intermediate electronically excited state.

In the first approach, treating the excited state coefficients  as optimization 
parameters, the optimization  can be performed for the odd
Legendre moments only, focussing on reproducing PECD, or for 
both odd and even Legendre 
moments, in order to reproduce the complete PAD.
The different experimental
uncertainties for odd and even Legendre coefficients~\cite{LuxCPC15} 
motivate such a two-step approach. Moreover, 
optimizing for the odd Legendre coefficients alone allows to quantify
the minimal requirements on the intermediate electronically excited 
state for reproducing PECD. 

In the second approach, when using the \textit{ab initio}
two-photon absorption tensors and expansion coefficients,
we need to account for the unavoidable error bars of the \textit{ab
  initio} results. To this 
end, we also utilize optimization, allowing the two-photon tensor
matrix elements to vary, whereas the 
excited state coefficients  
are taken as is from the reexpansion of the \textit{ab initio}
wavefunctions.

\subsection{Fenchone}
\label{subsec:fenchone}
We start by addressing the question of how many partial waves are
required in the intermediate electronically 
excited state to yield odd Legendre coefficients with $\mathcal{L}>1$, as
observed experimentally. 
To this end, we consider the expansion of the  intermediate electronically 
excited state, cf. Eq.~\eqref{eq:exc_state}, with
$L_{o,\mathrm{max}}=2$ and $L_{o,\mathrm{max}}=3$,
i.e., up to $d$ and $f$ waves, for the
states B and C, and employ 
the two-photon tensor elements from the CCSD/Rydberg-TZ calculations,
cf. Table~\ref{tab:twophotonA}.  
\begin{table*}[tbp]
 \caption{\label{tab:fenchone:opt1} 
   Legendre coefficients for the
   PAD of fenchone (calculated at a photoelectron energy
   of $0.56\,$eV and normalized with respect to $c_0$), obtained by
   fitting to the experimental 
   values with the excited state coefficients
   $a^{\ell_o}_{m_o}$ as free parameters. Only odd (top) and 
   both odd and even (bottom)  contributions were accounted for in the fitting procedure. 
   The Rydberg states B, C1, C2 and C3 of fenchone are  characterized by
   their two-photon absorption tensor, cf. Tab.~\ref{tab:twophotonA}.
}
  \begin{tabular}{c !{\vrule width -4pt}c !{\vrule width -4pt}c !{\vrule width -4pt}c!{\vrule width -4pt}c !{\vrule width -4pt}c !{\vrule width -4pt}c !{\vrule width -4pt}c !{\vrule width -4pt}c !{\vrule width -4pt}c !{\vrule width -4pt}c !{\vrule width -4pt}c !{\vrule width -4pt}c !{\vrule width -4pt}c !{\vrule width -4pt}
   c !{\vrule width -4pt}c !{\vrule width -4pt}c !{\vrule width -4pt}c !{\vrule width -4pt}c !{\vrule width -4pt} 
    }
    \toprule[1.0pt]
   \addlinespace[0.05cm]
   & & \phantom{abc}       & & \multicolumn{3}{c}{state B} & &\multicolumn{3}{c}{state C1} & &\multicolumn{3}{c}{state C2}  & &\multicolumn{3}{c}{state C3}\\
   coeffs.  &  \phantom{abc}                     &exp.~\cite{LuxCPC15}   & \phantom{abcdef}   &$d$ waves &\phantom{ab}& $f$ waves &\phantom{abcd}&$d$ waves &\phantom{ab}&$f$ waves  &\phantom{abcd}&$d$ waves &\phantom{ab}&$f$ waves  &\phantom{abcd}&$d$ waves &\phantom{ab}&$f$ waves  \\
  \addlinespace[0.05cm]
   \cmidrule[0.1pt]{1-1}
   \cmidrule[0.1pt]{3-3}
   \cmidrule[0.1pt]{5-7}
   \cmidrule[0.1pt]{9-11}
   \cmidrule[0.1pt]{13-15}
   \cmidrule[0.1pt]{17-19}
   
  \addlinespace[0.15cm]
  $c_1$          &&$-0.067$&    & $-0.067$&     & $-0.067 $         &&     $-0.067   $&     & $-0.067$     &&     $-0.067   $&     & $-0.067$   &&     $-0.067   $&     & $-0.067$   \\  
  $c_3$          &&$+0.008$&    & $+0.080$&     & $+0.080 $         &&     $+0.008   $&     & $+0.008$     &&     $+0.008   $&     & $+0.008$   &&     $+0.008   $&     & $+0.008$   \\  
  $c_5$          &&$+0.004$&    & $     -$&     & $+0.0005$ &&     $-        $&     & $+0.004$     &&     $-        $&     & $+0.004$   &&     $-        $&     & $+0.004$    \\  
  \addlinespace[0.1cm]
  \cmidrule[0.1pt]{1-19}
  \addlinespace[0.15cm]
                    $c_1$         & &$-0.067 $&    & $ -0.028  $&     & $-0.041   $  && $-0.045    $&     & $-0.036  $    &     & $-0.040  $  &     & $-0.048  $   &     & $-0.045  $ &     & $-0.046  $ \\  
                    $c_2$         & &$-0.580 $&    & $-0.076   $&     & $-0.102   $  && $ -0.274   $&     & $ -0.176 $    &     & $ -0.146 $  &     & $ -0.226 $   &     & $ -0.224 $ &     & $ -0.246 $ \\  
                    $c_3$         & &$+0.008 $&    & $ +0.006  $&     & $+0.005   $  && $+0.006    $&     & $ +0.008 $    &     & $ +0.003 $  &     & $ +0.004 $   &     & $ +0.006 $ &     & $ +0.005 $ \\  
                    $c_4$         & &$-0.061 $&    & $ -0.004  $&     & $-0.004   $  && $-0.021    $&     & $ -0.012 $    &     & $ -0.012 $  &     & $ -0.011 $   &     & $ -0.012 $ &     & $ -0.019 $ \\  
                    $c_5$         & &$+0.004 $&    & $ -       $&     & $+0.0001   $  && $ -        $&     & $+0.001  $    &     & $-       $  &     & $+0.002  $   &     & $  -     $ &     & $+0.001  $ \\  
                    $c_6$         & &$-0.008 $&    & $ +0.0002 $&     & $+0.0003  $  && $+0.0007    $&    & $+0.0001$     &     & $+0.0006$   &     & $+0.001   $   &     & $-0.002  $ &    & $-0.002  $ \\  
\bottomrule[1.0pt]
\addlinespace[0.1cm]
\end{tabular}
\end{table*}
The results are presented 
in Table~\ref{tab:fenchone:opt1}. 
Presence of $f$-waves is required to obtain a non-zero coefficient $c_5$, 
as expected from Table~\ref{table:all_in_one}.
Allowing for $f$ waves (with $n_0$=4) results in a perfect match for the odd coefficients
for states C1, C2 and C3,                           
cf. the upper part of Table~\ref{tab:fenchone:opt1}.           
In contrast, for state B, $c_3$ and $c_5$, while having the correct sign, are off by an order of magnitude. 
Modifying the optimization weights improves $c_5$ for state B, but
only at the expense of the agreement for $c_1$ and $c_3$. State B can therefore be ruled out as intermediate electronically excited state. This is further confirmed by the lower part 
of Table~\ref{tab:fenchone:opt1}, showing the results for both odd and even Legendre coefficients in the  optimization target.
For state B, the sign of $c_6$  does not match the experimental one.
Fitting both odd and even Legendre coefficients also allows to differentiate between the C states---only state C3 reproduces the correct sign of $c_6$. For all other Legendre moments, signs and order of magnitude of the coefficients match the experimental ones for all three C states. 
Fitting to all and not just the odd Legendre coefficients decreases the agreement 
between theoretical and experimental results for all C states. This
may indicate that the model, with a single $n_o$, is not capable of reproducing the full complexity of the process, 
or it may be due to different experimental error bars for even and odd Legendre coefficients. In our fitting procedure,  we have neglected the experimental error bars to keep
the calculations manageable. The experimental error bars for the even
Legendre coefficients are much larger than for the odd
ones~\cite{LuxCPC15}, and ignoring them may introduce a bias into the optimization procedure that could also explain the decreased agreement.

\begin{table*}[tbp]
 \caption{\label{tab:fenchone_opt_vs_nonopt} Legendre
  coefficients  for the PAD of fenchone (calculated at a photoelectron energy of $0.58$ eV
  and  normalized with  respect to $c_0$),   obtained by employing the excited state
  coefficients and two-photon tensors from the \textit{ab initio}
  calculations. When including error bars, the tensor
  elements are allowed to vary within
  $\pm$20\%.}                                      
  \begin{tabular}{c !{\vrule width -4pt}c !{\vrule width -4pt}c !{\vrule width -4pt}c!{\vrule width -4pt}c !{\vrule width -4pt}c !{\vrule width -4pt}c !{\vrule width -4pt}c !{\vrule width -4pt}c !{\vrule width -4pt}c !{\vrule width -4pt}c !{\vrule width -4pt}c !{\vrule width -4pt}c !{\vrule width -4pt}c !{\vrule width -4pt}
   c !{\vrule width -4pt}c !{\vrule width -4pt}c !{\vrule width -4pt}c !{\vrule width -4pt}c !{\vrule width -4pt} 
    }
    \toprule[1.0pt]
   \addlinespace[0.05cm]
   & & \phantom{abc}       & & \multicolumn{3}{c}{state B} & &\multicolumn{3}{c}{state C1} & &\multicolumn{3}{c}{state C2}  & &\multicolumn{3}{c}{state C3}\\
   coeffs.  &  \phantom{a}                     &exp.~\cite{LuxCPC15}   & \phantom{abcde}   &fixed &\phantom{ab}& error bars&\phantom{abcd}&fixed&\phantom{ab}&error bars&\phantom{abcd}&fixed&\phantom{ab}&error bars&\phantom{abcd}&fixed&\phantom{ab}&error bars  \\
  \addlinespace[0.05cm]
   \cmidrule[0.1pt]{1-1}
   \cmidrule[0.1pt]{3-3}
   \cmidrule[0.1pt]{5-7}
   \cmidrule[0.1pt]{9-11}
   \cmidrule[0.1pt]{13-15}
   \cmidrule[0.1pt]{17-19}
  \cmidrule[0.1pt]{1-19}
  \addlinespace[0.15cm]
                    $c_1$         & &$-0.067  $&    & $+0.003  $&     & $+0.003  $ && $-0.004  $&     & $-0.003$    &     & $-0.002$   &     & $-0.001$   &     & $-0.013$ &     & $-0.015 $ \\  
                    $c_2$         & &$-0.580  $&    & $-0.238  $&     & $-0.193 $  && $-0.272  $&     & $-0.217$    &     & $-0.409$   &     & $-0.358$   &     & $-0.250$ &     & $-0.213 $ \\  
                    $c_3$         & &$+0.008  $&    & $-0.039  $&     & $-0.029 $  && $+0.050  $&     & $+0.038$    &     & $+0.033$   &     & $+0.025$   &     & $+0.008$ &     & $+0.010 $ \\  
                    $c_4$         & &$-0.061  $&    & $-0.095  $&     & $-0.113 $  && $-0.084  $&     & $-0.105$    &     & $+0.010$   &     & $-0.015$   &     & $-0.023$ &     & $-0.048 $ \\  
                    $c_5$         & &$+0.004  $&    & $-0.001  $&     & $-0.001 $  && $+0.003  $&     & $+0.002$    &     & $-0.004$   &     & $+0.003$   &     & $-0.0004$ &     & $-0.00004$ \\ 
                    $c_6$         & &$-0.008  $&    & $-0.003  $&     & $-0.005 $  && $+0.003  $&     & $-0.001$    &     & $-0.004$   &     & $-0.017$   &     & $-0.013$ &     & $-0.007 $ \\  
 \bottomrule[1.0pt]
 \addlinespace[0.1cm]
\end{tabular}
 \end{table*}
While already Table~\ref{tab:fenchone:opt1} suggests that C3 is likely the intermediate electronically excited state state probed in the 2+1 photoexcitation process, 
the ultimate test consists in using \textit{ab initio} results for all parameters in Eq.~\eqref{eq:PADfinal}, i.e., the excited state expansion coefficients and the two-photon tensor elements, and compare the resulting Legendre coefficients to the  experimental data. The results are shown in Table~\ref{tab:fenchone_opt_vs_nonopt}
(``fixed tensor elements''). 
Choosing a slightly larger photoelectron energy, specifically 
  $0.58\,$eV instead of $0.56\,$eV, with the shift of $0.02\,$eV well within the error bars of the calculated excitation energies, 
considerably improves the agreement between  
  theoretical and experimental values, in particular for the $c_1$ coefficient. 
Additionally, we allow the tensor elements to vary within a range
of $\pm 20\%$ to account for unavoidable errors in the electronic structure calculations. The best tensor elements within the error range are obtained by minimization. The corresponding functional is defined as
\begin{eqnarray}
\label{eq:minimization_functional}
\Gamma &=& \dfrac{1}{\Gamma^{(0)}}\, \sum^6_{j=1}\omega_j\left(\dfrac{{c}_j - {c}^{\text{exp}}_j}{{c}^{\text{exp}}_j}\right)^2,
\end{eqnarray}
where $\omega_j$ are optimization weights  and 
$\Gamma^{(0)}$ is the value of the functional using the fixed tensor elements.
Table~\ref{tab:fenchone_opt_vs_nonopt} confirms 
state B to be ruled out, since it does not reproduce correctly even a single sign of the odd coefficients. For all states C, the correct signs are obtained for the lower order Legendre coefficients, up to $c_4$. State C1 yields the correct sign of $c_6$ only if the tensor elements are allowed to vary within $\pm 20\%$; the same holds for C2 and the sign of $c_5$. C3 does not reproduce the correct sign of $c_5$, but the value of $c_5$ is very small and close to zero when accounting for the error bars. In terms of PECD, the most important coefficient for fenchone is $c_1$, since its experimental value is an order of magnitude larger than that of the other odd coefficients. For $c_1$, the best agreement is obtained for state C3, differing from the experimental value by a factor of five. In contrast, the difference is by a factor of about twenty for state C1, and even larger for state C2. While $c_1$ is too small by more than an order of magnitude for states C1 and C2, $c_3$ is overestimated by a factor of five for C1 and a factor of three for C2. For states C1 and C2, the largest odd Legendre coefficient is thus $c_3$, unlike the experimental result where it is $c_1$. In contrast, the theoretical result for $c_3$ is in quantitative agreement for state C3 which therefore yields the correct ordering of the odd Legendre coefficients in terms of their magnitude. We thus conjecture that 
for fenchone, state C3 is most likely the intermediate electronically state probed in the experiment, despite the fact that $c_5$ is very close to zero. 
The reason for the discrepancy exclusively 
for $c_5$, while all other coefficients match the experimental ones at least qualitatively, is not entirely clear. A necessary condition for  non-vanishing $c_5$ is, according to Table~\ref{table:all_in_one}, 
that the $d$-wave contribution of the intermediate state to be
non-vanishing. The results shown in Table~\ref{tab:fenchone_opt_vs_nonopt} thus suggest that our calculations underestimate the $d$-wave character of C3. 
This may be caused by an improper description of 
long-range interaction between the photoelectron and the remaining ion, i.e., by the fact that the true potential felt by the photoelectron is neither central nor point-like, or by the interaction between the laser field and the photoelectron whose time dependence is neglected in our model. 
Finally, the error bars of the two-photon tensor elements may be larger than 20\%. Indeed, allowing error bars
of $\pm$50\% in the two-photon absorption tensor elements removes the
disagreement for $c_5$ and state C3. At the same time, these error bars do not significantly improve the agreement for the other two states. For example, 
the coefficient $c_1$ is $-0.0061$ for state C1 and  $-0.0045$ for state C2, leaving the conclusion that  state C3 is 
the intermediate resonance unchanged.

\begin{table}[ht]
\caption{\label{table:fenchone_50_percent_for_c5} Legendre
  coefficients  for the PAD of fenchone (calculated at a photoelectron energy of $0.58\,$eV and normalized with
  respect to $c_0$),   obtained by employing the excited state
  coefficients and two-photon tensor elements from the \textit{ab initio} for state C3 and increasing error bars of the two-photon tensor elements.
 Minimization of the functional in Eq.~\eqref{eq:minimization_functional} carried out with equal (top) and unequal (bottom, $\omega_5 = 10\omega, \omega_{j=1,\ldots,4,6}=\omega$) optimization weights.  
}
  \begin{tabular}{c !{\vrule width 0pt}c !{\vrule width -4pt}c !{\vrule width -12pt}c!{\vrule width -4pt}c !{\vrule width -4pt}c !{\vrule width -4pt}c !{\vrule width -4pt}c !{\vrule width -4pt} c!{\vrule width -4pt}c!{\vrule width -4pt}c!{\vrule width -4pt} }
   \bottomrule[0.8pt]
   \addlinespace[0.05cm]
           &exp.~\cite{LuxCPC15}   & \phantom{ab}   &fixed  &\phantom{ab}& $\pm 20\%$ &\phantom{ab}& $\pm 30\%$ &\phantom{ab}&$\pm 50\%$ &\phantom{a} \\
  \addlinespace[0.05cm]
  \cmidrule[0.1pt]{1-11}
  \addlinespace[0.15cm]
                     $c_1$           &$-0.067 $&    & $-0.012 $     && $-0.015  $  && $-0.016 $  && $-0.016 $ &\\  
                     $c_2$           &$-0.580 $&    & $+0.250 $     && $-0.213  $  && $-0.210 $  && $-0.212 $ &\\  
                     $c_3$           &$+0.008 $&    & $+0.008 $     && $+0.010  $  && $+0.010 $  && $+0.010 $ &\\  
                     $c_4$           &$-0.061 $&    & $-0.023 $     && $-0.045  $  && $-0.048 $  && $-0.048 $ &\\  
                     $c_5$           &$+0.004 $&    & $-0.0004$     && $-0.00004$  && $-0.00001$ && $+0.00002$ &\\  
                     $c_6$           &$-0.008 $&    & $-0.013 $     && $-0.007  $  && $ -0.007 $ && $-0.007 $ &\\ 
  \addlinespace[0.08cm]
  \cmidrule[0.05pt]{1-11}
  \addlinespace[0.08cm]
                    \multicolumn{2}{l}{$\Gamma$ (equal $\omega_j$)} &    & $1.0$     && $0.714 $  && $0.711 $ && $0.705 $ &\\ 
  \addlinespace[0.05cm]
  \bottomrule[0.8pt]
  \bottomrule[0.8pt]
  \addlinespace[0.05cm]
                       &exp.~\cite{LuxCPC15}   & \phantom{ab}   &fixed  &\phantom{ab}& $\pm 20\%$ &\phantom{ab}& $\pm 30\%$ &\phantom{ab}&$\pm 50\%$ &\phantom{a} \\
  \addlinespace[0.05cm]
  \cmidrule[0.1pt]{1-11}
  \addlinespace[0.15cm]
                     $c_1$           &$-0.067 $&    & $-0.012 $     && $-0.015 $  && $-0.018 $ && $-0.022 $ &\\  
                     $c_2$           &$-0.580 $&    & $+0.250 $     && $-0.223 $  && $-0.227 $ && $-0.268 $ &\\  
                     $c_3$           &$+0.008 $&    & $+0.008 $     && $+0.010 $  && $+0.011 $ && $+0.014 $ &\\  
                     $c_4$           &$-0.061 $&    & $-0.023 $     && $-0.045 $  && $-0.0504$ && $-0.033 $ &\\  
                     $c_5$           &$+0.004 $&    & $-0.0004$     && $+0.00004$ && $+0.0004$ && $+0.001 $ &\\  
                     $c_6$           &$-0.008 $&    & $-0.013 $     && $-0.006 $  && $-0.001 $ && $-0.001 $ &\\  
  \addlinespace[0.05cm]
  \cmidrule[0.05pt]{1-11}
  \addlinespace[0.08cm]
    \multicolumn{2}{l}{$\Gamma$ (unequal $\omega_j$)}   &    & $1.0$     && $0.775$  && $0.710$ && $0.686$ &\\ 
    \multicolumn{2}{l}{$\Gamma$ (equal $\omega_j$)} &    & $1.0$     && $0.720$  && $0.917$ && $0.994$ &\\ 
  
  \addlinespace[0.05cm]
   \bottomrule[0.8pt]
\end{tabular}
\end{table}
A systematic increase of the two-photon tensor error bars for  state C3 is presented in Table~\ref{table:fenchone_50_percent_for_c5}.
We compare minimization of the functional~\eqref{eq:minimization_functional} with equal weights for all Legendre coefficients (upper part of  Table~\ref{table:fenchone_50_percent_for_c5}) to that with a ten times larger weight of $c_5$ (lower part of  Table~\ref{table:fenchone_50_percent_for_c5}). 
The movitation behind the second choice is to see whether the
correct sign can be obtained for $c_5$ without the need to increase the error bars to a very high value.
When increasing the error bars of the two-photon tensor elements, while using the same optimization weights in Eq.~\eqref{eq:minimization_functional}, the value of $c_5$ is increased until it changes sign. The overall value of the functional decreases monotonically, as expected. When the optimization weight of $c_5$ is taken 10 times larger than those of all other Legendre coefficients, assuming an error range of $\pm$20\% for the two-photon tensor elements of state C3 already yields the correct sign for all Legendre coefficients. Increasing the error range in this case further improves the magnitude of $c_5$, until it differs from the experimental one by a factor of four for error bars of $\pm$50\%. However, this comes at the expense of the agreement for all other Legendre coefficients except $c_1$. It is quantified by evaluating $\Gamma$ in Eq.~\eqref{eq:minimization_functional} 
with equal weights, using the optimized two-photon tensor elements obtained with unequal weights. 

Overall, already the two-photon tensor elements taken directly from the \textit{ab initio} calculations yield a satisfactory agreement for the PAD between theory and experiment for state C3. The agreement is further improved by allowing the two-photon tensor elements to vary within a range of $\pm 20\%$
to account for the error bars of the \textit{ab initio} calculations. 
All Legendre coefficients except $c_3$ are sensitive to a variation
within this range. Except for $c_5$, i.e., underestimation of the excite state $f$-wave character, 
a surprisingly good agreement between theoretical and experimental values is obtained,
with the numerical values differing from the experimental ones up to a factor of five. 
\begin{figure}[tb]
\centering
\includegraphics[width=0.9\linewidth]{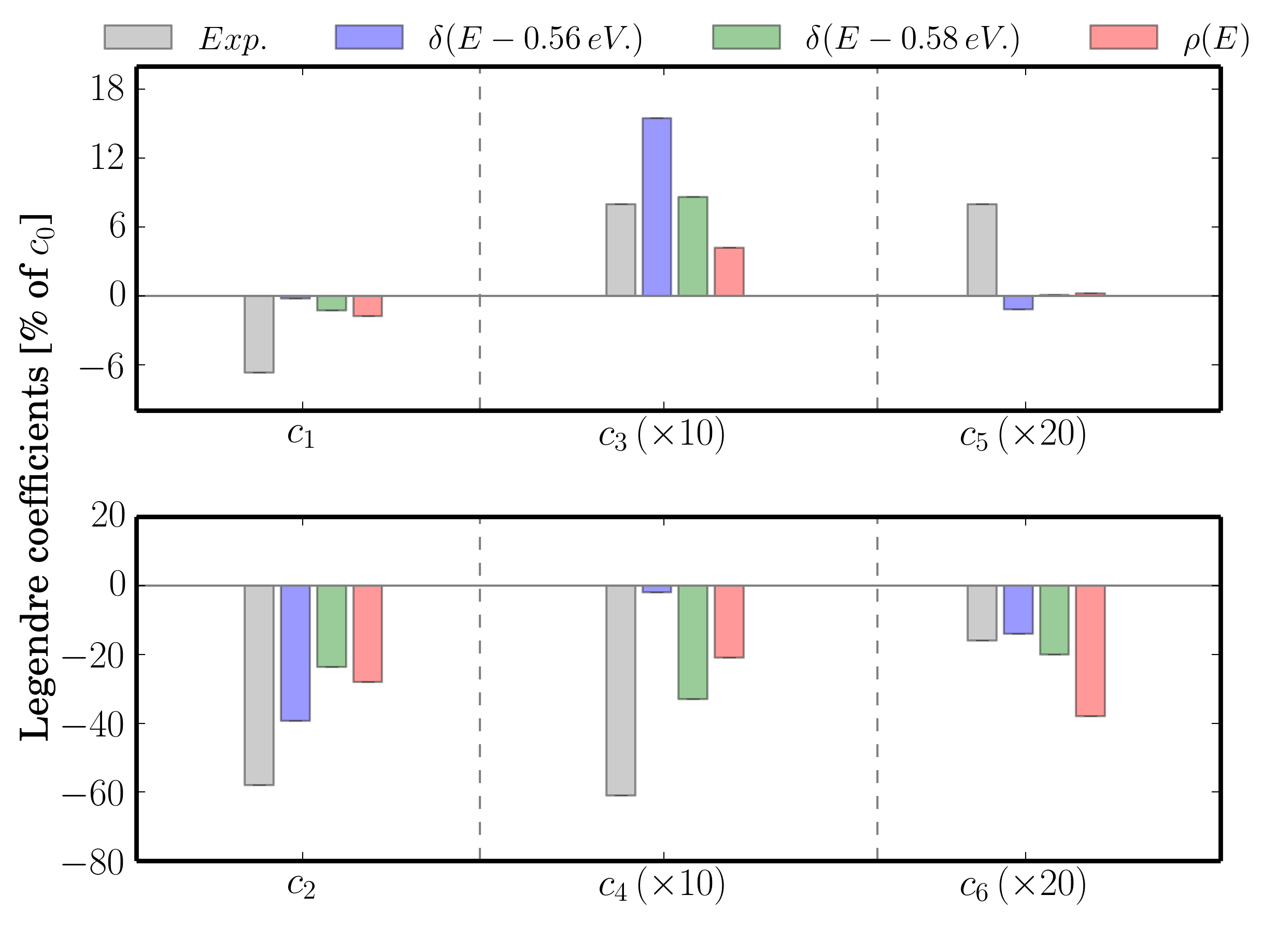}
\caption{Comparison of experimentally obtained 
  and theoretically  calculated Legendre coefficients in the PAD for
  $S$-$(+)$-fenchone, using state C3 and 
  right circular polarization.
  The calculations were carried out for a fixed photoelectron energy of
  $0.56\,$eV, respectively
  $0.58\,$eV,  as well as integrating over a Gaussian distribution
  of photoelectron energies (denoted by $\rho(E)$) centered at $0.56\,$eV with a FWHM 
  of 200$\,$meV.
}
\label{fig:fenchone:ci}  
\end{figure}
The semi-quantitative agreement between theory and experiment 
is further illustrated in Fig.~\ref{fig:fenchone:ci} where we compare 
calculation results for two specific photoelectron energies, 
$0.56\,$eV and $0.58\,$eV, to the experimentally obtained
Legendre coefficients.
The differences for the Legendre coefficients for $0.56\,$eV and
$0.58\,$eV indicates the dependence of our results on the error
bar of the calculated excitation energy of the intermediate
electronically excited state.
Additionally, Fig.~\ref{fig:fenchone:ci} also shows the result
of integrating over a normal distribution of photoelectron energies
centered at $0.56\,$eV with a full width at half maximum (FWHM) of
200$\,$meV. This accounts for the experimental averaging over
photoelectron energies~\cite{LuxCPC15}. The disagreement
between theoretical and experimental results amounts  to a factor
of about two which translates into a ``mean'' PECD 
of 3\% and 4\% for the fixed and $\pm 20\%$ adjustable tensor elements, respectively,
compared to the experimental value of 
10.1\%~\cite{LuxCPC15}. 
\begin{figure}[tb]
\centering
\includegraphics[width=0.9\linewidth]{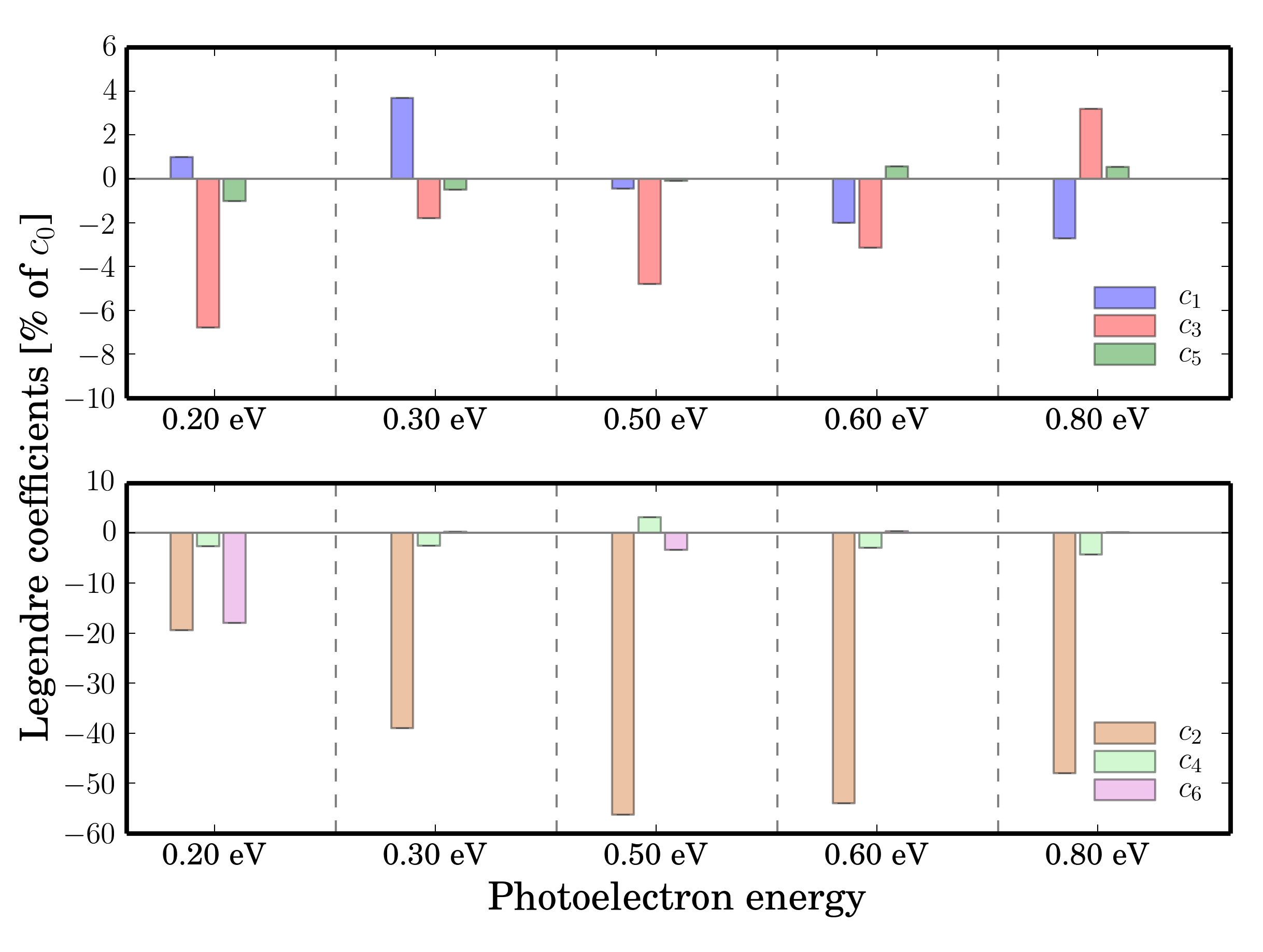}
\caption{Dependence of the calculated Legendre coefficients on photoelectron energy for the PAD of state C3 for $S$-$(+)$-fenchone , using right
  circular polarization.
}
\label{fig:fenchone:civsE}  
\end{figure}
The dependence of the calculated Legendre coefficients on the
photoelectron energy is investigated in more detail in 
Fig.~\ref{fig:fenchone:civsE}. A non-monotonic behavior is observed
for all orders. Such a non-monotonic behavior of the Legendre
coefficients as a function of the photoelectron energy has
already been reported for $c_1$ in the one-photon ionization of randomly oriented
molecules~\cite{HardingChemPhys2005}. It reflects the dependence of
the Legendre coefficients on the radial part of the photoelectron
wavefunction.  
\begin{table*}[tb]
\caption{\label{table:kummer_vs_plane_waves} 
  Legendre coefficients in the PAD of fenchone for state C3 and different
  photoelectron energies, obtained with
  hydrogenic continuum functions which include the Coulomb interaction
  between photoelectron and photoion and plane waves where this
  interaction is neglected. $\rho(E)$ stands for integration over a
  Gaussian distribution of photoelectron energies centered at
  0.56$\,$eV with a FWHM of 200$\,$meV.
}
\centering
\begin{minipage}{1\linewidth}
  \begin{tabular}{c !{\vrule width 0pt}c !{\vrule width -4pt}c !{\vrule width -4pt}c!{\vrule width -4pt}c !{\vrule width -4pt}c !{\vrule width -4pt}c !{\vrule width -4pt}c !{\vrule width -4pt}c !{\vrule width -4pt}c !{\vrule width -4pt}c !{\vrule width -4pt}c !{\vrule width -4pt}c !{\vrule width -4pt}c !{\vrule width -4pt}c !{\vrule width -4pt}c}\toprule[1.0pt]
 & & & \multicolumn{5}{c}{
hydrogenic continuum 
 functions} &\phantom{abc}&  & & \multicolumn{5}{c}{plane waves}\\
& &\phantom{abc}&\multicolumn{5}{c}{photoelectron energy~(eV)}&\phantom{abc}&&\phantom{abc}&\multicolumn{5}{c}{photoelectron energy~(eV)}\\
coeffs.\phantom{a } &exp.~\cite{LuxCPC15} &\phantom{abcd} &\multicolumn{1}{c}{$\hspace{0.7cm}0.36\hspace{0.5cm}$}&\multicolumn{1}{c}{$0.58$}& &\multicolumn{1}{c}{$\hspace{0.7cm}0.75\hspace{0.7cm}$}&\multicolumn{1}{c}{$\rho(E)$} &&\phantom{a} & &\multicolumn{1}{c}{$\hspace{0.7cm}0.36\hspace{0.5cm}$}& \multicolumn{1}{c}{$0.58$}&&\multicolumn{1}{c}{$\hspace{0.7cm}0.75\hspace{0.5cm}$}& \multicolumn{1}{c}{$\rho(E)$}\\
\hline
$c_1$   &  $-0.061$    &  &     $-0.002$ &$-0.012$ && $-0.058$   &$-0.037$ & & 
&&         $+0.002$ &$+0.006$ && $+0.002$&$-0.017$\\  

$c_2$ &    $-0.580$ &  &$-0.341$ &$-0.250$ && $-0.385$  &$-0.411$ & & 
&&         $+0.034$ &$+0.012$ && $-0.029$&$-0.126$\\  

$c_3$ &$+0.008$ &  &$-0.008$ &$+0.008 $ && $+0.170$  &$+0.005$ && 
&& $-0.006$ &$-0.061$ && $-0.012$&$+0.009$\\  

$c_4$ &$-0.061$ &  &$+0.002$ &$-0.023$ && $-0.008$ &$-0.030$ & &
&& $+0.114$ &$-0.178$ && $-0.001$&$-0.051$\\  

$c_5$ &$+0.004$ &  &$-0.001$ &$-0.0004$ && $+0.192$ &$-0.00003$  & &
&& $ +0.0001$ &$-0.004$ && $-0.001$&$+0.00001$\\  
$c_6$ &$-0.008$ &  &$-0.004$ &$-0.007$ && $+0.001$&$-0.004$ & & && $+0.001$ &$-0.013$ && $+0.006$&$-0.004$\\  
\bottomrule[1.0pt]
\addlinespace[0.1cm]
\end{tabular}
\end{minipage}
\end{table*}
This dependence                                                         
is studied further in Table~\ref{table:kummer_vs_plane_waves}, where we
compare the Legendre coefficients obtained with the Kummer confluent
functions, i.e., the hydrogenic continuum wavefunctions defined Appendix~\ref{subsec:H-cont}, to those obtained with plane waves.  
The latter completely neglect the Coulomb
interaction between photoelectron and photoion. The plane waves clearly fail to reproduce the experimentally observed PECD, see in particular the values for $0.58\,$eV.
Moreover, their values vary drastically with photoelectron energy. This difference is most likely explained by the highly oscillatory nature of plane waves even at short distances, in contrast to the hydrogenic scattering functions. 
Our finding is in line with the observation of Ref.~\cite{LeinPRA2014} for the strong field approximation where plane waves fail completely to produce any PECD. In our model, non-zero odd Legendre coefficients are obtained, but a description of the photoelectron continuum that accounts for the Coulomb interaction between photoelectron and photoion provides clearly better results.

\subsection{Camphor}
\label{subsec:camphor}
\begin{table*}[tbp]
 \caption{\label{tab:camphor:opt1} 
   Legendre coefficients for the
   PAD of camphor (calculated at a photoelectron energy
   of $0.52\,$eV and normalized with respect to $c_0$), obtained by
   fitting to the experimental 
   values~\cite{LuxCPC15} with the excited state coefficients
   $a^{\ell_o}_{m_o}$ as free parameters. Only odd (top) and 
   both odd and even (bottom)  contributions were accounted for in the fitting procedure. 
   The Rydberg states B, C1, C2 and C3 of camphor are  characterized by
   their two-photon absorption tensor, cf. Tab.~\ref{tab:twophotonA}.
}
  \begin{tabular}{c !{\vrule width -4pt}c !{\vrule width -4pt}c !{\vrule width -4pt}c!{\vrule width -4pt}c !{\vrule width -4pt}c !{\vrule width -4pt}c !{\vrule width -4pt}c !{\vrule width -4pt}c !{\vrule width -4pt}c !{\vrule width -4pt}c !{\vrule width -4pt}c !{\vrule width -4pt}c !{\vrule width -4pt}c !{\vrule width -4pt}
   c !{\vrule width -4pt}c !{\vrule width -4pt}c !{\vrule width -4pt}c !{\vrule width -4pt}c !{\vrule width -4pt} 
    }
    \toprule[1.0pt]
   \addlinespace[0.05cm]
   & & \phantom{abc}       & & \multicolumn{3}{c}{state B} & &\multicolumn{3}{c}{state C1} & &\multicolumn{3}{c}{state C2}  & &\multicolumn{3}{c}{state C3}\\
   coeffs.  &  \phantom{abc}                     &exp.~\cite{LuxCPC15}   & \phantom{abcdef}   &$d$ waves &\phantom{ab}& $f$ waves &\phantom{abcd}&$d$ waves &\phantom{ab}&$f$ waves  &\phantom{abcd}&$d$ waves &\phantom{ab}&$f$ waves  &\phantom{abcd}&$d$ waves &\phantom{ab}&$f$ waves  \\
  \addlinespace[0.05cm]
   \cmidrule[0.1pt]{1-1}
   \cmidrule[0.1pt]{3-3}
   \cmidrule[0.1pt]{5-7}
   \cmidrule[0.1pt]{9-11}
   \cmidrule[0.1pt]{13-15}
   \cmidrule[0.1pt]{17-19}
   
  \addlinespace[0.15cm]
  $c_1$          &&$+0.026$&    & $+0.026 $&     & $+0.024 $         &&     $+0.028   $&     & $+0.026$     &&     $+0.020   $&     & $+0.027$   &&     $+0.025   $&     & $+0.026$   \\  
  $c_3$          &&$-0.053$&    & $+0.038 $&     & $-0.025 $         &&     $-0.038   $&     & $-0.040$     &&     $-0.032   $&     & $-0.042$   &&     $-0.042   $&     & $-0.047$   \\  
  $c_5$          &&$+0.008$&    & $     -$&      & $+0.004$          &&     $-        $&     & $+0.006$     &&     $-        $&     & $+0.006$   &&     $-        $&     & $+0.005$    \\  
  \addlinespace[0.1cm]
  \cmidrule[0.1pt]{1-19}
  \addlinespace[0.15cm]
                    $c_1$         & &$+0.026 $&    & $+0.099   $&     & $+0.096   $  && $+0.051    $&     & $+0.054$    &     & $+0.054$    &    & $+0.041 $   &     & $+0.040  $ &     & $+0.048 $ \\  
                    $c_2$         & &$-0.670 $&    & $-0.198   $&     & $-0.248   $  && $-0.130    $&     & $-0.209$    &     & $-0.135$   &     & $-0.170 $   &     & $-0.193  $ &     & $-0.230 $ \\  
                    $c_3$         & &$-0.053 $&    & $-0.034   $&     & $-0.022   $  && $-0.023    $&     & $-0.020$    &     & $+0.037$   &     & $+0.043 $   &     & $+0.028  $ &     & $+0.013 $ \\  
                    $c_4$         & &$+0.012 $&    & $+0.013   $&     & $+0.013   $  && $+0.014    $&     & $+0.013$    &     & $+0.017$   &     & $+0.018 $   &     & $+0.011  $ &     & $+0.019 $ \\  
                    $c_5$         & &$+0.008 $&    & $ -       $&     & $+0.001   $  && $ -        $&     & $+0.001$    &     & $ -     $  &     & $+0.002 $   &     & $ -      $ &     & $ +0.002$ \\  
                    $c_6$         & &$-0.001 $&    & $-0.001   $&     & $-0.001   $  && $-0.001    $&     & $-0.001 $    &     & $-0.003$   &     & $-0.002 $   &     & $-0.001  $ &     & $-0.003 $ \\  
\bottomrule[1.0pt]
\addlinespace[0.1cm]
\end{tabular}
\end{table*}
We now turn to camphor, for which the experimentally recorded 
photoelectron spectrum peaks at $0.52\,$eV~\cite{LuxCPC15}. Analogously to our discussion for fenchone, we first investigate 
possible candidates for the intermediate resonance by considering 
the respective two-photon tensor alone and treating the excited state expansion coefficients as free optimization parameters.  The results are displayed in Table~\ref{tab:camphor:opt1}, comparing the optimization that targets only the odd Legendre coefficients to that considering both odd and even $c_j$. For all states, a non-zero $c_5$ coefficient is only obtained by including $f$-waves in the electronically excited state (corresponding to $n_o=4$), as expected. When expanding up to $f$-waves, all four candidates allow for odd Legendre coefficients close to the experimental ones, unlike the case of fenchone, where state B could already be ruled out at this stage. However, states C2 and C3 do not allow for the correct sign of $c_3$, when the optimization targets both odd and even Legendre coefficients.  

\begin{table*}[tbp]
 \caption{\label{table:reexpansion_camphore_other_states_0.52} 
  Legendre coefficients for the PAD of camphor (calculated at a photoelectron
  energy of $0.52\,$eV and normalized with respect 
  to $c_0$), obtained by employing the excited state
  coefficients and two-photon tensor elements from the \textit{ab
    initio} calculations. When including error bars, the tensor
  elements are allowed to vary within
  $\pm$20\%.} 
  \begin{tabular}{c !{\vrule width 0pt}c !{\vrule width -4pt}c !{\vrule width -4pt}c!{\vrule width -4pt}c !{\vrule width -4pt}c !{\vrule width -4pt}c !{\vrule width -4pt}c !{\vrule width -4pt}c !{\vrule width -4pt}c !{\vrule width -4pt}c !{\vrule width -4pt}c !{\vrule width -4pt}c !{\vrule width -4pt}c !{\vrule width -4pt}
   c !{\vrule width -4pt}c !{\vrule width -4pt}c !{\vrule width -4pt}c !{\vrule width -4pt}c !{\vrule width -4pt} 
    }
    \toprule[1.0pt]
   \addlinespace[0.05cm]
   & & \phantom{abc}       & & \multicolumn{3}{c}{state B} & &\multicolumn{3}{c}{state C1} & &\multicolumn{3}{c}{state C2}  & &\multicolumn{3}{c}{state C3}\\
   coeffs.  &  \phantom{a}                     &exp.~\cite{LuxCPC15}   & \phantom{abcde}   &fixed &\phantom{ab}& error bars&\phantom{abcd}&fixed&\phantom{ab}&error bars&\phantom{abcd}&fixed&\phantom{ab}&error bars&\phantom{abcd}&fixed&\phantom{ab}&error bars  \\
  \addlinespace[0.05cm]
   \cmidrule[0.1pt]{1-1}
   \cmidrule[0.1pt]{3-3}
   \cmidrule[0.1pt]{5-7}
   \cmidrule[0.1pt]{9-11}
   \cmidrule[0.1pt]{13-15}
   \cmidrule[0.1pt]{17-19}
  \cmidrule[0.1pt]{1-19}
  \addlinespace[0.15cm]
                    $c_1$         & &$+0.026  $&    & $+0.003  $&     & $+0.002  $  && $+0.002   $&     & $+0.001 $    &     & $-0.002 $   &     & $-0.002 $   &     & $-0.001 $ &     & $-0.001 $ \\  
                    $c_2$         & &$-0.678  $&    & $-0.384  $&     & $-0.383  $  && $-0.389   $&     & $-0.401 $    &     & $-0.395 $   &     & $-0.395 $   &     & $-0.421 $ &     & $-0.425 $ \\  
                    $c_3$         & &$-0.053  $&    & $-0.025  $&     & $-0.022  $  && $-0.020   $&     & $-0.017 $    &     & $+0.005 $   &     & $+0.008 $   &     & $+0.004 $ &     & $+0.003 $ \\  
                    $c_4$         & &$+0.012  $&    & $-0.066  $&     & $-0.050  $  && $+0.020   $&     & $+0.023 $    &     & $+0.004 $   &     & $-0.002 $   &     & $-0.008 $ &     & $+0.0001$ \\  
                    $c_5$         & &$+0.008  $&    & $-0.002  $&     & $-0.001  $  && $+0.0001  $&     & $+0.0001$    &     & $+0.001 $   &     & $+0.001 $   &     & $+0.0003$ &     & $+0.001 $ \\  
                    $c_6$         & &$-0.001 $&     & $+0.043  $&     & $+0.035  $  && $-0.026   $&     & $-0.023 $    &     & $-0.008 $   &     & $-0.001 $   &     & $+0.005 $ &     & $-0.0004$ \\  
 \bottomrule[1.0pt]
 \addlinespace[0.1cm]
 \end{tabular}
 \end{table*}
 \begin{table*}[tbp]
 \caption{\label{table:reexpansion_camphore_other_states_0.58} 
The same as Table~\ref{table:reexpansion_camphore_other_states_0.52} but for a photoelectron energy of 0.58$\,$eV.
 } 
  \begin{tabular}{c !{\vrule width 0pt}c !{\vrule width -4pt}c !{\vrule width -4pt}c!{\vrule width -4pt}c !{\vrule width -4pt}c !{\vrule width -4pt}c !{\vrule width -4pt}c !{\vrule width -4pt}c !{\vrule width -4pt}c !{\vrule width -4pt}c !{\vrule width -4pt}c !{\vrule width -4pt}c !{\vrule width -4pt}c !{\vrule width -4pt}
   c !{\vrule width -4pt}c !{\vrule width -4pt}c !{\vrule width -4pt}c !{\vrule width -4pt}c !{\vrule width -4pt} 
    }
    \toprule[1.0pt]
   \addlinespace[0.05cm]
   & & \phantom{abc}       & & \multicolumn{3}{c}{state B} & &\multicolumn{3}{c}{state C1} & &\multicolumn{3}{c}{state C2}  & &\multicolumn{3}{c}{state C3}\\
   
   coeffs.  &  \phantom{a}                     &exp.~\cite{LuxCPC15}   & \phantom{abcde}   &fixed &\phantom{ab}& error bars&\phantom{abcd}&fixed&\phantom{ab}&error bars&\phantom{abcd}&fixed&\phantom{ab}&error bars&\phantom{abcd}&fixed&\phantom{ab}&error bars  \\
  \addlinespace[0.05cm]
   \cmidrule[0.1pt]{1-1}
   \cmidrule[0.1pt]{3-3}
   \cmidrule[0.1pt]{5-7}
   \cmidrule[0.1pt]{9-11}
   \cmidrule[0.1pt]{13-15}
   \cmidrule[0.1pt]{17-19}
  \cmidrule[0.1pt]{1-19}
  \addlinespace[0.15cm]
                    $c_1$         & &$+0.026  $&    & $+0.033  $&     & $+0.030  $  && $+0.026   $&     & $+0.027 $    &     & $-0.005 $   &     & $-0.009 $   &     & $-0.004 $ &     & $-0.002 $ \\  
                    $c_2$         & &$-0.678  $&    & $-0.450  $&     & $-0.498  $  && $-0.477   $&     & $-0.502 $    &     & $-0.431 $   &     & $-0.427 $   &     & $-0.432 $ &     & $-0.437 $ \\  
                    $c_3$         & &$-0.053  $&    & $-0.029  $&     & $-0.031  $  && $-0.024   $&     & $-0.022 $    &     & $-0.003 $   &     & $-0.0002$   &     & $+0.001 $ &     & $-0.003 $ \\  
                    $c_4$         & &$+0.012  $&    & $-0.074  $&     & $-0.034  $  && $+0.003   $&     & $+0.009 $    &     & $-0.022 $   &     & $-0.036 $   &     & $-0.026 $ &     & $-0.018 $ \\  
                    $c_5$         & &$+0.008  $&    & $-0.001  $&     & $-0.001  $  && $+0.0001  $&     & $+0.0001$    &     & $+0.0002$   &     & $+0.001 $   &     & $+0.0002$ &     & $+0.0001$ \\  
                    $c_6$         & &$-0.001 $&     & $+0.030  $&     & $+0.024  $  && $-0.015   $&     & $-0.011 $    &     & $-0.020 $   &     & $-0.010 $   &     & $+0.0001$ &     & $+0.003 $ \\  
 \bottomrule[1.0pt]
 \addlinespace[0.1cm]
 \end{tabular}
 \end{table*}
Once again, the ultimate test to rule out a given state consists in
using both two-photon tensor elements and excited state expansion coefficients obtained from the \textit{ab initio} calculations. The corresponding 
results 
are shown in Table~\ref{table:reexpansion_camphore_other_states_0.52}. First of all, Table~\ref{table:reexpansion_camphore_other_states_0.52} confirms that states C2 and C3 are not the intermediate resonance probed in the experiment, since both states yield the wrong sign for both $c_1$ and $c_3$. Comparing the remaining two candidates, states B and C1, a much better agreement is observed for C1 which yields the correct signs for all Legendre coefficients. In contrast, state B only yields correct signs for the lower orders, $c_1$, $c_2$, and $c_3$. When accounting for the error bars in the two-photon tensor, a  correct sign is additionally obtained for $c_4$, but the signs for
$c_5$ and $c_6$ still cannot properly be reproduced with state B as intermediate resonance. As to the state C1, not only all signs but also 
the correct order of magnitude for $c_2$, $c_3$ and $c_4$ is observed, whereas the values are too small by one order of magnitude  for $c_1$ and by two orders for $c_5$ and too large by one order of magnitude for $c_6$. Allowing the
two-photon absorption tensor for state C1 to vary within an error range of $\pm 20\%$ does not yield any significant improvement. It therefore does not seem to be the unavoidable error in the two-photon tensor elements that is important. 

A second source of error in the \textit{ab initio} calculations is found in the excitation energy of the intermediate electronically excited state. This is reflected in the photoelectron energy. We thus present results for a second photoelectron energy, 0.58$\,$eV in 
Table~\ref{table:reexpansion_camphore_other_states_0.58}.  For 
state C1, all signs still match, and the correct order of magnitude is now obtained for $c_1$ to $c_4$. In particular, $c_1$ is now in quantitative agreement with the experimental value, and $c_2$ and $c_3$ differ by less than factor of 1.5, respectively 2.5.
Despite the disagreement in the numerical values for $c_5$ and $c_6$, C1 is clearly the state the best matches the experimental data---the results obtained for states 
B, C2 and C3 show the same deficiencies as in Table~\ref{table:reexpansion_camphore_other_states_0.52}. 

\begin{table}[ht]
\caption{\label{table:camphore_50_percent_for_c6} Legendre
  coefficients  for the PAD of camphor (calculated at a photoelectron energy of $0.58$ eV and normalized with
  respect to $c_0$), obtained by employing the excited state
  coefficients and two-photon tensor elements from the \textit{ab initio}
  calculations for state C3 and increasing error bars of the two-photon tensor elements. 
 Minimization of the functional $\Gamma$ in Eq.~\eqref{eq:minimization_functional} is  carried out with equal  optimization weights.  }
  \begin{tabular}{c !{\vrule width 0pt}c !{\vrule width -4pt}c !{\vrule width -4pt}c!{\vrule width -4pt}c !{\vrule width -4pt}c !{\vrule width -4pt}c !{\vrule width -4pt}c !{\vrule width -4pt} c!{\vrule width -4pt}c!{\vrule width -4pt}c!{\vrule width -4pt} }
   \toprule[0.8pt]
  \addlinespace[0.03cm]
   coeffs.                    &exp.~\cite{LuxCPC15}   & \phantom{ab}   &fixed  &\phantom{ab}& $\pm 20\%$ &\phantom{ab}& $\pm 30\%$ &\phantom{ab}&$\pm 50\%$ &\phantom{a} \\
  \addlinespace[0.05cm]
  \cmidrule[0.1pt]{1-11}
  \addlinespace[0.15cm]
                     $c_1$           &$+0.026 $&    & $+0.026 $     && $+0.027 $  && $+0.026 $  && $+0.022 $ &\\  
                     $c_2$           &$-0.678 $&    & $-0.477 $     && $-0.502 $  && $-0.515 $  && $-0.529 $ &\\  
                     $c_3$           &$-0.053 $&    & $-0.024 $     && $-0.022 $  && $-0.020 $  && $-0.014 $ &\\  
                     $c_4$           &$+0.012 $&    & $+0.003 $     && $+0.009 $  && $+0.012 $  && $+0.012 $ &\\  
                     $c_5$           &$+0.008 $&    & $+0.0001$     && $+0.0001$  && $+0.0001$  && $+0.0003 $ &\\  
                     $c_6$           &$-0.001 $&    & $-0.015 $     && $-0.011 $  && $-0.008 $  && $-0.001 $ &\\ 
  \addlinespace[0.15cm]
  \cmidrule[0.05pt]{1-11}
  \addlinespace[0.08cm]
                     $\Gamma$      &$       $&    & $1.0$     && $0.50$  && $0.26$ && $0.01$ &\\ 
  \addlinespace[0.08cm]
   \bottomrule[0.9pt]
\end{tabular}
\end{table}
The agreement with the experimental data obtained for state C1 can be further improved by allowing for larger error bars in the two-photon tensor elements. This is demonstrated in Table~\ref{table:camphore_50_percent_for_c6}. In fact, the agreement can be made fully quantitative, except for $c_5$, when increasing the error bars up to $\pm$50\%, as indicated by the small value of the optimization functional. 
In comparison to fenchone, cf. Table~\ref{table:fenchone_50_percent_for_c5}, 
minimization  results in significantly smaller values for  $\Gamma$, as 
the error range is increased. Also, 
the higher order Legendre coefficients are found to be more sensitive to
modifications of the two-photon tensor elements than the lower
ones. This is not surprising since the higher order coefficients
depend more strongly on the anisotropy induced by the two-photon absorption.  
Analogously to fenchone, $c_5$ has the correct sign but remains too small by one order of magnitude. This indicates once more that we underestimate significantly the $d$-wave contribution to the intermediate electronically excited state. It amounts to just 6\% for both
fenchone and camphor in our calculations.

\begin{figure}[tb]
\centering
\includegraphics[width=0.9\linewidth]{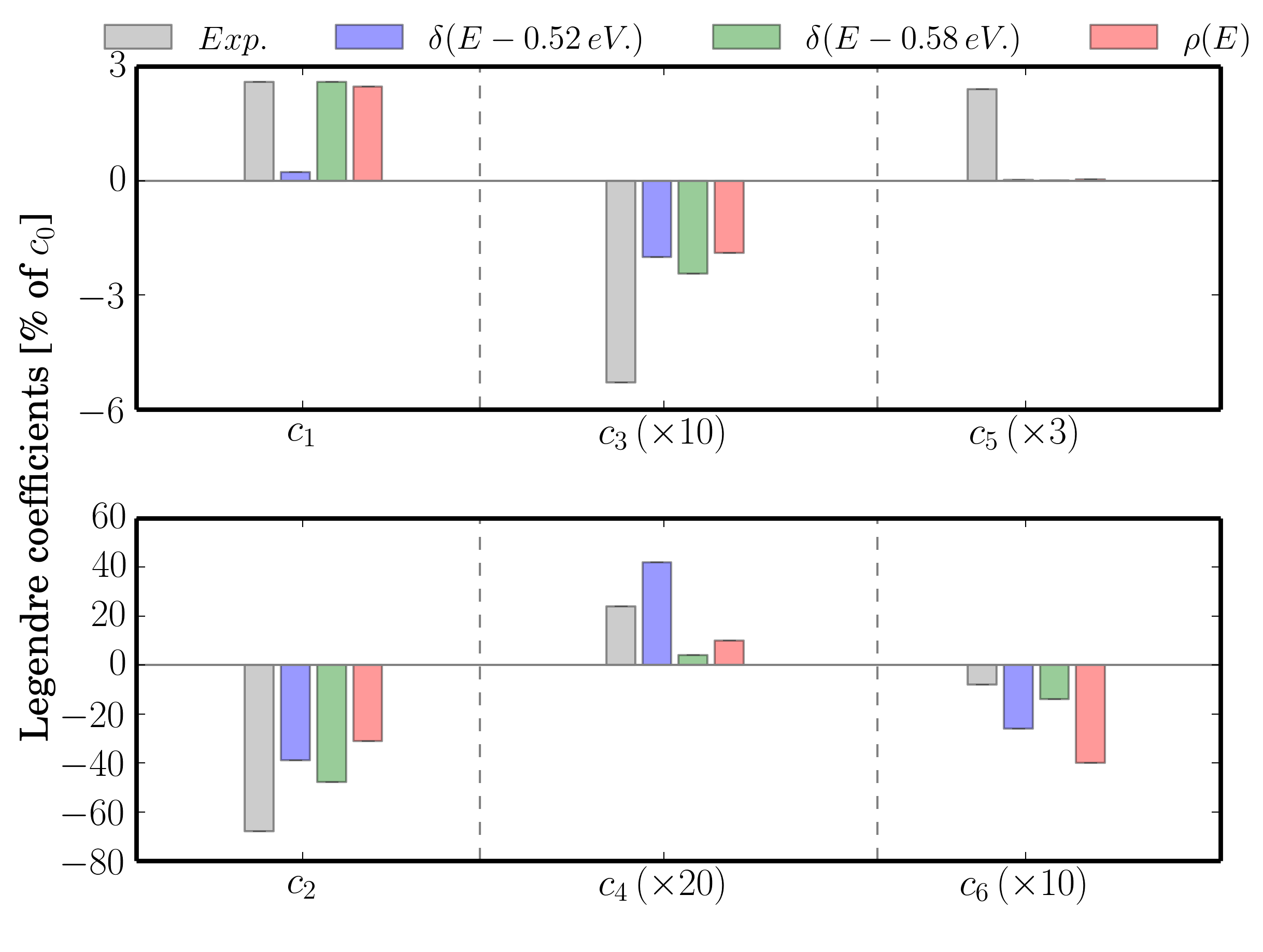}
\caption{Comparison of experimentally obtained 
  and theoretically
  calculated Legendre coefficients in the PAD for $R$-$(+)$-camphor ,
  using  state C1 and  right circular polarization.
  The calculations considered fixed photoelectron energies of
  $0.52\,$eV and 
  $0.58\,$eV as well as an integration over a Gaussian distribution
  of energies centered at $0.58\,$eV with a FWHM
  of 200$\,$meV. 
}
\label{fig:camphor:ci}  
\end{figure}
\begin{figure}[tb]
\centering
\includegraphics[width=0.9\linewidth]{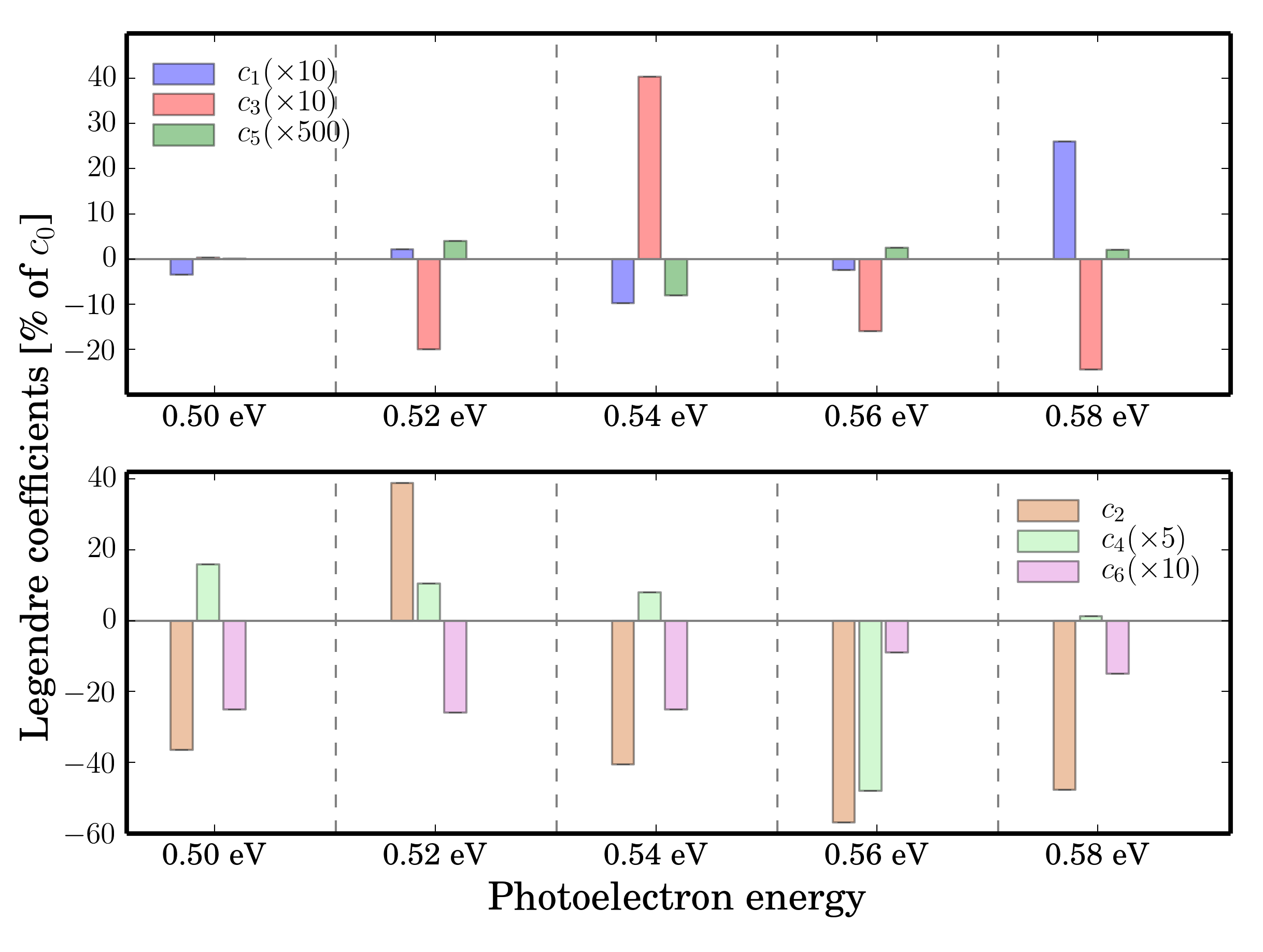}
\caption{Dependence of the calculated Legendre coefficients in the PAD
  of camphor, state C1, on the photoelectron energy  within the range of
  0.50$\,$eV to 0.58$\,$eV using right circularly polarized light.}
\label{fig:camphor:civsE}  
\end{figure}
The discussion above is summarized and illustrated in Fig.~\ref{fig:camphor:ci} which shows, besides the Legendre coefficients for photoelectron energies of 0.52$\,$eV and 0.58$\,$eV, those obtained when integrating over a normal distribution of photoelectron energies, centered at 0.52$\,$eV, with a FWHM of 200$\,$meV. The latter mimicks the spectral bandwidth in the experiment.
Introducing a distribution of photoelectron energies slightly  worsens the agreement between theory and experiment. 
This can be attributed  to the striking sensitivity of 
the Legendre coefficients on photoelectron energy, as shown  in Fig.~\ref{fig:camphor:civsE}.
A further improvement of the theoretical model would thus require experimental data for more than one photoelectron energy and with better energy resolution.

\subsection{Discussion and Summary}
\label{subsec:discussion}
Before concluding our paper, we briefly summarize our main findings. 
Our model describing the one-photon photoionization of an "initial" state that is prepared by non-resonant, orientation-dependent  two-photon absorption using a single-center approximation of the photoelectron continuum and ideas from optimal control 
allows for PECD as defined in Eq.~\eqref{eq:opt_func} of up to 35\%. This is, within
our model, the maximum PECD that could be expected for an ensemble of randomly
oriented chiral molecules. The upper limit is below 100\% is due to the random
orientation of the molecules and, possibly, due to the underlying approximations
made within our model.
One might thus speculate whether a better treatment of e.g. static exchange or contributions from the magnetic dipole interaction would allow for raising this limit even higher. It is, at any rate, already  significantly higher than the largest PECD observed experimentally so far~\cite{LuxACIE12,LehmannJCP13,Janssen2014,LuxCPC15,FanoodPCCP15}. This encourages studies of molecules beyond bicyclic ketones, both experimentally and theoretically. 

Our model accounts for the electronic structure of the experimentally investigated examples of fenchone and camphor in terms of their two-photon absorption tensor and intermediate electronically excited state based on \textit{ab initio} calculations. In both cases, there are several candidate electronic states which could serve as the intermediate resonance. For fenchone, knowledge of the two-photon tensors of the candidate states alone already suggests state C3 to be the intermediate resonance. Calculations 
employing both two-photon tensors  and excited state wavefunctions confirm this conjecture, in particular if the calculations account for error bars in the two-photon tensor. Compared to the other electronically excited states that could be accessed by the two-photon excitation, state C3 has a much larger $d$-wave component than all other states. The largest disagreement is observed in the Legendre coefficient $c_5$, suggesting that our model underestimates the $f$-wave component of state C3. For the lower order Legendre coefficients, a semi-quantitative agreement between theoretical and experimental values is obtained.

We find proper account of the Coulomb interaction between photoelectron and photoion to be crucial. When replacing, in our expansion of the photoelectron continuum wavefunction, hydrogenic basis functions by plane waves, no agreement with the experimental values is obtained. This is in line with an earlier study of PECD using the strong-field approximation~\cite{LeinPRA2014}, where plane waves completely fail to produce any PECD.

In contrast to fenchone, knowledge of the two-photon tensors for camphor is not sufficient to point to a single state as the intermediate resonance. However, calculations accounting for the \textit{ab initio} two-photon absorption matrix elements and excited state wavefunctions strongly suggest state C1 to be the intermediate resonance, in particlar when including error bars of the two-photon absorption tensor. The agreement is found to depend very strongly on the photoelectron energy, with semi-quantitative agreement found for a slightly larger value than the experimental one. Such an energy shift could be explained by the error bars of the calculated excitation energy or by the dynamic Stark shift, which is neglected in our model.

\section{Conclusions \& Outlook}
\label{sec:conclusions}
We have derived a theoretical model to study PECD after (2+1)
resonantly enhanced multi-photon ionization in randomly oriented
chiral molecules. The model is based on a perturbative treatment of
the light-matter interaction within the electric dipole
approximation and combines an \textit{ab initio}
description  of the non-resonant two-photon absorption with a
single-center expansion of the 
photoelectron wavefunction into hydrogenic continuum functions. This
allows to account for the Coulomb interaction between photoelectron
and photoion as well as electronic correlations in the transition to
the intermediate
electronically excited state. It neglects static exchange and 
dynamic correlations in the
interaction of the photoelectron with the parent ion as well as the 
time-dependence of the laser pulse and the possible multi-center
character of the continuum wavefunction.
The model
correctly reproduces the basic symmetry behavior expected under exchange of
handedness and exchange of light helicity.

Making use of the fundamental selection rules for two-photon absorption
and one-photon ionization, we have shown which Legendre coefficients
may be expected in the photoelectron angular distributions, depending
on the basic geometric properties in the electronic structure of the
molecules as well as the possible combinations of polarization for
two-photon absorption and one-photon ionization. 
We have identified the role of the two-photon absorption tensor and
intermediate 
state wavefunction---it is the partial wave decomposition of the
latter which determines PECD whereas the two-photon absorption tensor
(in the electronic dipole approximation)
merely introduces an anisotropic distribution of photoexcited
molecules. Notably, the anisotropy is achieved by selection and not by
rotational dynamics which would occur on a much slower timescale than
that of femtosecond laser excitation. 

We have applied our theoretical framework to fenchone and camphor,
which have been studied extensively in recent
experiments~\cite{LuxACIE12,LehmannJCP13,Janssen2014,LuxCPC15,FanoodPCCP15}.   
The \textit{ab initio} calculations employed the coupled cluster
method as well 
as density functional theory. Due to the Rydberg-like character of the
intermediate electronically excited state, diffuse basis functions
needed to be added to the standard basis sets. This has allowed to
reach a reasonable agreement with 
experimental values for the excited state energies. 

We have used the electronic structure data to calculate the
photoionization cross section. 
Accounting for the basic structure of the two-photon absorption tensor
alone has already  allowed us to qualitatively reproduce the
experimental results for fenchone and camphor. The minimal requirement
was identified to be a 
contribution of $d$-waves in the intermediate electronically excited
state. Such a contribution can be expected if the two-photon
absorption tensor is anisotropic. 
Employing the \textit{ab initio} data in the calculation of the
photoelectron angular distribution, we have obtained a
semi-quantitative agreement between theoretical and experimental
Legendre coefficients characterizing the photoelectron angular
distribution.

The satisfactory agreement of our model with the experimental data
encourages a number of follow-up studies. First of all, 
a fully time-dependent description should be employed, following the
lines of Ref.~\cite{SeidemanPRA2001}, because the photoelectron angular
distributions depend on the polarization  
as well as the dynamics~\cite{HardingChemPhys2005}. Based on the model
developed here, an extension to time-dependent studies is
straightforward, but will require substantial numerical effort. Such
an extension will allow to investigate the dependence of the
photoelectron angular distribution on the laser parameters, including
intensity, central frequency, spectral bandwidth and
varying polarization. The latter would be a first step towards the coherent
control of PECD. 

In parallel to accounting for time-dependent effects, the electronic
structure treatment may be improved. In particular, the
  multi-center character of the continuum wavefunction can be
  accounted for by employing Dyson orbitals
  in the calculation of the photoionization cross
  section~\cite{OanaJCP07,OanaJCP09,HumeniukJCP13}. 
Moreover, a perturbative
treatment of the static exchange for the photoelectron and extension
to beyond the electric dipole approximation should be
straightforward. The former would allow for a detailed  study of the
dependence of the angular distribution on the photoelectron energy,
including low photoelectron kinetic energies. It would thus open the
way toward investigating the role of the chiral ionic core in the dynamics
leading to  the photoelectron angular distributions.  
An extension to beyond the electric dipole approximation would allow for a unified theoretical treatment of further observables beyond PECD, such as 
circular dichroism in laser mass spectrometry of photoions~\cite{BoeslCPC06,LiJCP06,BreunigCPC09}, as well as comparison with different levels of electronic structure theory~\cite{KroenerPCCP15}.

\begin{acknowledgments}
  We would like to thank Christian Lux and Thomas Baumert for 
  discussions as well as Sebastian Marquardt and Hauke Westemeier for help
  and discussions.
  Financial support by the State Hessen Initiative for the
  Development of Scientific and Economic Excellence (LOEWE) within the
  focus project Electron Dynamic of Chiral Systems (ELCH) is
  gratefully acknowledged.
\end{acknowledgments}

\appendix

\section{Wavefunctions and rotation matrices}
In the following, we summarize for completeness the properties of the
continuum wavefunctions, rotation matrices and complex spherical
harmonics in Secs.~\ref{subsec:H-cont}, \ref{subsec:rotmat}
and~\ref{subsec:real_spherical} that were used in the calculations in
the main body of the paper.

\subsection{Radial continuum wavefunctions of the hydrogen atom}
\label{subsec:H-cont}
An explicit expression of the radial continuum wavefunctions is given
in terms of the Kummer confluent hypergeometric functions~\cite{Hans1957}, 
\begin{equation}
\label{eq:G_k}
G_{k,\ell}(r) = C_{E,\rm\ell}\, (2kr)^{\ell}\,e^{-\mathrm{i}kr} 
               F_1(\ell+1+\mathrm{i}/k,2\ell+2,2\mathrm{i}kr)\,.
\end{equation}
The factor 
\[
  C_{E,\rm\ell}\equiv\sqrt{\frac{2\mu k}{\pi\hbar^2}}
  \frac{|\Gamma(\ell+1-\mathrm{i}/k)|}{(2\ell+1)!}e^{\pi/2k}\,,
\]
where $\Gamma(\cdot)$ refers to the Euler Gamma function,
ensures proper normalization such that 
\[
  \int^{\infty}_0 G_{E,\rm\ell}(r)G_{E^\prime,\rm\ell}(r)r^2 dr
  = \delta(E-E^\prime)\,.
\]
In order to avoid numerical instabilities when generating the radial
continuum wavefunctions, Eq.~\eqref{eq:G_k} may be written
in integral form~\cite{Aslangul08},
\begin{eqnarray}
\label{eq:G_k2}
G_{k,\ell}(r) &=& \sqrt{\frac{2\mu k}{\pi\hbar^2}}\,
|\Gamma(\ell+1-\mathrm{i}/k)|^{-1}\, e^{\pi/2k}\,
(2kr)^{\ell}\,e^{-\mathrm{i}kr}\nonumber\\ 
&&\times \int^1_0
s^{\ell+\mathrm{i}/k}(1-s)^{\ell-\mathrm{i}/k}\,
e^{2\mathrm{i}krs}\, ds\,.
\end{eqnarray}

\subsection{Bound state wavefunctions of the hydrogen atom}
\label{subsec:H-bound}

As for the radial part of bound states for hydrogenic wavefunctions, $R^{n_o}_{\ell_o}(r)$, cf.
Eq.~\eqref{eq:exited_state}, they can also be expressed in terms of the 
Kummer confluent hypergeometric functions~\cite{Aslangul08},
\begin{subequations}
\begin{eqnarray}
R^{n_o}_{\ell_o}(r) &=&
  \left( 4k^3_{n_o}\dfrac{(n_o+\ell_o-1)!}{\left[(n_o+\ell_o)!\right]^3}\right)^{1/2} \left(
2k_{n_o} r\right)^l  \\\nonumber
  & & 
\times F_1(\ell_o+1-n_o,2\ell_o+1,2k_{n_o} r)\, e^{-k_{n_o}r}   \,,
\label{radial:excitedstates}
\end{eqnarray}
with
\begin{eqnarray}
  k_{n_o} &\equiv& \dfrac{1}{1+\dfrac{m_e}{M_n}} \dfrac{1}{n_o\,a_o}\approx \dfrac{1}{n_o\, a_o} 
\end{eqnarray}
where $m_e$, $M_n$ and $a_o$ refer to the masses of the electron and that of the
nucleous and the Bohr's radius, respectively.
\end{subequations}

\subsection{Rotation matrices}
\label{subsec:rotmat}
We summarize here  some useful properties that are utilized in the
derivation of the photoionization cross section, following the
standard  angular momentum algebra as found in 
Refs.~\cite{edmonds,irreducible,Hans1957,Rose1967,Varshalovich1988}.
Any irreducible tensor field $f^{k}_{m_k}$ of rank $k$ is transformed
from the molecular frame to the laboratory frame as
follows~\cite{edmonds,irreducible}: 
\begin{eqnarray}
  \label{eq:transformation_rot}
  {f}^{k}_{m_k}(\mathbf{r}^\prime) &=&
D(\alpha\beta\gamma)f^{k}_{m_k}(\mathbf{r})\nonumber\\  &=& 
\sum_{m^\prime_k=-k}^{+k} f^{k}_{m^\prime_k}(\mathbf{r})
\mathcal{D}^{(k)}_{m^\prime_k,m_k}(\alpha\beta\gamma)\,,
\label{eq:eq1}
\end{eqnarray}
where $\mathcal{D}^{(j)}_{m^\prime_j,m_j}(\alpha\beta\gamma) = 
\langle j,m^\prime\vert D(\alpha\beta\gamma) 
\vert j,m\rangle$ refers to the Wigner rotation matrix of rank $j$,
and the subscripts $m_k$ and $m^\prime_k$ stand for the projection of
the total angular  momentum $k$ onto the $z$ axis in the molecular,
respectively laboratory, frame.
Conversely, the inverse of the transformation~\eqref{eq:eq1}
is given by 
\begin{eqnarray}
  {f}^{k}_{m_k}(\mathbf{r}) &=&
  D^{-1}(\alpha\beta\gamma)f^{k}_{m_k}(\mathbf{r}^\prime)\nonumber\\  
  &=& 
  \sum_{m^\prime_k=-k}^{+k} f^{k}_{m^\prime_k}(\mathbf{r}^\prime)
  \mathcal{D}^{\dagger,\rm (k)}_{m^\prime_k,m_k}(\alpha\beta\gamma)\,.
  \label{eq:eq1_inverse}
\end{eqnarray}
We express all vector quantities in spherical coordinates,
\begin{eqnarray}\label{eq:rprime}
{\mathbf{r}}^\prime &=&\sqrt{\frac{4\pi}{3}}\, r\sum_{\mu=0,\pm 1}(-1)^{\mu} Y^1_{\mu}(\Omega_{\mathbf{r}^\prime})\epsilon^{\prime}_{-\mu}\,,
\end{eqnarray}
where $\epsilon^{\prime}_{-\mu}$ refers to the spherical unit vector
in the laboratory frame, 
and $\mu=0,\pm 1$ denotes linear, left and right unit
components, respectively. The correspondence between the components of a arbitrary vector operator $\mathcal{V}$ in spherical
and cartesian basis is given by~\cite{edmonds,irreducible,Abramowitz1972},
\begin{eqnarray}
  \label{eq:cartesian_spherical}
  \begin{array}{ccl}
  \mathcal{V}_{-1} &=&\dfrac{1}{\sqrt{2}}\left(\mathcal{V}_{x}-
\mathrm{i}\mathcal{V}_{y} \right)\\
  \mathcal{V}_{0} &=& \mathcal{V}_{z}\\ 
  \mathcal{V}_{+1} &=&-\dfrac{1}{\sqrt{2}}\left(\mathcal{V}_{x}+
\mathrm{i}\mathcal{V}_{y} \right)
\end{array}
\end{eqnarray}
Transforming the spherical components $\mathbf{r}_q$, with $q=\pm 1,0$ in
to the Cartesian basis using
Eq.~\eqref{eq:cartesian_spherical} and Eq.~\eqref{tanmoments}, we find the two-photon absoption
tensor in the spherical basis, 
\begin{eqnarray}
  \label{eq:tensor_def2}
  \begin{array}{ccl}
  T_{-1, -1}&=&
  \dfrac{1}{2}\left(T_{xx}-2\mathrm{i}T_{xy}-T_{yy} \right)\\ 
  T_{-1, 0}&=&
  \dfrac{1}{\sqrt{2}}\left(T_{xz}-\mathrm{i}T_{yz}\right)\\ 
  T_{-1, +1}&=&
  -\dfrac{1}{2}\left(T_{xx}+T_{yy}\right)\\ 
  T_{0, 0}&=&
  T_{zz}\\
  T_{0,+1}&=&
  -\dfrac{1}{\sqrt{2}}\left(T_{zx}+\mathrm{i}T_{zy}\right)\\ 
  T_{+1,+1}&=&
  \dfrac{1}{2}\left(T_{xx}+2\mathrm{i}T_{xy}-T_{yy}\right)\\ 
\end{array}
\end{eqnarray}
Because $T_{\alpha,\beta} = T_{\beta,\alpha}$, with $\alpha,\beta = x,y,z$,
cf. Eq.~\eqref{tanmoments}, it can be straightforwardly shown, using Eq.~\eqref{eq:cartesian_spherical}, 
that $T_{q_1,q_2}=T_{q_2,q_1}$. 

In the derivations we make heavily use of the product rule 
for two Wigner rotations matrices of ranks $k$ and $k^\prime$,
\begin{subequations}
\label{subeq:properties}
\begin{eqnarray}
\label{subeq:property1}
\mathcal{D}^{(k)}_{\mu,\nu}(\omega)\mathcal{D}^{(k^\prime)}_{\mu^\prime,\nu^\prime}(\omega)
&=&  \sum^{k+k^\prime}_{J=|k-k^\prime|}
(2J+1)\mathcal{D}^{*(J)}_{-\mu-\mu^\prime,\rm
-\nu-\nu^\prime}(\omega)\nonumber\\
&&\times
\begin{pmatrix} k & k^\prime & J \vspace*{0.33cm} \\ \mu & \mu^\prime & -\mu-\mu^\prime 
\end{pmatrix}
\begin{pmatrix} k & k^\prime & J \vspace*{0.33cm} \\ \nu & \nu^\prime &-\nu-\nu^\prime 
\end{pmatrix},\nonumber\\
\end{eqnarray}
together with the following symmetry property,
\begin{eqnarray}
  \label{subeq:property2}
  \mathcal{D}^{(k)}_{\mu,\nu} = (-1)^{\mu-\nu}\mathcal{D}^{*(k)}_{-\mu,-\nu}(\omega)\,,
\end{eqnarray}
  \label{subeq:properties1:2}
\end{subequations}
where $(^*)$ denotes the complex conjugate.

\subsection{Conversion to complex spherical harmonics}

\label{subsec:real_spherical}
The standard complex spherical harmonics $Y^\ell_m(\Omega)$
are related to the real
spherical harmonics $\Upsilon_{\ell,|m|}(\Omega)$ by
\begin{subequations}
\begin{eqnarray}
  Y^\ell_m(\Omega) = \begin{cases}
  \frac{1}{\sqrt{2}}\Big(\Upsilon_{\ell,|m|}(\Omega)
-\mathrm{i} \Upsilon_{\ell,-|m|}(\Omega)\Big) &  \mathrm{if}\quad m\leq 0\,, \\
    \vspace*{0.1cm}
    \Upsilon_{\ell,0}(\Omega) &  \mathrm{if}\quad m= 0\,, \\\nonumber
    \vspace*{0.1cm}
    \frac{(-1)^\ell}{\sqrt{2}}\Big(\Upsilon_{\ell,|m|}(\Omega)
+\mathrm{i} \Upsilon_{\ell,-|m|}(\Omega)\Big) & \mathrm{if}\quad m\geq 0 \,.
 \end{cases}
\end{eqnarray}
Therefore the excited state expansion coefficients 
$a^{\ell_o}_{m_o}(n)$, defined in Eq.~\eqref{eq:exited_state}, are
connected to the coefficients in the basis of real spherical harmonics
by  
\begin{eqnarray}
  a^{\ell_o}_{m_o}(n) = \left\{
\begin{array}{ll}
  \frac{1}{\sqrt{2}}\Big(\tilde{a}^{\ell_o}_{m_o}(n)
+\mathrm{i} \tilde{a}^{\ell_o}_{m_o}(n)\Big) & \qquad \mathrm{if}\quad m\leq 0\,, \\\nonumber
  \tilde{a}^{\ell_o}_{0}(n) & \qquad \mathrm{if}\quad m= 0 \,,\\
  \frac{(-1)^{\ell_o}}{\sqrt{2}}\Big(\tilde{a}^{\ell_o}_{m_o}(n)
-\mathrm{i} \tilde{a}^{\ell_o}_{m_o}(n)\Big) & \qquad \mathrm{if}\quad m\geq 0 \,.\\
 \end{array}
\right.
\end{eqnarray}
\end{subequations}
The naming of the real spherical harmonics used in the reexpansion of the molecular wavefunctions is explained in Table~\ref{tab:coef_def}.
\begin{table}[tb]
\caption{\label{tab:coef_def} Definition of the  non-normalized real
  spherical harmonics 
   in Cartesian coordinates.}
\begin{tabular}{ccc}
\toprule[0.8pt]
\addlinespace[0.1cm]
\multicolumn{1}{c}{designation} & \phantom{abc} &
\multicolumn{1}{c}{real spherical harmonic}\\ 
\toprule[0.5pt]
\addlinespace[0.1cm]
$S 0  $ & & $1$ \\
$P Z  $ & & $z$\\
$P Y  $ & & $y$\\ 
$P X  $ & & $x$\\ 
$D 0  $ & & $(-x^2-y^2+2z^2)/\sqrt{12}$ \\
$D 1a $ & & $xz$                    \\
$D 1b $ & & $yz$                    \\
$D 2a $ & & $xy$                    \\
$D 2b $ & & $(x^2-y^2)/2$            \\ 
$F 0  $ & & $(-3x^3-3y^3+2z^3)/\sqrt{60}$ \\
$F 1a $ & & $(-x^3-xy^2+4xz^2)/\sqrt{40}$  \\
$F 1b $ & & $(-y^3-x^2y+4yz^2)/\sqrt{40}$  \\
$F 2a $ & & $xyz                        $  \\   
$F 2b $ & & $(x^2z-y^2z)/2              $  \\    
$F 3a $ & & $(x^3-3xy^3)/\sqrt{24}      $  \\      
$F 3b $ & & $(y^3-3x^2y)/\sqrt{24}      $  \\ 
\bottomrule[0.8pt]
\end{tabular}
\end{table} 

\section{Derivations}
\label{sec:derivation}

Here, we provide details of the
derivation of the one-photon transition rate, two-photon
absorption tensor and the 
photoionization cross section in Secs.~\ref{subsec:X1P}
to~\ref{subsec:X2+1} as well as the behavior of the Legendre
coefficients under change of helicity in the one-photon
photoionization and two-photon absorption processes 
in Secs.~\ref{subsec:ci:0} to~\ref{subsec:ci2}. 

\subsection{One-photon transition rate}
\label{subsec:X1P}
This section is devoted to deriving the rate for the
photoionization transition 
from the intermediate electronically excited 
state to the continuum, driven by an electric field with polarization 
$\epsilon^{\prime}_{\varrho_2}$. The starting point is the  doubly
differential cross section 
in the molecular frame given in Eq.~\eqref{eq:photoelectron}.
It contains  the laboratory-frame product
$\epsilon^\prime_{\varrho_2}\cdot\mathbf{r}^\prime$, which, using
Eq.~\eqref{eq:rprime},  becomes 
\begin{eqnarray}
  \mathbf{\epsilon}^\prime_{\varrho_2}\cdot\mathbf{r}^\prime &=&
\sqrt{\frac{4\pi}{3}}\,r\,Y^{1}_{\varrho_2}(\Omega_{\mathbf{r}^\prime})\equiv
\mathbf{r}^{\prime}_{\varrho_2} \,.
\end{eqnarray}
This is rotated into the molecular frame, employing 
Eq.~\eqref{eq:eq1}, resulting in 
\begin{eqnarray}\label{eq:prod2}
  \mathbf{\epsilon}^\prime_{\varrho_2}\cdot\mathbf{r}^\prime &=&
  \sqrt{\frac{4\pi}{3}}\,r\,\sum_{q=0,\pm
    1}\mathcal{D}^{(1)}_{q,\rm\varrho_2}(\omega)
  Y^{1}_{q}(\Omega_{\mathbf{r}}) \,.
\end{eqnarray}
Inserting Eq.~\eqref{eq:prod2}
into Eq.~\eqref{eq:photoelectron} yields the photoionization cross
section in the molecular frame as a function of the 
Euler angles $\omega\equiv(\alpha,\beta,\gamma)$,
cf. Eq.~\eqref{eq:diffXa}. 
Evaluating Eq.~\eqref{eq:diffXa}  requires evaluation of the product 
$\langle\Psi_{\mathbf{k}}| \mathbf{r}_q |\Psi_o\rangle\langle\Psi_{\mathbf{k}}| \mathbf{r}_{q^\prime} |\Psi_o\rangle^{*}$. 
Inserting Eqs.~\eqref{eq:exited_state2} and~\eqref{eq:photoelectron}
yields, for a fixed polarization direction $q$, 
\begin{eqnarray*}
  \langle\Psi_{\mathbf{k}}| \mathbf{r}_q |\Psi_o\rangle &=&
  \sum_{\substack{\ell,m \\
n_o,\ell_o,m_o}}(-\mathrm{i})^{\ell}e^{\mathrm{i}\delta_{\ell}}
  I^{n_o}_{k}(\ell,\ell_o)S^{\ell,m}_{\ell_o,m_o}(q)\nonumber\\
  &&\times 
  a^{\ell_o}_{m_o}(n_o)\,Y^{\ell}_{m}(\Omega_{\mathbf{k}})
\end{eqnarray*}
with $I^{n_o}_{k}(\ell,\ell_o)$ and $S^{\ell,m}_{\ell_o,m_o}(q)$ defined in
Eqs.~\eqref{eq:Ik} and~\eqref{eq:S} such that
Eq.~\eqref{eq:diffXa} comprises the product 
$Y^{\ell}_{m}(\Omega_{\mathbf{k}}) Y^{*\ell^\prime}_{m^\prime}(\Omega_{\mathbf{k}})$.
Using the symmetry properties of the spherical harmonics, we can write 
\begin{widetext}
\begin{subequations}
    \begin{eqnarray}
      Y^{\ell}_{m}(\Omega_{\mathbf{k}}) Y^{*\ell^\prime}_{m^\prime}(\Omega_{\mathbf{k}}) &=&
      (-1)^{m^\prime} Y^{\ell}_{m}(\Omega_{\mathbf{k}}) Y^{\ell^\prime}_{-m^\prime}(\Omega_{\mathbf{k}})\nonumber\\
      &=&
      (-1)^{-m}\sum^{\ell+\ell^{\prime}}_{\mathcal{L}=|\ell-\ell^{\prime}|} 
      \tilde{\gamma}(\ell,\ell^{\prime},\mathcal{L}) 
      \begin{pmatrix} \ell & \ell^{\prime} & \mathcal{L}
        \vspace*{0.33cm} \\
        m & -m^{\prime} & m^{\prime}-m 
      \end{pmatrix}
      \begin{pmatrix} 
        \ell & \ell^{\prime} & \mathcal{L} \vspace*{0.33cm} \\ 
        0 & 0 &        0 
      \end{pmatrix}
       Y^{\mathcal{L}}_{m-m^{\prime}}(\Omega_{\mathbf{k}})
      \label{eq:YY_mol}
    \end{eqnarray}    
  with 
  \begin{eqnarray}
    \label{eq:gamma:def}
    \tilde{\gamma}(\ell,\ell^{\prime},\mathcal{L})=
    \sqrt{(2\ell+1)(2\ell^{\prime}+1)(2\mathcal{L}+1)/4\pi}\quad
  \end{eqnarray}
\end{subequations}
and $\Omega_{\mathbf{k}}=(\vartheta_{\mathbf{k}}, \phi_{\mathbf{k}})$
refering to polar and azimuthal angles of the momentum vector in the
molecular frame 
of reference. In order to express the photoionization direction in the
laboratory frame, we need to
apply the inverse transformation~\eqref{eq:eq1_inverse}
to $Y^{\mathcal{L}}_{m-m^{\prime}}(\Omega_{\mathbf{k}^\prime})$, i.e., 
\begin{subequations}
\begin{eqnarray}
  \label{eq:product_YY}
  Y^{\mathcal{L}}_{m-m^{\prime}}(\Omega_{\mathbf{k}})
  &=&
  D^{-1}(\omega)\,Y^{\mathcal{L}}_{m-m^\prime}(\Omega_{\mathbf{k}^\prime})
  =\sum^{\mathcal{L}}_{\mu=-\mathcal{L}}\mathcal{D}^{\dagger(\mathcal{L})}_{\mu,m-m^\prime}(\omega)
  Y^{\mathcal{L}}_{\mu}(\Omega_{\mathbf{k}^\prime})
  =\sum^{\mathcal{L}}_{\mu=-\mathcal{L}} (-1)^{m^\prime-m-\mu}\mathcal{D}^{(\mathcal{L})}_{m^\prime-m,-\mu}(\omega)
  Y^{\mathcal{L}}_{\mu}(\Omega_{\mathbf{k}^\prime})\nonumber\\
  &=&\sum^{\mathcal{L}}_{\mu=-\mathcal{L}} 
  \sqrt{\frac{(2\mathcal{L}+1)}{4\pi}\frac{(\mathcal{L}-\mu)!}{(\mathcal{L}+\mu)!}}
  (-1)^{m^\prime-m}\,  \mathcal{D}^{(\mathcal{L})}_{m^\prime-m,-\mu}(\omega)
  P^{\mu}_{\mathcal{L}}(\cos\vartheta^\prime_{k})\,e^{i\mu\varphi^\prime_k}
\end{eqnarray}
Using Eq.~\eqref{eq:product_YY}, Eq.~\eqref{eq:YY_mol} then becomes,
  \begin{eqnarray}
    \label{eq:lambda:def}
    Y^{\ell}_{m}(\Omega_{\mathbf{k}}) Y^{*\ell^\prime}_{m^\prime}(\Omega_{\mathbf{k}}) &=&
    (-1)^{m^{\prime}}
    \sum^{\ell+\ell^{\prime}}_{\mathcal{L}=|\ell-\ell^{\prime}|} 
    (2\mathcal{L}+1)\,\varsigma^\mu_{\mathcal{L}}(\ell,\ell^\prime) 
    \begin{pmatrix} 
      \ell & \ell^{\prime} & \mathcal{L} \vspace*{0.33cm} \\ 
      m & -m^{\prime} & m^{\prime}-m 
    \end{pmatrix}
    \begin{pmatrix} 
      \ell & \ell^{\prime} & \mathcal{L} \vspace*{0.33cm} \\ 
      0 & 0 & 0 
    \end{pmatrix}\nonumber\\
    &&\quad\quad\quad
    \times \sum^{\mathcal{L}}_{\mu=-\mathcal{L}} 
    \mathcal{D}^{(\mathcal{L})}_{m^\prime-m,-\mu}(\omega)
     P^{\mu}_{\mathcal{L}}(\cos\vartheta^\prime_{k})\,e^{i\mu\varphi^\prime_k}
   \end{eqnarray}  
     \label{eq:YY_lab}
 \end{subequations}
 \end{widetext}
 with 
\begin{eqnarray} 
  \label{eq:varsigma:def}
  \varsigma^\mu_{\mathcal{L}}(\ell,\ell^\prime)=\sqrt{\dfrac{(2\ell+1)(2\ell^\prime+1)}{16\pi^2}\dfrac{(\mathcal{L}-\mu)!}{(\mathcal{L}+\mu)!}} 
\end{eqnarray} 
In Eq.\eqref{eq:YY_lab},
we have used the equality between spherical harmonics and associate
Legendre polynomials, including the Condon-Shortley phase
convention~\cite{edmonds,Hans1957,Abramowitz1972},
\begin{eqnarray}
\label{eq:associate}
Y^{\mathcal{L}}_{\mu}(\vartheta^\prime_k,\varphi^\prime_k) &=&(-1)^{\mu}
\sqrt{\frac{(2\mathcal{L}+1)}{4\pi}\frac{(\mathcal{L}-\mu)!}{(\mathcal{L}+\mu)!}}\,
P^{\mu}_{\mathcal{L}}(\cos\vartheta^\prime_k)\, e^{\mathrm{i}\mu\varphi^\prime_k}\,.\nonumber\\
\end{eqnarray}
Inserting  Eq.~\eqref{eq:YY_lab} into Eq.~\eqref{eq:diffX2}, 
we obtain the differential one-photon cross section in the laboratory
frame of reference for a fixed molecular orientation defined in  Eq.~\eqref{eq:differential2}.
\subsection{Two-photon absorption tensor}
\label{subsec:2Ptensor}
The probability of two-photon absorption, Eq.~\eqref{eq:density2},  of
a molecule that is oriented with
angles $\omega=(\alpha,\beta,\gamma)$ with respect to the laboratory
frame of reference contains the product
$\mathcal{D}^{(1)}_{q_1,\rm\varrho_1}(\omega)
\mathcal{D}^{(1)}_{q_2,\rm\varrho_1}(\omega)
\mathcal{D}^{*(1)}_{q_3,\rm\varrho_1}(\omega)
\mathcal{D}^{*(1)}_{q_4,\rm\varrho_1}(\omega)$.
Using Eqs.~\eqref{subeq:properties}, we obtain
\begin{widetext}
\begin{subequations}\label{subeq:Dprod}
\begin{eqnarray}\label{subeq:part1}
\mathcal{D}^{(1)}_{q_1,\rm\varrho_1}(\omega)\mathcal{D}^{(1)}_{q_2,\rm\varrho_1}(\omega)=
(-1)^{q_1+q_2}\sum^2_{Q=0}(2Q+1)\mathcal{D}^{(Q)}_{q_1+q_2,\rm
  2\sigma_1}(\omega)
\begin{pmatrix} 1 & 1 & Q \vspace*{0.33cm} \\ q_1 & q_2 & -q_1-q_2 
\end{pmatrix}
\begin{pmatrix} 1 & 1 & Q \vspace*{0.33cm} \\ \varrho_1 & \varrho_1 &-2\varrho_1\\ 
\end{pmatrix}\,,
\end{eqnarray}
and analogously for 
$\mathcal{D}^{*(1)}_{q_3,\rm\varrho_1}(\omega)
\mathcal{D}^{*(1)}_{q_4,\rm\varrho_1}(\omega)$,
\begin{eqnarray}\label{subeq:part2}
\mathcal{D}^{*(1)}_{q_3,\rm\varrho_1}(
\omega)\mathcal{D}^{*(1)}_{q_4,\rm\varrho_1}(\omega)=
  \sum^2_{Q^\prime=0}(2Q^\prime+1)\mathcal{D}^{(Q^\prime)}_{-q_3-q_4,\rm -2\varrho_1}(\omega)
\begin{pmatrix} 1 & 1 & Q^\prime \vspace*{0.33cm} \\ q_3 & q_4 & -q_3-q_4 
\end{pmatrix}
\begin{pmatrix} 1 & 1 & Q \vspace*{0.33cm} \\ \varrho_1 & \varrho_1
  &-2\varrho_1  \end{pmatrix}\,.
\end{eqnarray}
\end{subequations}
Inserting Eqs.~\eqref{subeq:Dprod} into~\eqref{eq:density2}
and using 
\begin{eqnarray*}
\mathcal{D}^{(Q)}_{q_1+q_2,2\varrho_1}(
\omega)\mathcal{D}^{(Q^\prime)}_{-q_3-q_4,-2\varrho_1}(\omega)=
  \sum^4_{K=0}(2K+1)\mathcal{D}^{*(K)}_{s,0}(\omega)
\begin{pmatrix} Q & Q^\prime & K \vspace*{0.33cm} \\ q_1+q_2 & -q_3-q_4 & -s 
\end{pmatrix}
\begin{pmatrix} Q & Q^\prime & K \vspace*{0.33cm} \\ 2\varrho_1 & -2\varrho_1 &0 
\end{pmatrix}
\end{eqnarray*}
with $s=q_1+q_2-q_3-q_4$, the orientation-dependent probability of
two-photon absorption becomes, 
\begin{eqnarray}
\rho_{2\mathrm{P}}(\omega) &=& 
\sum_{q_1,q_2}T_{q_1,q_2} \sum_{q_3,q_4} T^{*}_{q_3,q_4}(-1)^{q_3+q_4} 
\sum^2_{Q=0}(2Q+1)
\begin{pmatrix} 1 & 1 & Q \vspace*{0.33cm} \\ q_1 & q_2 & -q_1-q_2 
\end{pmatrix}
\begin{pmatrix} 1 & 1 & Q \vspace*{0.33cm} \\ \varrho_1 & \varrho_1
  &-2\varrho_1  
\end{pmatrix}\nonumber\\
&&\times
\sum^2_{Q^\prime=0}(2Q^\prime+1) 
\begin{pmatrix} 1 & 1 & Q^\prime \vspace*{0.33cm} \\ q_3 & q_4 & -q_3-q_4 
\end{pmatrix}
\begin{pmatrix} 1 & 1 & Q^\prime \vspace*{0.33cm} \\ \varrho_1 & \varrho_1
  &-2\varrho_1  
\end{pmatrix}\nonumber\\
&&\times
\sum^4_{K=0}(2K+1)
\begin{pmatrix} Q & Q^\prime & K \vspace*{0.33cm} \\ q_1+q_2 & -q_3-q_4 & -s 
\end{pmatrix}
\begin{pmatrix} Q & Q^\prime & K \vspace*{0.33cm} \\ 2\varrho_1 & -2\varrho_1 &0 
\end{pmatrix}\mathcal{D}^{(K)}_{s,0}(\omega)\nonumber\\
&\equiv&
\sum_{q_1,q_2}T_{q_1,q_2} \sum_{q_3,q_4} (-1)^{q_3+q_4}\, T^{*}_{q_3,q_4}
\sum^{4}_{K=0} g^{(K)}_{q_1,q_2,q_3,q_4} \mathcal{D}^{(K)}_{s,0}(\omega)\,,
\label{eq:density_appendix}
\end{eqnarray}
\end{widetext}
cf. Eq.~\eqref{eq:rho}. Two useful properties of the Wigner $3j$ symbols utilized 
throughout this work,
involve odd permutations 
of two columns~\cite{Varshalovich1988}, 
\begin{eqnarray}
  \label{eq:permutation_law}
\begin{pmatrix} j & j^\prime & J \vspace*{0.33cm} \\ m & m^\prime & M \end{pmatrix}
                    &=& (-1)^{j+j^\prime+J} 
\begin{pmatrix} j^\prime & j        & J \vspace*{0.33cm} \\ m^\prime&
  m & M \end{pmatrix}\,, 
\end{eqnarray}
as well as the unitary condition for the Wigner rotation
matrices~\cite{Varshalovich1988},  
\begin{eqnarray}
\label{eq:wigner_unitary} 
\sum_{M=-J}^J \mathcal{D}^{(J)}_{M,\rm M^\prime}(\omega)\,
\mathcal{D}^{*(J)}_{M,\rm \tilde{M}^\prime}(\omega)
&=&
\delta_{M^\prime,\rm \tilde{M}^\prime}\,.
\end{eqnarray}

\subsection{Cross section for $(2+1)$ photoionization}
\label{subsec:X2+1}
In order to simplify the expression of the cross section for the (2+1)
REMPI process, we utilize  
the properties defined in Eq.~\eqref{subeq:properties1:2}, to the product
involving the first
and second Wigner $3j$ symbols in Eq.~\eqref{eq:differential2}, 
\begin{widetext}
\begin{eqnarray}
  \label{eq:condense_1}
  \mathcal{D}^{(1)}_{q,\varrho_2}(\omega)
  \mathcal{D}^{(1)}_{-q^\prime,-\varrho_2}(\omega) &=& (-1)^{q^\prime-q}\sum^{2}_{\nu=0}(2\nu+1)
  \,\mathcal{D}^{(\nu)}_{q-q^\prime,\rm 0}(\omega)            
  \begin{pmatrix} 1 & 1 & \nu \vspace*{0.33cm} \\ q & -q^\prime & q^\prime-q \end{pmatrix}
  \begin{pmatrix} 1 & 1 & \nu \vspace*{0.33cm} \\ \varrho_2 & -\varrho_2 & 0 
  \end{pmatrix}\,.            
\end{eqnarray}
This allows for exploiting, in the integration over the Euler angles, 
the well-known properties for integrating over a product of 
three Wigner $3j$ symbols. 
With Eq.~\eqref{eq:condense_1},
Eq.~\eqref{eq:differential2} takes the following form, 
\begin{eqnarray}
  \frac{d^2\sigma_{1\mathrm{P}}}{d\omega d\Omega_{{\mathbf{k}}^\prime}} &=& \nonumber
  c_o\, \sum_{\substack{\ell,m \\ \ell_o,m_o}}\, 
\sum_{\substack{\ell^\prime,m^\prime \\ \ell^\prime_o,m^\prime_o}}\, \sum_{q,q^\prime}
(-\mathrm{i})^{\ell-\ell^\prime}e^{\mathrm{i}(\delta_{\ell}-\delta_{\ell^\prime})}\, 
a^{\ell_o}_{m_o}a^{\ell^\prime_o}_{m^\prime_o} 
I_{_{k}}(\ell,\ell_o)\, I_{_{k}}(\ell^\prime,\ell^\prime_o) \mathcal{S}^{\ell,m}_{\ell_o,m_o}(q) 
\mathcal{S}^{\ell^\prime,m^\prime}_{\ell^\prime_o,m^\prime_o}(q^\prime)\,\\\nonumber 
&&\times \sum^{\ell+\ell^\prime}_{\mathcal{L}=\vert\ell-\ell^\prime\vert}
(2\mathcal{L}+1)\,
\begin{pmatrix} \ell & \ell^\prime & \mathcal{L}\vspace*{0.33cm}\\ 0 & 0 & 0 \end{pmatrix}
\begin{pmatrix} \ell & \ell^\prime & \mathcal{L} 
\vspace*{0.33cm} \\ m & -m^\prime & -(m-m^\prime) \end{pmatrix}
\sum_{\mu=-\mathcal{L}}^{\mathcal{L}}
\varsigma^\mu_{\mathcal{L}}(\ell,\ell^\prime)\,
(-1)^{m^\prime-q-\varrho_2}
P^{\mu}_{\mathcal{L}}(\cos\vartheta^\prime_{k})\,e^{\mathrm{i}\mu\varphi^\prime_k}\nonumber\\
&&\times
\sum^{2}_{\nu=0}(2\nu+1)
\begin{pmatrix} 1 & 1 & \nu\vspace*{0.33cm}\\ q & q^\prime & q^\prime-q \end{pmatrix}
\begin{pmatrix} 1 & 1 & \nu\vspace*{0.33cm}\\ \varrho_2 & -\varrho_2 & 0 \end{pmatrix}\,
\mathcal{D}^{(\nu)}_{q-q^\prime,\rm 0}(\omega)
\mathcal{D}^{{(\mathcal{L})}}_{m^\prime-m,-\mu}(\omega)\,.
\label{eq:differential3_appendix}
\end{eqnarray}
Inserting Eq.~\eqref{eq:density_appendix} and Eq.~\eqref{eq:differential3_appendix} 
into Eq.~\eqref{eq:weighted0}, the PAD measured in the laboratory frame, resulting from a fixed molecular
orientation $\omega$ reads, 
\begin{eqnarray}\label{eq:body_fixed2}
  \frac{d^2\sigma_{2+1}}{d\omega d\Omega_{{\mathbf{k}}^\prime}} &=&
  \mathcal{N}_0
  c_o\, \sum_{\substack{\ell,m \\ \ell_o,m_o}}\, 
\sum_{\substack{\ell^\prime,m^\prime \\ \ell^\prime_o,m^\prime_o}}\, \sum_{q,q^\prime}
(-\mathrm{i})^{\ell-\ell^\prime}e^{\mathrm{i}(\delta_{\ell}-\delta_{\ell^\prime})}\, 
a^{\ell_o}_{m_o}\,a^{*\ell^\prime_o}_{m^\prime_o}
I_{_{k}}(\ell,\ell_o)\, I_{_{k}}(\ell^\prime,\ell^\prime_o) \mathcal{S}^{\ell,m}_{\ell_o,m_o}(q) 
\mathcal{S}^{\ell^\prime,m^\prime}_{\ell^\prime_o,m^\prime_o}(q^\prime)\\ 
&&\times
\sum^{\ell+\ell^\prime}_{\mathcal{L}=\vert\ell-\ell^\prime\vert}
(2\mathcal{L}+1)
\begin{pmatrix} \ell & \ell^\prime & \mathcal{L}\vspace*{0.33cm}\\ 0 & 0 & 0 \end{pmatrix}
\begin{pmatrix} \ell & \ell^\prime & \mathcal{L} 
\vspace*{0.33cm} \\ m & -m^\prime & -(m-m^\prime) \end{pmatrix} 
\sum^{2}_{\nu=0}(2\nu+1)\nonumber
\begin{pmatrix} 1 & 1 & \nu\vspace*{0.33cm}\\ q & q^\prime & q^\prime-q \end{pmatrix}
\begin{pmatrix} 1 & 1 & \nu\vspace*{0.33cm}\\ \varrho_2 & -\varrho_2 & 0 \end{pmatrix}\nonumber\\
 &&\times
\sum_{q_1,q_2}T_{q_1,q_2} \sum_{q_3,q_4}(-1)^{q_3+q_4}\,T^{*}_{q_3,q_4}
\sum^{4}_{K=0} g^{(K)}_{q_1,q_2,q_3,q_4}
\sum_{\mu=-\mathcal{L}}^{\mathcal{L}}\,\varsigma^\mu_{\mathcal{L}}(\ell,\ell^\prime)
(-1)^{m^\prime-q-\varrho_2}\,
P^{\mu}_{\mathcal{L}}(\cos\vartheta^\prime_{k})\,e^{\mathrm{i}\mu\varphi^\prime_k}\nonumber\\
&&\times
\mathcal{D}^{(K)}_{s,0}(\omega)
\mathcal{D}^{(\nu)}_{q-q^\prime,\rm 0}(\omega)
\mathcal{D}^{{(\mathcal{L})}}_{m^\prime-m,-\mu}(\omega)\,,\nonumber
\end{eqnarray}
with $s=q_1+q_2-q_3-q_4$. Equation~\eqref{eq:body_fixed2} may be written in
the more compact form of Eqs.~\eqref{eq:diffXLF}, namely,
\begin{subequations}\label{eq:compact_form}
\begin{eqnarray}
\label{eq:compact1}
\frac{d^2\sigma_{2+1}}{d\omega d\Omega_{{\mathbf{k}}^\prime}} &=&
\mathcal{N}_0
c_o\,
\sum^{\infty}_{\mathcal{L}=0}
\sum^{+\mathcal{L}}_{\mu=-\mathcal{L}}
b^{\mu}_\mathcal{L}(\omega)
P^{\mu}_{\mathcal{L}}(\cos{\vartheta^\prime_{k}})\,e^{i\mu\phi^{\prime}_{k}}\,,
\end{eqnarray}
In Eq.~\eqref{eq:compact1},  the only orientation-dependent quantity, 
$b^{\mu}_\mathcal{L}(\omega)$, is given by 
\begin{eqnarray}
\label{eq:compact2}
b^{\mu}_\mathcal{L}(\omega) &=&
\sum_{\lambda}
\kappa(\lambda)\,\,
\mathcal{D}^{K}_{s,\rm 0}(\omega)
\mathcal{D}^{\nu}_{q-q^\prime,\rm 0}(\omega)
\mathcal{D}^{\mathcal{L}}_{m^\prime-m,\rm -\mu}(\omega)\,.\quad\quad
\end{eqnarray}
\end{subequations}
with $\kappa^\mu_{\mathcal{L}}(\lambda)$ defined as
 \begin{eqnarray}
 \label{eq:kappa_def}
 \kappa^\mu_{\mathcal{L}}(\lambda) &=& \nonumber
(-\mathrm{i})^{\ell-\ell^\prime}e^{\mathrm{i}(\delta_{\ell}-\delta_{\ell^\prime})}\, 
a^{\ell_o}_{m_o}a^{\ell^\prime_o}_{m^\prime_o}
I_{_{k}}(\ell,\ell_o)\, I_{_{k}}(\ell^\prime,\ell^\prime_o) \mathcal{S}^{\ell,m}_{\ell_o,m_o}(q) 
\mathcal{S}^{\ell^\prime,m^\prime}_{\ell^\prime_o,m^\prime_o}(q^\prime)\,\varsigma^\mu_{\mathcal{L}}(\ell,\ell^\prime)\\ 
&&\times
\begin{pmatrix} \ell & \ell^\prime & \mathcal{L}\vspace*{0.33cm}\\ 0 & 0 & 0 \end{pmatrix}
\begin{pmatrix} \ell & \ell^\prime & \mathcal{L} 
\vspace*{0.33cm} \\ m & -m^\prime & -(m-m^\prime) \end{pmatrix} 
\begin{pmatrix} 1 & 1 & \nu\vspace*{0.33cm}\\ q & q^\prime & q^\prime-q \end{pmatrix}
\begin{pmatrix} 1 & 1 & \nu\vspace*{0.33cm}\\ \varrho_2 & -\varrho_2 & 0 \end{pmatrix}\nonumber\\
 &&\times
T_{q_1,q_2}
(-1)^{q_3+q_4}\,
T^{*}_{q_3,q_4}
g^{(K)}_{q_1,q_2,q_3,q_4}
(2\nu+1)(2\mathcal{L}+1)
(-1)^{m^\prime-q-\varrho_2}
\end{eqnarray}
\end{widetext}
where $\lambda$ comprises all summation indices, except for
$\mathcal{L}$ and $\mu$, as described in Sec.~\ref{subsec:X2+1}. 
Next, according to Eq.~\eqref{eq:weighted}, we need to average over
all initial orientations, i.e., integrate
the doubly differential cross section over the Euler
angles. To this end, we utilize the following integration property involving
the product of three Wigner $3j$ symbols~\cite{edmonds,irreducible,Varshalovich1988},
\begin{widetext}
\begin{eqnarray}
  \label{eq:three_integral}
  \int\mathcal{D}^{(K)}_{s,\rm 0}(\omega)
    \mathcal{D}^{(\nu)}_{q-q^\prime,\rm 0}(\omega)
    \mathcal{D}^{(\mathcal{L})}_{m^\prime-m,\rm -\mu}(\omega)\,d^3\omega
    &=&
  \begin{pmatrix} K & \nu & \mathcal{L} \vspace*{0.33cm}\\ s & q-q^\prime & m^\prime-m \end{pmatrix}
  \begin{pmatrix} K & \nu & \mathcal{L}\vspace*{0.33cm}\\ 0 & 0 & -\mu \end{pmatrix}
\end{eqnarray}
\end{widetext}
with $d^3\omega\equiv d^3(\alpha,\beta,\gamma)=d\alpha\, d(\cos(\beta))\, d\gamma/8\pi^2$. Finally, following Eq.~\eqref{eq:weighted}, integration of Eq.~\eqref{eq:body_fixed2} over the Euler angles
$\omega\equiv(\alpha,\beta,\gamma)$, using Eq.~\eqref{eq:three_integral}, gives 
the expression of the laboratory frame PAD
resulting from a randomly ensemble of molecules in the context of a $(2+1)$ REMPI process,
defined in Eq.~\eqref{eq:PADfinal}. In particular, due 
to the second Wigner $3j$ symbol in Eq.~\eqref{eq:three_integral},
it is clear that the integral vanishes if $\mu\neq 0$.
As a consequence, this requirement translates into  cylindrical
symmetry of the PAD measured in the laboratory frame, as 
$\mu$ also appears in  the azimuthal angle dependent term 
$e^{\mathrm{i}\mu\varphi^\prime_k}$ in Eq.~\eqref{eq:body_fixed2}. Thus, we retrieve the
expression defined in Eq.~\eqref{eq:PADfinal}.
\subsection{Non-zero Legendre coefficients for two-photon absorption with
circularly polarized light and ionization with linear polarization}
\label{subsec:ci:0}
In this section, we show that a $(2+1)$ REMPI process for which the two-photon
absorption process is driven by circular
polarized light, followed by linearly polarized light for the radiative process,
lead within the electric dipole approximation exclusively to even Legendre coefficients. To this end, we exploit the
symmetry as well as invariance properties of Eq.~\eqref{eq:final_coeff}, by
making a change of variables for $q_1$, $q_2$, $q_3$ and $q_4$ in
Eq.~\eqref{eq:final_coeff} that preserves $c_{\mathcal{L}}(\varrho_1,\varrho_2)$
unchanged and also 
keeps $s=q_1+q_2-q_3-q_4$ invariant (in order to keep  the
fifth Wigner $3j$ symbol in Eq.~\eqref{eq:final_coeff} unchanged).
A  change of variables fulfilling this property reads, 
\vspace{-0.1448cm}              
\begin{eqnarray} 
  \label{eq:first_invariant_transformation:0}
 \begin{pmatrix} q^\prime_1\\q^\prime_2\\q^\prime_3\\q^\prime_4\end{pmatrix} &=&  
 \begin{pmatrix} 0 & 0 & -1 & 0  \\ 
                 0 & 0 &  0 & -1 \\
                 -1 & 0 &  0 & 0 \\
                 0 & -1 &  0 & 0 \end{pmatrix}
 \begin{pmatrix} q_1\\q_2\\q_3\\q_4\end{pmatrix} \,,
 \end{eqnarray}

i.e. it interchanges 
$q_1\rightleftharpoons -q_3$ and  $q_2\rightleftharpoons -q_4$.

For simplicity, we define the quantity, 

\begin{subequations}
\label{eq:gK_simplicity:0}
  \begin{eqnarray}
    \Lambda_{\theta}(\varrho_1)&=&\sum_K \sum_{q_1,q_2}\sum_{q_3,q_4} (-1)^{q_3+q_4} g^K_{q_1,q_2,q_3,q_4}(\varrho_1) T_{q_1,q_2}T^{*}_{q_3,q_4}\nonumber\\
 &&\times
    W_{\theta}(s)
  \end{eqnarray}
with
\begin{eqnarray}
  W_{\theta}(s)&=&
\begin{pmatrix} K & \nu & \mathcal{L}\vspace*{0.33cm}\\s & q-q^\prime & m^\prime-m\end{pmatrix}
\end{eqnarray}
\end{subequations}
Eqs.~\eqref{eq:gK_simplicity:0} appear in Eq.~\eqref{eq:final_coeff}. In Eq.~\eqref{eq:gK_simplicity:0}, $s=q_1+q_2-q_3-q_4$ and  $\theta$
stands for the indices $(\theta\equiv K,\nu,m,m^\prime,q,q^\prime,\mathcal{L})$.
Analogously, $\Lambda_{\theta}^\prime(\varrho_1)$ is defined using the primed dummy variables $q^\prime_k$, for
$k=1,\dots,4$ with the symmetry property
$\Lambda_{\theta}(\varrho_1)=\Lambda_{\theta}^\prime(\varrho_1)$. Of course, we
have,
\begin{eqnarray}
  \label{eq:invariance_prop:0}
  c_{\mathcal{L}}(\varrho_1,0) = c^\prime_{\mathcal{L}}(\varrho_1,0)
\end{eqnarray}
Using Eq.~\eqref{eq:first_invariant_transformation:0}, the tensor elements appearing in Eq.~\eqref{eq:final_coeff} then
transform according to 
\begin{subequations}
  \label{eq:transform_tensor1:0}
  \begin{eqnarray}
   T_{q^\prime_1,q^\prime_2}  =  T_{-q_3,-q_4}=(-1)^{q_3+q_4}T^{*}_{q_3,q_4}
 \end{eqnarray}
and 
\begin{eqnarray}
  T^{*}_{q^\prime_3,q^\prime_4}  =  T^{*}_{-q_1,-q_2} = (-1)^{q_1+q_2}T_{q_1,q_2}\,,\nonumber\\
\end{eqnarray}
\end{subequations}
Using Eq.~\eqref{eq:transform_tensor1:0},  Eq.~\eqref{eq:gK_simplicity:0} reads, upon transformation, 
\begin{eqnarray}
  \label{eq:Lambda_prime:def}
  \Lambda_{\theta}^\prime(\varrho_1) &=& \sum_{q_1,q_2}\sum_{q_3,q_4} (-1)^{q_3+q_4}\, g^K_{-q_3,-q_4,-q_1,-q_2}(\varrho_1)\nonumber\\
  &&\times T_{q_1,q_2}\,T^{*}_{q_3,q_4}\, W_{\theta}(s)\,,
\end{eqnarray}
with $\Lambda_{\theta}(\varrho_1)=\Lambda_{\theta}^\prime(\varrho_1)$.
Next, we  evaluate the quantity $g^K_{-q_3,-q_4,-q_1,-q_2}(\varrho_1)$ present
in Eq.~\eqref{eq:Lambda_prime:def} using Eq.~\eqref{eq:gK}, we find
\begin{widetext}
\begin{eqnarray}
\label{eq:gK_transformed1:0}
g^{(K)}_{-q_3,-q_4,-q_1,-q_2}(\varrho_1) &=&
\sum^2_{Q=0}
\sum^2_{Q^\prime=0}
\sum^{Q+Q^\prime}_{K=|Q-Q^\prime|}
\gamma^{(K)}_{Q,\rm Q^\prime}    
\begin{pmatrix} 1 & 1 & Q^\prime \vspace*{0.33cm} \\ -q_3 & -q_4 & q_3+q_4\end{pmatrix}
\begin{pmatrix} 1 & 1 & Q^\prime \vspace*{0.33cm} \\ \varrho_1 & \varrho_1 & -2\varrho_1 \end{pmatrix}\\\nonumber
&&\times
\begin{pmatrix} 1 & 1 & Q        \vspace*{0.33cm} \\ -q_1 & -q_2 &  q_1+q_2\end{pmatrix}
\begin{pmatrix} 1 & 1 & Q        \vspace*{0.33cm} \\ \varrho_1 & \varrho_1 & -2\varrho_1 \end{pmatrix}
\begin{pmatrix} Q^\prime& Q & K \vspace*{0.33cm} \\ -q_3-q_4 & q_1+q_2 & s\end{pmatrix}
\begin{pmatrix} Q^\prime& Q & K \vspace*{0.33cm} \\ 2\varrho_1 & -2\varrho_1 & 0\end{pmatrix}\,,
\end{eqnarray}
\end{widetext}
where we have interchanged the dummy indices $Q$ and $Q^\prime$.
Application of Eq.~\eqref{eq:wigner_prop1} to the first and third
Wigner $3j$ symbol in Eq.~\eqref{eq:gK_transformed1:0} gives,
\begin{subequations}
  \label{eq:demo1_step2:0}
  \begin{eqnarray}
\begin{pmatrix} 1 & 1 & Q^\prime \vspace*{0.33cm} \\ -q_3 & -q_4 & q_3+q_4\end{pmatrix}
                  &=&(-1)^{Q^\prime}
\begin{pmatrix} 1 & 1 & Q^\prime \vspace*{0.33cm} \\ q_3 & q_4 &
  -q_3-q_4\end{pmatrix} 
\quad\quad\quad
\end{eqnarray}
and 
\begin{eqnarray}
\begin{pmatrix} 1 & 1 & Q        \vspace*{0.33cm} \\ -q_1 & -q_2 &  q_1+q_2\end{pmatrix}
                  &=&(-1)^{Q}
\begin{pmatrix} 1 & 1 & Q        \vspace*{0.33cm} \\ q_1 & q_2 &
  -q_1-q_2\end{pmatrix},\quad\quad \quad
\end{eqnarray}
respectively. Next, we permute the first and second column in the fifth
Wigner $3j$ symbol in Eq.~\eqref{eq:gK_transformed1:0}, following Eq.~\eqref{eq:permutation_law}, which yields
\begin{eqnarray}
\begin{pmatrix} Q^\prime& Q & K \vspace*{0.33cm} \\ -q_3-q_4 & q_1+q_2 & s\end{pmatrix}
&=&
\begin{pmatrix} Q& Q^\prime & K \vspace*{0.33cm} \\ q_1+q_2 & -q_3-q_4 & s\end{pmatrix}\nonumber\\
                 & &\times  (-1)^{Q+Q^\prime+K}
\end{eqnarray}
\end{subequations}

Finally, inserting Eqs.~\eqref{eq:demo1_step2:0} 
into Eq.~\eqref{eq:gK_transformed1:0} together with the property
$\Lambda_{\theta}(\varrho_1)=\Lambda_{\theta}^\prime(\varrho)$, we find
\begin{widetext}
\begin{eqnarray}
\label{eq:gK_transform2:0}
\sum_K\sum_{q_1,q_2}\sum_{q_3,q_4} (-1)^{q_3+q_4} g^K_{q_1,q_2,q_3,q_4}(\varrho_1)\, T_{q_1,q_2}\,T^{*}_{q_3,q_4}W_{\theta}(s)
  &=&\sum_K\sum_{q_1,q_2}\sum_{q_3,q_4} (-1)^{q_3+q_4} g^K_{q_1,q_2,q_3,q_4}(\varrho_1)\, (-1)^K\, T_{q_1,q_2}\,T^{*}_{q_3,q_4}W_{\theta}(s)\nonumber\\ 
\end{eqnarray}
\end{widetext}
with $W_{\theta}(s)$ invariant as $s$ invariant, and where $g^K_{q_1,q_2,q_3,q_4}(\varrho_1)$ is defined in Eq.~\eqref{eq:gK}.
Eq.~\eqref{eq:gK_transform2:0}  means that the summations over $K$ and $q_{k}$ is invariant under the transformation $g^K\rightarrow (-1)^K g^K$. 
Using Eq.~\eqref{eq:gK_transform2:0}, we find for $\varrho_1=\pm 1$ and $\varrho_2=0$, 
\begin{widetext}
\begin{eqnarray}
\label{eq:final_coeff:demo1:0}
c^\prime_{\mathcal{L}}({\varrho_1},0) &=&\mathcal{N}_0 \tilde{c}_o\, \sum_{\substack{\ell,m \\ n_o,\ell_o,m_o}}\, 
\sum_{\substack{\ell^\prime,m^\prime \\ n^\prime_o\ell^\prime_o,m^\prime_o}}\, \sum_{q,q^\prime}
\sum_{\substack{q_1,q_2 \\ q_3,q_4}}
\sum^2_{\nu=0} \sum^4_{K=0} 
(-1)^{q3+q4}\,(2\nu+1)(2\mathcal{L}+1)
a^{\ell_o}_{m_o}(n_o)\, a^{*\ell^\prime_o}_{m^\prime_o}(n^\prime_o)\, T_{q_1,q_2} T^{*}_{q_3,q_4}\nonumber
\\         
&&\times (-\mathrm{i})^{\ell-\ell^\prime}\,
(-1)^{m^\prime-q-\varrho_2}\,e^{\mathrm{i}(\delta_{\ell}-\delta_{\ell^\prime})}\,\,g^{(K)}_{q_1,q_2,q_3,q_4}(\varrho_1)
\,\,I^{n_o}_{_{k}}(\ell,\ell_o)\,\, I^{n^\prime_o}_{_{k}}(\ell^\prime,\ell^\prime_o)\,\,\mathcal{S}^{\ell,m}_{\ell_o,m_o}(q) 
\,\,\mathcal{S}^{\ell^\prime,m^\prime}_{\ell^\prime_o,m^\prime_o}(q^\prime)\,\hat{\varsigma}(\ell,\ell^\prime)\nonumber\\ 
&&
\times
\begin{pmatrix} \ell & \ell^\prime & \mathcal{L}\vspace*{0.33cm}\\ m & -m^\prime & m^\prime-m\end{pmatrix}
\begin{pmatrix} \ell & \ell^\prime & \mathcal{L}\vspace*{0.33cm}\\ 0 & 0 & 0 \end{pmatrix}
\begin{pmatrix} 1 & 1 & \nu\vspace*{0.33cm}\\ q & -q^\prime & q^\prime-q \end{pmatrix}
\begin{pmatrix} 1 & 1 & \nu\vspace*{0.33cm}\\ 0 & 0 & 0 \end{pmatrix}
\begin{pmatrix} K & \nu & \mathcal{L}\vspace*{0.33cm}\\s & q-q^\prime & m^\prime-m\end{pmatrix}
\begin{pmatrix} K & \nu & \mathcal{L}\vspace*{0.33cm}\\0
                  & 0 & 0\end{pmatrix}\nonumber\\
                  &&\times\,(-1)^K\,(-1)^{K+\nu+\mathcal{L}}\nonumber\\
&=&(-1)^{\mathcal{L}}c_{\mathcal{L}}(\varrho_1,0)\,.
\end{eqnarray}
\end{widetext}
In Eq.~\eqref{eq:final_coeff:demo1:0}, the factors $(-1)^K$ and
$(-1)^{\nu+K+\mathcal{L}}$ 
arise from Eqs.~\eqref{eq:gK_transform2:0} and from application of the property
defined in Eq.~\eqref{eq:wigner_prop1} to the sixth Wigner $3j$ symbol in Eq.~\eqref{eq:final_coeff:demo1:0},
respectively. Furthermore, we used the property that $\nu$ is even, i.e. only even $\nu$
contribute to the summation, due to the triple zeros in the second row of the fourth Wigner $3j$ symbol.
Finally, using Eq.~\eqref{eq:invariance_prop:0}, it follows that for
$\varrho_2=0$, 
\begin{eqnarray}
\label{eq:demo_finished:0}
c_{\mathcal{L}}(\varrho_1,0) = (-1)^{\mathcal{L}}\,
c_{\mathcal{L}}(\varrho_1,0).
\end{eqnarray} 
Because no assumptions have been made on the polarization direction $\varrho_1$,  Eq.~\eqref{eq:demo_finished:0} 
shows that only even Legendre coefficients
are present in the PAD if the  
radiative photoabsorption is driven by linearly polarized light, i.e. $\varrho_2=0$,
independently of the polarization direction, $\varrho_1$, driving the non-resonant two-photon 
absorption process. As a consequence, only even Legendre orders contribute to the PAD if
$\varrho_1=\pm 1,0$ and $\varrho_2=0$, translating into a vanishing PECD.
\subsection{Behavior of Legendre coefficients when changing the
  helicity of the one-photon photoionization}
\label{subsec:ci}
The easiest way to prove Eq.~\eqref{eq:ci_rho2} consists of making 
the change of variables defined in Eq.~\eqref{eq:first_invariant_transformation:0},
and evaluate $c^\prime_{\mathcal{L}}(\varrho_1,-\varrho_\varrho)$, using the
property 
\begin{eqnarray}
\label{eq:invariance_prop2:2}
c_{\mathcal{L}}(\varrho_1,-\varrho_2) = c^\prime_{\mathcal{L}}(\varrho_1,-\varrho_2),
\end{eqnarray}
where the unprimed (primed) quantities in Eq.~\eqref{eq:invariance_prop2:2} refer 
to the Legendre coefficients before (after) the change of variables, respectively.

Keeping $\epsilon_{\varrho_1}$ is fixed while changing the
polarization direction $\varrho_2$ 
transforms the fourth Wigner $3j$ symbol in Eq.~\eqref{eq:final_coeff} according
to,
\begin{eqnarray}
  \label{eq:nu_changed1}
\begin{pmatrix} 1 & 1 & \nu\vspace*{0.33cm}\\ -\varrho_2 & \varrho_2 & 0 \end{pmatrix}
                  &=& (-1)^\nu
\begin{pmatrix} 1 & 1 & \nu\vspace*{0.33cm}\\ \varrho_2 & \varrho_2 & 0 \end{pmatrix}\,,
\end{eqnarray}

where we used Eq.~\eqref{eq:wigner_prop1}. 
Inserting Eqs.~\eqref{eq:gK_transform2:0} and~\eqref{eq:nu_changed1} to 
Eq.~\eqref{eq:final_coeff}, for
$c^\prime_{\mathcal{L}}(\varrho_1,-\varrho_2)$  gives,
\begin{widetext}

\begin{eqnarray}
\label{eq:final_coeff:demo1}
c^\prime_{\mathcal{L}}({\varrho_1},{-\varrho_2}) &=&\mathcal{N}_0 \tilde{c}_o\, \sum_{\substack{\ell,m \\ n_o,\ell_o,m_o}}\, 
\sum_{\substack{\ell^\prime,m^\prime \\ n^\prime_o\ell^\prime_o,m^\prime_o}}\, \sum_{q,q^\prime}
\sum_{\substack{q_1,q_2 \\ q_3,q_4}}
\sum^2_{\nu=0} \sum^4_{K=0} 
(-1)^{q3+q4}\,(2\nu+1)(2\mathcal{L}+1)
a^{\ell_o}_{m_o}(n_o)\, a^{*\ell^\prime_o}_{m^\prime_o}(n^\prime_o)\, T_{q_1,q_2} T^{*}_{q_3,q_4}\nonumber
\\         
&&\times (-\mathrm{i})^{\ell-\ell^\prime}\,
(-1)^{m^\prime-q-\varrho_2}\,e^{\mathrm{i}(\delta_{\ell}-\delta_{\ell^\prime})}\,\,g^{(K)}_{q_1,q_2,q_3,q_4}(\varrho_1)
\,\,I^{n_o}_{_{k}}(\ell,\ell_o)\,\, I^{n^\prime_o}_{_{k}}(\ell^\prime,\ell^\prime_o)\,\,\mathcal{S}^{\ell,m}_{\ell_o,m_o}(q) 
\,\,\mathcal{S}^{\ell^\prime,m^\prime}_{\ell^\prime_o,m^\prime_o}(q^\prime)\,\hat{\varsigma}(\ell,\ell^\prime)\nonumber\\ 
&&
\times
\begin{pmatrix} \ell & \ell^\prime & \mathcal{L}\vspace*{0.33cm}\\ m & -m^\prime & m^\prime-m\end{pmatrix}
\begin{pmatrix} \ell & \ell^\prime & \mathcal{L}\vspace*{0.33cm}\\ 0 & 0 & 0 \end{pmatrix}
\begin{pmatrix} 1 & 1 & \nu\vspace*{0.33cm}\\ q & -q^\prime & q^\prime-q \end{pmatrix}
\begin{pmatrix} 1 & 1 & \nu\vspace*{0.33cm}\\ \varrho_2 & -\varrho_2 & 0 \end{pmatrix}
\begin{pmatrix} K & \nu & \mathcal{L}\vspace*{0.33cm}\\s & q-q^\prime & m^\prime-m\end{pmatrix}
\begin{pmatrix} K & \nu & \mathcal{L}\vspace*{0.33cm}\\0
                  & 0 & 0\end{pmatrix}\nonumber\\
                  &&\times\,(-1)^K\,(-1)^\nu\,(-1)^{K+\nu+\mathcal{L}}\nonumber\\
&=&
(-1)^{\mathcal{L}}\,c_{\mathcal{L}}({\varrho_1},{+\varrho_2})
\,.
\end{eqnarray}
\end{widetext}
In Eq.~\eqref{eq:final_coeff:demo1}, the factors $(-1)^K$ and $(-1)^\nu$ 
arise from the invariance property defined in Eq.~\eqref{eq:gK_transform2:0} 
for the transformation defined in Eq.~\eqref{eq:first_invariant_transformation:0}, and~\eqref{eq:nu_changed1},
respectively.  Application of the property defined in Eq.~\eqref{eq:wigner_prop1} to the sixth Wigner $3j$ symbol
in Eq.~\eqref{eq:final_coeff:demo1} gives rise to 
the factor $(-1)^{K+\nu+\mathcal{L}}$. The terms in $K$ and $\nu$ compensates, giving rise
to the factor in $(-1)^{\mathcal{L}}$ alone. Finally, using~\eqref{eq:invariance_prop2:2} and
comparing Eq.~\eqref{eq:final_coeff} for $\varrho_1$ and $\varrho_2$ and Eq.~\eqref{eq:final_coeff:demo1}
for $\varrho_1$ and $-\varrho_2$, determines the proof for Eq.~\eqref{eq:ci_rho2}, i.e.,
\begin{eqnarray}
  \label{eq:final_proof2}
  c_{\mathcal{L}}({\varrho_1},{-\varrho_2})&=& (-1)^{\mathcal{L}}\,c_{\mathcal{L}}({\varrho_1},{+\varrho_2})
\end{eqnarray}
\subsection{Behavior of Legendre coefficients when changing the
  helicity of the two-photon absorption process }
\label{subsec:ci2}
In this section, we present the proof of Eq.~\eqref{eq:main_sym_result2}.
To verify that it is the polarization
direction of the ionizing field alone which imposes the sign for all
odd Legendre 
coefficients, whereas the polarization direction of the two-photon
absorption plays no role, we define the
following transformation 
 \begin{eqnarray} 
   \label{eq:transformation2}
 \begin{pmatrix} q^\prime_1\\q^\prime_2\\q^\prime_3\\q^\prime_4\end{pmatrix} &=&  
 \begin{pmatrix} 0 & 0 &  0 & -1  \\ 
                 0 & 0 &  -1 & 0 \\
                 0 & -1 &  0 & 0 \\
                -1 &  0 &  0 & 0 \end{pmatrix}
 \begin{pmatrix} q_1\\q_2\\q_3\\q_4\end{pmatrix}   
 \end{eqnarray}
which interchanges the indices $q_1\rightleftharpoons -q_4$ and
$q_2\rightleftharpoons -q_3$  
while keeping Eq.~\eqref{eq:final_coeff} unchanged and $s$ invariant. In particular, the 
tensor elements appearing in Eq.~\eqref{eq:final_coeff} then transform according to,
\begin{subequations}
  \begin{eqnarray}
    T_{q^\prime_1,q^\prime_2} =  T_{-q_4,-q_3}&=&(-1)^{q_3+q_4}T^{*}_{q_4,q_3}\nonumber\\
    &=&(-1)^{q_3+q_4}T^{*}_{q_3,q_4}
  \end{eqnarray}
  and 
  \begin{eqnarray}
  T^{*}_{q^\prime_3,q^\prime_4}  =  T^{*}_{-q_2,-q_1} &=& (-1)^{q_1+q_2}T_{q_2,q_1}\nonumber\\
   &=&(-1)^{q_1+q_2}T_{q_1,q_2},
 \end{eqnarray}
\end{subequations}
where we have made use of the correspondence between the components of a vector operator in
spherical and cartesian basis, defined in Eq.~\eqref{eq:cartesian_spherical} 
in Appendix~\ref{subsec:rotmat}, in  $T_{q_k,q_{k^\prime}}$,
for $q_k,q_{k^\prime}=\pm 1,0$, together with the fact that the two-photon absorption tensor is
symmetric in cartesian coordinates, i.e., $T_{i,j}=T_{j,i}$ for $i,j=(x,y,z)$.

We define $\Lambda_{\theta}(\varrho_1)$, according 
Eq.~\eqref{eq:gK_simplicity:0}
and we study the symmetry properties of $\Lambda^\prime_{\theta}(\varrho_1)$
upon transformation defined in Eq.~\eqref{eq:transformation2}. 
In particular, because the quantity given by,
\begin{eqnarray}
  (-1)^{q^\prime_1+q^\prime_3} T_{q^\prime_1,q^\prime_2}T_{q^\prime_3,q^\prime_4}W_{\theta}(s^\prime)\,,
\end{eqnarray}  
is (as for the earlier transformation defined in Eq.~\eqref{eq:first_invariant_transformation:0}) invariant under transformation defined in
Eq.~\eqref{eq:transformation2}, we may neglect it in the following, avoiding
cumbersome notations. We outline, however, that a full notation was 
used in Section~\ref{subsec:ci:0}. Therefore, given such invariance properties, we may  consider the
behavior of $g^K$ under exchange $\varrho_1\rightarrow -\varrho_1$ alone,  and neglect
the extra terms depending on $K,q_1,\dots,q_4$ in the expression for $\Lambda^\prime_{\theta}(s)$.
Because $\varrho_1$ is changed to $-\varrho_1$ while $\varrho_2$ is kept
fixed, we consider 
$g^K(-\varrho_1)$ which becomes, upon
transformation defined in Eq.~\eqref{eq:transformation2}, 

\begin{widetext}
\begin{eqnarray}
\label{eq:gK_transformed2}
g^{(K)}_{q^\prime_1,q^\prime_2,q^\prime_3,q^\prime_4}(-\varrho_1) &=&
\sum^2_{Q=0}
\sum^2_{Q^\prime=0}
\sum^{Q+Q^\prime}_{K=|Q-Q^\prime|}
\gamma^{(K)}_{Q,\rm Q^\prime}    
\begin{pmatrix} 1 & 1 & Q^\prime \vspace*{0.33cm} \\ -q_4 & -q_3 & q_4+q_3\end{pmatrix}
\begin{pmatrix} 1 & 1 & Q^\prime \vspace*{0.33cm} \\ -\varrho_1 & -\varrho_1 & +2\varrho_1 \end{pmatrix}\\\nonumber
&&\times
\begin{pmatrix} 1 & 1 & Q        \vspace*{0.33cm} \\  -q_2 & -q_1 & q_2+q_1\end{pmatrix}
\begin{pmatrix} 1 & 1 & Q        \vspace*{0.33cm} \\ -\varrho_1 & -\varrho_1 & +2\varrho_1 \end{pmatrix}
\begin{pmatrix} Q^\prime& Q & K \vspace*{0.33cm} \\ -q_4-q_3 & q_3+q_2 & s\end{pmatrix}
\begin{pmatrix} Q^\prime& Q & K \vspace*{0.33cm} \\ -2\varrho_1 & +2\varrho_1 & 0\end{pmatrix}\,,
\end{eqnarray}
\end{widetext}
where we have interchanged the indexes $Q$ and $Q^\prime$.
Next we apply the symmetry property given in Eq.~\eqref{eq:wigner_prop1},
followed by an odd permutation of the first and second columns, according to 
Eq.~\eqref{eq:permutation_law}, 
to the first Wigner $3j$ symbol in Eq.~\eqref{eq:gK_transformed2}. We find,
\begin{subequations}
  \label{eq:manips}
  \begin{eqnarray}
\label{eq:manip1}
\begin{pmatrix} 1 & 1 & Q^\prime \vspace*{0.33cm} \\ -q_4 & -q_3 & q_4+q_3\end{pmatrix}
                  &=&
\begin{pmatrix} 1 & 1 & Q^\prime \vspace*{0.33cm} \\ q_3 & q_4 & -q_3-q_4\end{pmatrix}\,.\quad\quad
\end{eqnarray} 
The same procedure is applied to the third symbol in
Eq.~\eqref{eq:gK_transformed2}, i.e., 
\begin{eqnarray}
\label{eq:manip2}
\begin{pmatrix} 1 & 1 & Q        \vspace*{0.33cm} \\  -q_2 & -q_1 & q_2+q_1\end{pmatrix}
                  &=&
\begin{pmatrix} 1 & 1 & Q        \vspace*{0.33cm} \\  q_1 & q_2 & -q_1-q_2\end{pmatrix}\,.\quad\quad
\end{eqnarray} 
Next, odd permutation of the first and second columns in the fifth 
Wigner $3j$ symbol gives, 
\begin{eqnarray}
\label{eq:manip3}
\begin{pmatrix} Q^\prime& Q & K \vspace*{0.33cm} \\ -q_4-q_3 & q_2+q_1 & s\end{pmatrix}
                        &=&
\begin{pmatrix} Q& Q^\prime & K \vspace*{0.33cm} \\  q_1+q_2& -q_4-q_3 & s\end{pmatrix}\nonumber\\
                 &&\times(-1)^{Q+Q^\prime+K}\,.
\end{eqnarray} 
Application of Eq.~\eqref{eq:wigner_prop1}, followed by permutation of the first two
rows leaves the sign of the second Wigner $3j$ symbol unchanged for all $Q^\prime$, namely 
\begin{eqnarray}
\label{eq:manip3.5}
\begin{pmatrix} 1       & 1 & Q^\prime \vspace*{0.33cm} \\ -\varrho_1 & -\varrho_1 & -2\varrho_1\end{pmatrix}
                        &=&
\begin{pmatrix} 1       & 1 & Q^\prime \vspace*{0.33cm} \\ +\varrho_1 & +\varrho_1 & +2\varrho_1\end{pmatrix}
\end{eqnarray} 
and analogously for the fourth Wigner symbol involving $Q$. It is to note that,
the left side of Eq.~\eqref{eq:manip3.5} is related to $g^K(-\varrho_1)$ while the 
right side is related to $g^K(+\varrho_1)$.

Permuting the first two rows of the fifth Wigner symbol in Eq.~\eqref{eq:gK_transformed2} gives, 
\begin{eqnarray}
\label{eq:manip4}
\begin{pmatrix} Q^\prime& Q & K \vspace*{0.33cm} \\ -2\varrho_1 & +2\varrho_1 & 0\end{pmatrix}
                        &=&(-1)^{Q+Q^\prime+K}
\begin{pmatrix} Q& Q^\prime & K \vspace*{0.33cm} \\ 2\varrho_1 & -2\varrho_1 & 0\end{pmatrix}\,.\nonumber\\
\end{eqnarray} 
\end{subequations}
Inserting the symmetry transformations~\eqref{eq:manips} into
Eq.~\eqref{eq:gK_transformed2}, leads to
a compensation of the terms $(-1)^{Q+Q^\prime+K}$ in Eqs.~\eqref{eq:manip3} and~\eqref{eq:manip4}. 
Finally, comparing Eq.~\eqref{eq:gK_transformed2} and Eq.~\eqref{eq:gK} gives
the following property,
\begin{eqnarray}
\label{eq:main_sym_result}
\sum_{K}\sum_{\substack{q_1,q_2 \\ q_3,q_4}}\, 
g^K_{q_1,q_2,q_3,q_4}(-\varrho_1)
&=&
\sum_{K}\sum_{\substack{q_1,q_2 \\ q_3,q_4}}\, 
g^K_{q_1,q_2,q_3,q_4}(+\varrho_1)\,,\nonumber\\
\end{eqnarray}
which implies $c_{\mathcal{L}}(-\varrho_1,\varrho_2) = c_{\mathcal{L}}(\varrho_1,\varrho_2)$ according to
Eq.~\eqref{eq:final_coeff}, cf. Eq.~\eqref{eq:main_sym_result2}.

\section{Evaluation of the two-photon transition moments 
  in the framework of coupled cluster theory} 
\label{appen_twophoton}
The rotationally averaged two-photon transition strength strength $\tilde{\delta}^{\rm TP}$ (in a.u.)                                                                                                        
and the two-photon transition probability rate constant $K_{go}$ are defined in 
units of $\rm{cm^4 s}$ as follows~\cite{chattig:1998,Monson:1970,christiansen:1998,sd1994}
\begin{subequations}
  \begin{eqnarray}\label{eq:deltaTP}
    \tilde{\delta}^{\rm TP}&=&a_0^4E_\mathrm{h}^{-2}(F\delta_F+G\delta_G+H\delta_H),\\
    K_{go}&=&\hbar^2t_0(2\pi)^2\alpha^2\omega_{\mathrm{ph},1}\omega_{\mathrm{ph},2}\tilde{\delta}^{TP}\,,
             \label{eq:Kgo}
  \end{eqnarray}
 \label{defintion}
\end{subequations}
where $a_0$ is the Bohr radius, $t_0 = \hbar/E_{\mathrm{h}}$ is the atomic unit
of time, $\alpha$ the fine structure constant 
and $\omega_{\mathrm{ph},1}$ and $\omega_{\mathrm{ph},2}$ the photon energies. 
$F$, $G$ and $H$ are parameters depending on the arrangement 
and polarization of the laser used in the experiment~\cite{chattig:1998,Monson:1970,christiansen:1998,sd1994}. 
In Eq.~\eqref{eq:deltaTP}, the parameters $\delta_F$, $\delta_H$ and $\delta_G$ read~\cite{chattig:1998,christiansen:1998}
\begin{align}
 \delta_F&=\frac{1}{30}\sum_{\alpha\beta} S^{go}_{\alpha\alpha,\beta\beta}\nonumber\\
 \delta_G&=\frac{1}{30}\sum_{\alpha\beta} S^{go}_{\alpha\beta,\alpha\beta}\nonumber\\
 \delta_H&=\frac{1}{30}\sum_{\alpha\beta} S^{go}_{\alpha\beta,\beta\alpha},
 \label{defintion1}
\end{align}
where $\alpha,\beta=x,y$ and $z$.
Here $g$ and $o$ refer to the ground and excited states. In the above relations, $S^{go}_{\alpha\beta,\gamma\delta}$,  
the so-called transition strength, is defined as follows~\cite{chattig:1998,christiansen:1998}
\begin{eqnarray} 
  S^{go}_{\alpha\beta,\gamma\delta}(\omega_{\mathrm{ph}})&=&\frac{1}{2}[T_{go}^{\alpha\beta}(-\omega_{\mathrm{ph}})T_{og}^{\gamma\delta}(\omega_{\mathrm{ph}})\nonumber\\
  &&\,\,\,+  T_{go}^{\gamma\delta}(-\omega_{\mathrm{ph}})^*T_{og}^{\alpha\beta}(\omega_{\mathrm{ph}})^*]\nonumber\\
  &=& T_{go}^{\alpha\beta}(-\omega_{\mathrm{ph}})T_{og}^{\gamma\delta}(\omega_{\mathrm{ph}}),
 \label{smatdefin}
\end{eqnarray}
where the $T_{og}^{\alpha\beta}(\omega_{\mathrm{ph}})$ and $T_{go}^{\alpha\beta}(\omega_{\mathrm{ph}})$ are called the 
two-photon transition matrix elements.
These tensors read~\cite{chattig:1998,christiansen:1998} 
 \begin{align}
   T_{og}^{\alpha\beta}(\omega_{\mathrm{ph},2})&=\sum_n
   \Big[\frac{\langle \psi_o \vert \beta \vert n\rangle \langle n \vert \alpha \vert \psi_g\rangle}{E_g-E_n+\hbar\omega_{\mathrm{ph},1}}
   +\frac{\langle \psi_o \vert \alpha \vert n\rangle \langle n \vert \beta \vert \psi_g\rangle}{E_g-E_n+\hbar\omega_{\mathrm{ph},2}}\Big]\nonumber\\
   &=T_{go}^{\alpha\beta}(-\omega_{\mathrm{ph},2})^*,
  \label{tanmoments}
 \end{align}
where $\alpha$ and $\beta$ are Cartesian components of the position
operator ($\alpha,\beta=x,y$ and $z$). 
$\hbar\omega_{\mathrm{ph},1}$ and $\hbar\omega_{\mathrm{ph},2}$ are the photon energies which satisfy the matching condition 
$\hbar\omega_{\mathrm{ph},1}+\hbar\omega_{\mathrm{ph},2}=\hbar\omega_{og}=E_o-E_g$. For variational {\it ab initio} methods, the two-photon 
absorption tensor is symmetric with respect not only to the permutation of the operators $\alpha$ 
and $\beta$ (assuming that $\omega_{\mathrm{ph},2}$ is replaced by $\omega_{og}-\omega_{\mathrm{ph},2}$) but also to complex 
conjugation combined with a simultaneous inversion of the frequencies and exchange of the initial and 
final states~\cite{chattig:1998,christiansen:1998}. In coupled cluster response theory, the 
two-photon absorption tensors $T_{og}^{\alpha\beta}(\omega_{\mathrm{ph},2})$ and $T_{go}^{\alpha\beta}(-\omega_{\mathrm{ph},2})$ 
are are in general not each other complex conjugate {\it i.e.} $T_{og}^{\alpha\beta}(\omega_{\mathrm{ph},2}) \neq T_{go}^{\alpha\beta}(-\omega_{\mathrm{ph},2})^*$, 
whereas for the transition strengths, 
which are calculated as a symmetrized product of right 
$T_{og}^{\alpha\beta}(\omega_{\mathrm{ph}})$ and left $T_{go}^{\alpha\beta}(\omega_{\mathrm{ph}})$ two-photon absorption 
tensors as shown in Eq.~(\ref{smatdefin}), we have~\cite{chattig:1998,christiansen:1998}
 \begin{align}
   S^{go}_{\alpha\beta,\gamma\delta}(\omega_{\mathrm{ph}})&=S^{go}_{\alpha\beta,\gamma\delta}(-\omega_{\mathrm{ph}})^*\nonumber\\
                                                 &=S^{go}_{\gamma\delta,\alpha\beta}(\omega_{\mathrm{ph}})^*=S^{go}_{\beta\alpha,\delta\gamma}(\omega_{og}-\omega_{\mathrm{ph}})^*.
 \end{align}
 These two-photon absorption tensors $T_{og}^{\alpha\beta}(\omega_{\mathrm{ph}})$ and $T_{go}^{\alpha\beta}(\omega_{\mathrm{ph}})$ are 
called right and left two-photon absorption tensor from the ground state $g$ to the excited state $o$, 
respectively~\cite{chattig:1998,christiansen:1998}. As a side remark, the imaginary part of two-photon
absorption tensors calculated using the CC method vanishes in the limit of complete cluster expansion and thus 
it does not influence the results of the two-photon absorption tensor~\cite{chh1998}.

We should mention that Eq.~(\ref{tanmoments}) is presented in a general form and in the Cartesian basis. 
However, we interest in the special case two photons with same polarization and energy values 
({\it i.e.} $\omega_{\mathrm{ph},1}=\omega_{\mathrm{ph},2}=\omega_\text{ph}$). If one uses the inverse relations of Eq.~(\ref{eq:tensor_def2}) 
and inserts them into Eq.~(\ref{tanmoments}), it will give Eq.~(\ref{eq:tensordef}), which is two-absorption 
tensor in the spherical basis. 

The left and right two-photon absorption tensors change under a rotation $R$, whereas the transition
strength $S^{go}_{\alpha\beta,\gamma\delta}(\omega_{\mathrm{ph}})$ remain unchanged 
($S^{go}_{\alpha\beta,\gamma\delta}(\omega_{\mathrm{ph}})=R S^{go}_{\alpha\beta,\gamma\delta}(\omega_{\mathrm{ph}}) R^\dagger$).
The left and right two-photon absorption tensors for fenchone and camphor (calculated at  
the rotated arrangement (see Fig.~\ref{fig:optimiz}) such that the origin is at the center 
of mass and  principal axes of inertia are along coordinate axes) are shown in 
Tables~\ref{tab11a} and \ref{tab11b}. In Fig.~\ref{fig:optimiz}, the eigenvectors of the 
left (red vectors) and right (blue vectors) two-photon absorption tensor corresponding to the third
electronically excited states of fenchone and camphor are shown. The corresponding eigenvalues of the left and right
two-photon absorption tensor are $(-10.96,0.20,13.38)$ and $(-5.58,0.10,6.83)$, respectively for fenchone and 
$(-11.06,-0.92, 10.51)$ and $(-5.73,-0.47,5.40)$, respectively for camphor. From this information, 
the rhombicity ($T_\mathrm{r}$), axialty ($T_\mathrm{a}$) and the ratio
($R=T_\mathrm{r}/T_\mathrm{a}$) of these symmetric tensors can be calculated using the 
following relations:
\begin{align}
 T_\mathrm{r}&=a_0^2~E_\mathrm{h}^{-1}\frac{2}{3}(b-e)\nonumber\\
 T_\mathrm{a}&=a_0^2~E_\mathrm{h}^{-1}(-b-e)\nonumber\\
 R&=\frac{T_\mathrm{r}}{T_\mathrm{a}},
 \label{constrohm}
\end{align}
where b and e are 
\begin{align}
 b&=T_{xx}^h-T_0^h\nonumber\\
 e&=T_{yy}^h-T_0^h\nonumber\\
 T_0^h&=\frac{1}{3}(T_{xx}^h+T_{yy}^h+T_{zz}^h)
\end{align}
Here $h$ refers to the left and right two-photon absorption tensors and $T_{xx}^h$ and $T_{yy}^h$ refers to the 
diagonal elements of the the left and right two-photon absorption tensors. Based on Eq.~(\ref{constrohm}),
the corresponding numerical values for the rhombicity ($T_\mathrm{r}$), axiality ($T_\mathrm{a}$) and
an their ratio ($R$) are shown in Table.~\ref{tab:rohombic}.
\begin{table}[tb]
\caption{\label{tab:rohombic} Axiality and rhombicity for fenchone and camphor. $T_\mathrm{r}$ and $T_\mathrm{a}$ 
are given in units of $a_0^2~E_\mathrm{h}^{-1}$.}
\begin{ruledtabular}
\begin{tabular}{@{}l*{5}{D{.}{.}{2}}@{}}
& &\multicolumn{1}{c}{fenchone} &\multicolumn{2}{c}{camphor} \\
\cline{2-3} \cline{4-5} 
    &\multicolumn{1}{c}{left}& \multicolumn{1}{c}{right}
    &\multicolumn{1}{c}{left}& \multicolumn{1}{c}{right}  \\
\hline
$T_\mathrm{r}$&-7.44 &-3.78 &-6.76 &-3.50  \\
$T_\mathrm{a}$&12.50 & 6.38 &11.00 & 5.68  \\
$R           $&-0.59 &-0.59 &-0.61 &-0.61  \\
\end{tabular}
\end{ruledtabular}
\end{table}
As inferred from Table.~\ref{tab:rohombic}, these values for fenchone and camphor are close to each other. 
Furthermore, we report the parameters $\delta_F$, $\delta_G$, $\delta_H$, $\tilde{\delta}^{\rm TP}$ and $K_{go}$ for different 
types of polarisations in Tables~\ref{tab10}, \ref{tab10a}, \ref{tab11} and \ref{tab11a}.

\begin{table*}
\caption{\label{tab11a} Left ($T_{go}^{\alpha \beta}$) and right ($T_{og}^{\alpha \beta}$) two-photon absorption
tensors in units of $a_0^4~E_\mathrm{h}^{-2}$ for fenchone as obtained with the CCSD method.}
\begin{ruledtabular}
  \begin{tabular}{@{}l*{12}{D{.}{.}{0}}@{}}
     \toprule
    States&T_{go}^{xx}&T_{og}^{xx}&T_{go}^{xy}&T_{og}^{xy}&T_{go}^{xz}&T_{og}^{xz}&T_{go}^{yy}&T_{og}^{yy}&T_{go}^{yz}&T_{og}^{yz}&T_{go}^{zz}&T_{og}^{zz}\\
\hline
\hline
    \midrule
    $\text{A}$    &-$0.15$&-$0.09$&-$0.05$&-$0.02$&$0.130$&$0.05$&-$0.37$&-$0.20$&$0.27$&$0.14$&-$0.36$&-$0.19$\\
    $\text{B}$    &$2.21$&$1.14$&$23.70$&$12.34$&$10.39$&$5.40$&-$2.31$&-$1.20$&-$0.34$&-$0.17$&-$3.39$&-$1.80$\\
    $\text{C}_1$  &-$0.30$&-$0.15$&-$10.60$&-$5.40$&-$5.74$&-$2.93$&$1.57$&$0.82$&$1.43$&$0.73$&$1.35$&$0.69$\\
    $\text{C}_2$  &-$29.42$&-$15.34$&$7.58$&$3.90$&-$1.87$&-$0.93$&-$8.39$&-$4.29$&-$2.62$&-$1.33$&-$2.77$&-$1.47$\\
    $\text{C}_3$  &-$39.74$&-$20.68$&-$2.18$&-$1.10$&$5.69$&$2.95$&-$11.02$&-$5.63$&$0.03$&$0.08$&-$9.28$&-$4.82$\\
    \bottomrule
  \end{tabular}
\end{ruledtabular}
\end{table*}

\begin{table*}
\caption{\label{tab11b} Left ($T_{go}^{\alpha \beta}$) and right ($T_{og}^{\alpha \beta}$) two-photon absorption 
tensors in units of $a_0^4~E_\mathrm{h}^{-2}$ for camphor as obtained with the CCSD method.}
\begin{ruledtabular}
  \begin{tabular}{@{}l*{12}{D{.}{.}{0}}@{}}
     \toprule
    States&T_{go}^{xx}&T_{og}^{xx}&T_{go}^{xy}&T_{og}^{xy}&T_{go}^{xz}&T_{og}^{xz}&T_{go}^{yy}&T_{og}^{yy}&T_{go}^{yz}&T_{og}^{yz}&T_{go}^{zz}&T_{og}^{zz}\\
\hline
\hline
    \midrule
    $\text{A}$    &-$0.46$&-$0.27$&-$0.35$&-$0.20$&-$0.67$&-$0.34$&$0.58$&$0.29$&-$0.04$&-$0.02$&-$1.62$&-$0.86$\\
    $\text{B}$    &$1.66$&$1.00$&$12.91$&$6.80$&$17.38$&$9.18$&$9.15$&$4.73$&$6.50$&$3.36$&$10.43$&$5.46$\\
    $\text{C}_1$  &$10.42$&$5.37$&$0.61$&$0.27$&$1.22$&$0.55$&-$4.83$&-$2.48$&-$4.90$&-$2.54$&-$7.06$&-$3.70$\\
    $\text{C}_2$  &$4.39$&$2.14$&$0.35$&$0.22$&-$5.72$&-$2.94$&$5.76$&$2.92$&$2.65$&$1.39$&-$8.13$&-$4.26$\\
    $\text{C}_3$  &-$29.68$&-$15.55$&$1.46$&$0.65$&$4.10$&$1.96$&-$2.69$&-$1.41$&-$1.59$&-$0.82$&-$1.03$&-$0.63$\\
    \bottomrule
  \end{tabular}
\end{ruledtabular}
\end{table*}

\begin{table*}
  \caption{\label{tab10} $\tilde{\delta}^{\rm TP}$ and $K_{go}$ referring to the rotationally averaged 
two-photon transition strength and the two-photon-transition probability rate constant, respectively for fenchone.
$\delta_F$, $\delta_G$, $\delta_H$ are calculated by using Eq.~(\ref{defintion1}). $\tilde{\delta}^{\rm TP}$ is given 
in units of $a_0^4~E_\mathrm{h}^{-2}$ and $K_{go}$ in units of $\mathrm{cm}^4~\mathrm{s}$. 
}
\begin{ruledtabular}
\begin{tabular}{@{}l*{5}{D{.}{.}{9}}@{}}
   \toprule
   states  & \delta_F &\delta_G& \delta_H & \tilde{\delta}^{\rm TP}\footnote{both photon circularly polarized {\it i.e.},
     $F=-\frac{1}{4},G=\frac{7}{2}$ and $H=-\frac{1}{4}$}&K_{go}\footnote{both photon circularly polarized {\it i.e.},
     $F=-\frac{1}{4},G=\frac{7}{2}$ and $H=-\frac{1}{4}$}\\
 \hline
 \hline
     \midrule
     $\text{A}$    &$0.01$&$0.00$&$0.00$&$0.02$&3.61\times 10^{-56}\\
     $\text{B}$    &$0.22$&$23.62$&$23.62$&$141.31$&4.59\times 10^{-52}\\
     $\text{C}_1$  &$0.12$&$5.08$&$5.08$&$30.27$&1.09\times 10^{-52}\\
     $\text{C}_2$  &$28.54$&$18.70$&$18.70$&$55.12$&2.03\times 10^{-52}\\
     $\text{C}_3$  &$62.31$&$32.24$&$32.24$&$68.81$&2.55\times 10^{-52}\\
     \bottomrule
   \end{tabular}
 \end{ruledtabular}
 \end{table*}

  \begin{table*}
    \caption{\label{tab10a} $\tilde{\delta}^{\rm TP}$ and $K_{go}$ referring to the rotationally averaged 
two-photon transition strength and the two-photon-transition probability rate constant, respectively for fenchone.
$\tilde{\delta}^{\rm TP}$ is given in units of $a_0^4~E_\mathrm{h}^{-2}$ and $K_{go}$ in units of $\mathrm{cm}^4~\mathrm{s}$.}
\begin{ruledtabular}
\begin{tabular}{@{}l*{5}{D{.}{.}{2}}@{}}
   \toprule
   states  &  \tilde{\delta}^{\rm TP}\footnote{both photons polarized linearly with parallel polarization {\it i.e.},
     $F=G=H=2$}&
     K_{go}\footnote{both photons polarized linearly with parallel polarization {\it i.e.},
     $F=G=H=2$}& 
       \tilde{\delta}^{\rm TP}\footnote{both photons polarized linearly with perpendicular polarization {\it i.e.},
     $F=-1,G=4$ and $H=-1$}&
     K_{go}\footnote{both photons polarized linearly with perpendicular polarization {\it i.e.},
     $F=-1,G=4$ and $H=-1$}\\
 \hline
 \hline
     \midrule
     $\text{A}$    &$0.06$&1.04\times 10^{-55}&$0.02$&4.62\times 10^{-55}\\
     $\text{B}$    &$94.92$&3.80\times 10^{-52}&$117.90$&3.82\times 10^{-52}\\
     $\text{C}_1$  &$20.57$&7.43\times 10^{-53}&$25.30$&9.14\times 10^{-53}\\
     $\text{C}_2$  &$131.87$&4.86\times 10^{-52}&$64.96$&2.39\times 10^{-52}\\
     $\text{C}_3$  &$253.58$&9.39\times 10^{-52}&$98.88$&3.66\times 10^{-52}\\
     \bottomrule
   \end{tabular}
 \end{ruledtabular}
 \end{table*}
 
 \begin{table*}
\caption{\label{tab11} Same as Table.~\ref{tab10} but for camphor.}
\begin{ruledtabular}
\begin{tabular}{@{}l*{5}{D{.}{.}{9}}@{}}
   \toprule
   states  & \delta_F &\delta_G& \delta_H & \tilde{\delta}^{\rm TP}\footnote{both photon circularly polarized {\it i.e.},
     $F=-\frac{1}{4},G=\frac{7}{2}$ and $H=-\frac{1}{4}$}
     &K_{go}\footnote{both photon circularly polarized {\it i.e.},
     $F=-\frac{1}{4},G=\frac{7}{2}$ and $H=-\frac{1}{4}$}\\
 \hline
 \hline
     \midrule
     $\text{A}$    &0.04 &0.08 &0.08 &0.37  &6.02 \times 10^{-55}\\
     $\text{B}$    &7.92 &21.34&21.34&112.22&3.80\times10^{-52}\\
     $\text{C}_1$  &0.04 &4.02 &4.02 &24.05 &9.21\times10^{-53}\\
     $\text{C}_2$  &0.05 &3.40 &3.40 &20.31 &7.84\times10^{-53}\\
     $\text{C}_3$  &19.58&16.21&16.21&58.13 &2.26\times10^{-52}\\
     \bottomrule
   \end{tabular}
 \end{ruledtabular}
 \end{table*}

   \begin{table*}                                                     
                                                 
\caption{\label{tab12a} The same as Table.~\ref{tab10a} but for camphor. }
\begin{ruledtabular}
\begin{tabular}{@{}l*{5}{D{.}{.}{2}}@{}}
   \toprule
   states  &  \tilde{\delta}^{\rm TP}\footnote{both photons polarized linearly with parallel polarization {\it i.e.},
     $F=G=H=2$}&
     K_{go}\footnote{both photons polarized linearly with parallel polarization {\it i.e.},
     $F=G=H=2$}& 
       \tilde{\delta}^{\rm TP}\footnote{both photons polarized linearly with perpendicular polarization {\it i.e.},
     $F=-1,G=4$ and $H=-1$}&
     K_{go}\footnote{both photons polarized linearly with perpendicular polarization {\it i.e.},
     $F=-1,G=4$ and $H=-1$}\\
 \hline
 \hline
     \midrule
     $\text{A}$    &$0.39$&6.29\times 10^{-55}&$0.33$&5.47\times 10^{-55}\\
     $\text{B}$    &$10.21$&3.42\times 10^{-52}&$98.80$&3.34\times 10^{-52}\\
     $\text{C}_1$  &$16.16$&6.18\times 10^{-52}&$20.06$&7.68\times 10^{-53}\\
     $\text{C}_2$  &$13.72$&5.29\times 10^{-53}&$16.96$&6.54\times 10^{-53}\\
     $\text{C}_3$  &$104.02$&4.04\times 10^{-52}&$61.49$&2.39\times 10^{-52}\\
     \bottomrule
   \end{tabular}
 \end{ruledtabular}
 \end{table*}
As indicated, there are two two-photon transition matrices
obtained when we employ the coupled cluster method. 
This is  problematic in the calculation of photoelectron angular distributions of the 
molecules under investigations, because the model constructed for this purpose (see Sec.~\ref{sec:model}) 
depends on only a single two-photon transition tensor. Thus, the computational 
procedure based on the CC calculation would not work for the evaluation of photoelectron angular 
distributions, unless the left and right two-photon transition tensors are combined such that the 
two-photon transition strength and the total cross section remain unchanged when compared to the
conventionally chosen recipe for coupled cluster calculations.

 The effective two-photon transition matrix element can be written as follows,
\begin{align} 
  \tilde{T}^{\alpha\beta}_{og}(\omega_{ph})=
  \sqrt{2}\sign(j)
  \sqrt{\frac{T_{go}^{\alpha\beta}(-\omega_{ph})T_{og}^{\alpha\beta}(\omega_{ph})}{2}}
  \label{newsmat}
\end{align}
and the $\sign(j)$ of $\tilde{T}^{\alpha\beta}_{og}(\omega_{ph})$ being
the same as the signs of the left $T_{go}^{\alpha\beta}$
and right $T_{og}^{\alpha\beta}$ two-photon absorption tensors
for each electronic state as shown in Tables~\ref{tab11a} and \ref{tab11b}. 

Employing Eq.~(\ref{newsmat}) 
leaves the transition strength of $S^{go}_{\alpha\beta,\gamma\delta}$ and 
the two-photon transition probability rate constant unchanged. Thus all parameters $\delta_F$, $\delta_G$, $\delta_H$, $\tilde{\delta}^{\rm TP}$ as 
well as $K_{go}$ in Eqs.~(\ref{defintion}) and (\ref{defintion1}) are the same as before combining 
the right and left transition moments. Thus, employing Eq.~(\ref{newsmat}) provides exactly the same 
reported values in Tables~\ref{tab10}, \ref{tab10a}, \ref{tab11} and \ref{tab11a}. 
The lower part of Tables~\ref{tab:twophotonA} and
\ref{tab:twophotonCam} presents the (symmetric) effective two-photon
transition matrix elements (transition moments) based on
Eq.~(\ref{newsmat}) for fenchone and camphor. 

\bibliography{mybib}

\begin{thebibliography}{67}%
\makeatletter
\providecommand \@ifxundefined [1]{%
 \@ifx{#1\undefined}
}%
\providecommand \@ifnum [1]{%
 \ifnum #1\expandafter \@firstoftwo
 \else \expandafter \@secondoftwo
 \fi
}%
\providecommand \@ifx [1]{%
 \ifx #1\expandafter \@firstoftwo
 \else \expandafter \@secondoftwo
 \fi
}%
\providecommand \natexlab [1]{#1}%
\providecommand \enquote  [1]{``#1''}%
\providecommand \bibnamefont  [1]{#1}%
\providecommand \bibfnamefont [1]{#1}%
\providecommand \citenamefont [1]{#1}%
\providecommand \href@noop [0]{\@secondoftwo}%
\providecommand \href [0]{\begingroup \@sanitize@url \@href}%
\providecommand \@href[1]{\@@startlink{#1}\@@href}%
\providecommand \@@href[1]{\endgroup#1\@@endlink}%
\providecommand \@sanitize@url [0]{\catcode `\\12\catcode `\$12\catcode
  `\&12\catcode `\#12\catcode `\^12\catcode `\_12\catcode `\%12\relax}%
\providecommand \@@startlink[1]{}%
\providecommand \@@endlink[0]{}%
\providecommand \url  [0]{\begingroup\@sanitize@url \@url }%
\providecommand \@url [1]{\endgroup\@href {#1}{\urlprefix }}%
\providecommand \urlprefix  [0]{URL }%
\providecommand \Eprint [0]{\href }%
\providecommand \doibase [0]{http://dx.doi.org/}%
\providecommand \selectlanguage [0]{\@gobble}%
\providecommand \bibinfo  [0]{\@secondoftwo}%
\providecommand \bibfield  [0]{\@secondoftwo}%
\providecommand \translation [1]{[#1]}%
\providecommand \BibitemOpen [0]{}%
\providecommand \bibitemStop [0]{}%
\providecommand \bibitemNoStop [0]{.\EOS\space}%
\providecommand \EOS [0]{\spacefactor3000\relax}%
\providecommand \BibitemShut  [1]{\csname bibitem#1\endcsname}%
\let\auto@bib@innerbib\@empty
\bibitem [{\citenamefont {Lux}\ \emph {et~al.}(2012)\citenamefont {Lux},
  \citenamefont {Wollenhaupt}, \citenamefont {Bolze}, \citenamefont {Liang},
  \citenamefont {K\"ohler}, \citenamefont {Sarpe},\ and\ \citenamefont
  {Baumert}}]{LuxACIE12}%
  \BibitemOpen
  \bibfield  {author} {\bibinfo {author} {\bibfnamefont {C.}~\bibnamefont
  {Lux}}, \bibinfo {author} {\bibfnamefont {M.}~\bibnamefont {Wollenhaupt}},
  \bibinfo {author} {\bibfnamefont {T.}~\bibnamefont {Bolze}}, \bibinfo
  {author} {\bibfnamefont {Q.}~\bibnamefont {Liang}}, \bibinfo {author}
  {\bibfnamefont {J.}~\bibnamefont {K\"ohler}}, \bibinfo {author}
  {\bibfnamefont {C.}~\bibnamefont {Sarpe}}, \ and\ \bibinfo {author}
  {\bibfnamefont {T.}~\bibnamefont {Baumert}},\ }\href {\doibase
  10.1002/anie.201201229} {\bibfield  {journal} {\bibinfo  {journal} {Angew.
  Chem. Int. Ed.}\ }\textbf {\bibinfo {volume} {51}},\ \bibinfo {pages} {4755}
  (\bibinfo {year} {2012})}\BibitemShut {NoStop}%
\bibitem [{\citenamefont {Lehmann}\ \emph {et~al.}(2013)\citenamefont
  {Lehmann}, \citenamefont {Ram}, \citenamefont {Powis},\ and\ \citenamefont
  {Janssen}}]{LehmannJCP13}%
  \BibitemOpen
  \bibfield  {author} {\bibinfo {author} {\bibfnamefont {C.~S.}\ \bibnamefont
  {Lehmann}}, \bibinfo {author} {\bibfnamefont {N.~B.}\ \bibnamefont {Ram}},
  \bibinfo {author} {\bibfnamefont {I.}~\bibnamefont {Powis}}, \ and\ \bibinfo
  {author} {\bibfnamefont {M.~H.~M.}\ \bibnamefont {Janssen}},\ }\href
  {\doibase http://dx.doi.org/10.1063/1.4844295} {\bibfield  {journal}
  {\bibinfo  {journal} {J. Chem. Phys.}\ }\textbf {\bibinfo {volume} {139}},\
  \bibinfo {eid} {234307} (\bibinfo {year} {2013})}\BibitemShut {NoStop}%
\bibitem [{\citenamefont {Janssen}\ and\ \citenamefont
  {Powis}(2014)}]{Janssen2014}%
  \BibitemOpen
  \bibfield  {author} {\bibinfo {author} {\bibfnamefont {M.~H.~M.}\
  \bibnamefont {Janssen}}\ and\ \bibinfo {author} {\bibfnamefont
  {I.}~\bibnamefont {Powis}},\ }\href {\doibase 10.1039/C3CP53741B} {\bibfield
  {journal} {\bibinfo  {journal} {Phys. Chem. Chem. Phys.}\ }\textbf {\bibinfo
  {volume} {16}},\ \bibinfo {pages} {856} (\bibinfo {year} {2014})}\BibitemShut
  {NoStop}%
\bibitem [{\citenamefont {Lux}\ \emph {et~al.}(2015)\citenamefont {Lux},
  \citenamefont {Wollenhaupt}, \citenamefont {Sarpe},\ and\ \citenamefont
  {Baumert}}]{LuxCPC15}%
  \BibitemOpen
  \bibfield  {author} {\bibinfo {author} {\bibfnamefont {C.}~\bibnamefont
  {Lux}}, \bibinfo {author} {\bibfnamefont {M.}~\bibnamefont {Wollenhaupt}},
  \bibinfo {author} {\bibfnamefont {C.}~\bibnamefont {Sarpe}}, \ and\ \bibinfo
  {author} {\bibfnamefont {T.}~\bibnamefont {Baumert}},\ }\href {\doibase
  10.1002/cphc.201402643} {\bibfield  {journal} {\bibinfo  {journal}
  {ChemPhysChem}\ }\textbf {\bibinfo {volume} {16}},\ \bibinfo {pages} {115}
  (\bibinfo {year} {2015})}\BibitemShut {NoStop}%
\bibitem [{\citenamefont {Rafiee~Fanood}\ \emph {et~al.}(2015)\citenamefont
  {Rafiee~Fanood}, \citenamefont {Janssen},\ and\ \citenamefont
  {Powis}}]{FanoodPCCP15}%
  \BibitemOpen
  \bibfield  {author} {\bibinfo {author} {\bibfnamefont {M.~M.}\ \bibnamefont
  {Rafiee~Fanood}}, \bibinfo {author} {\bibfnamefont {M.~H.~M.}\ \bibnamefont
  {Janssen}}, \ and\ \bibinfo {author} {\bibfnamefont {I.}~\bibnamefont
  {Powis}},\ }\href {\doibase 10.1039/C5CP00583C} {\bibfield  {journal}
  {\bibinfo  {journal} {Phys. Chem. Chem. Phys.}\ }\textbf {\bibinfo {volume}
  {17}},\ \bibinfo {pages} {8614} (\bibinfo {year} {2015})}\BibitemShut
  {NoStop}%
\bibitem [{\citenamefont {Ritchie}(1976{\natexlab{a}})}]{Ritchie1976}%
  \BibitemOpen
  \bibfield  {author} {\bibinfo {author} {\bibfnamefont {B.}~\bibnamefont
  {Ritchie}},\ }\href {\doibase 10.1103/PhysRevA.13.1411} {\bibfield  {journal}
  {\bibinfo  {journal} {Phys. Rev. A}\ }\textbf {\bibinfo {volume} {13}},\
  \bibinfo {pages} {1411} (\bibinfo {year} {1976}{\natexlab{a}})}\BibitemShut
  {NoStop}%
\bibitem [{\citenamefont {Powis}(2008)}]{PowisAdvCP08}%
  \BibitemOpen
  \bibfield  {author} {\bibinfo {author} {\bibfnamefont {I.}~\bibnamefont
  {Powis}},\ }\enquote {\bibinfo {title} {Photoelectron circular dichroism in
  chiral molecules},}\ in\ \href {\doibase 10.1002/9780470259474.ch5} {\emph
  {\bibinfo {booktitle} {Advances in Chemical Physics}}}\ (\bibinfo
  {publisher} {John Wiley \& Sons, Inc.},\ \bibinfo {year} {2008})\ pp.\
  \bibinfo {pages} {267--329}\BibitemShut {NoStop}%
\bibitem [{\citenamefont {Nahon}\ and\ \citenamefont
  {Powis}(2010)}]{Nahon2010}%
  \BibitemOpen
  \bibfield  {author} {\bibinfo {author} {\bibfnamefont {L.}~\bibnamefont
  {Nahon}}\ and\ \bibinfo {author} {\bibfnamefont {I.}~\bibnamefont {Powis}},\
  }in\ \href@noop {} {\emph {\bibinfo {booktitle} {Chiral Recognition in the
  Gas Phase}}},\ \bibinfo {editor} {edited by\ \bibinfo {editor} {\bibfnamefont
  {A.}~\bibnamefont {Zehnacker}}}\ (\bibinfo  {publisher} {CRC Press},\
  \bibinfo {year} {2010})\BibitemShut {NoStop}%
\bibitem [{\citenamefont {Dixit}\ and\ \citenamefont
  {Lambropoulos}(1983)}]{DixitPRA1983}%
  \BibitemOpen
  \bibfield  {author} {\bibinfo {author} {\bibfnamefont {S.~N.}\ \bibnamefont
  {Dixit}}\ and\ \bibinfo {author} {\bibfnamefont {P.}~\bibnamefont
  {Lambropoulos}},\ }\href {\doibase 10.1103/PhysRevA.27.861} {\bibfield
  {journal} {\bibinfo  {journal} {Phys. Rev. A}\ }\textbf {\bibinfo {volume}
  {27}},\ \bibinfo {pages} {861} (\bibinfo {year} {1983})}\BibitemShut
  {NoStop}%
\bibitem [{\citenamefont {Monson}\ and\ \citenamefont
  {McClain}(1970{\natexlab{a}})}]{pmw1970}%
  \BibitemOpen
  \bibfield  {author} {\bibinfo {author} {\bibfnamefont {P.~R.}\ \bibnamefont
  {Monson}}\ and\ \bibinfo {author} {\bibfnamefont {W.~M.}\ \bibnamefont
  {McClain}},\ }\href {\doibase http://dx.doi.org/10.1063/1.1673778} {\bibfield
   {journal} {\bibinfo  {journal} {J. Chem. Phys.}\ }\textbf {\bibinfo {volume}
  {53}},\ \bibinfo {pages} {29} (\bibinfo {year}
  {1970}{\natexlab{a}})}\BibitemShut {NoStop}%
\bibitem [{\citenamefont {Monson}\ and\ \citenamefont
  {McClain}(1970{\natexlab{b}})}]{Monson1970}%
  \BibitemOpen
  \bibfield  {author} {\bibinfo {author} {\bibfnamefont {P.~R.}\ \bibnamefont
  {Monson}}\ and\ \bibinfo {author} {\bibfnamefont {W.~M.}\ \bibnamefont
  {McClain}},\ }\href {\doibase http://dx.doi.org/10.1063/1.1673778} {\bibfield
   {journal} {\bibinfo  {journal} {J. Chem. Phys.}\ }\textbf {\bibinfo {volume}
  {53}},\ \bibinfo {pages} {29} (\bibinfo {year}
  {1970}{\natexlab{b}})}\BibitemShut {NoStop}%
\bibitem [{\citenamefont {McClain}(1972)}]{McClain1972}%
  \BibitemOpen
  \bibfield  {author} {\bibinfo {author} {\bibfnamefont {W.~M.}\ \bibnamefont
  {McClain}},\ }\href {\doibase http://dx.doi.org/10.1063/1.1678579} {\bibfield
   {journal} {\bibinfo  {journal} {J. Chem. Phys.}\ }\textbf {\bibinfo {volume}
  {57}},\ \bibinfo {pages} {2264} (\bibinfo {year} {1972})}\BibitemShut
  {NoStop}%
\bibitem [{\citenamefont {McClain}(1974)}]{wm1974}%
  \BibitemOpen
  \bibfield  {author} {\bibinfo {author} {\bibfnamefont {W.~M.}\ \bibnamefont
  {McClain}},\ }\href {\doibase 10.1021/ar50077a001} {\bibfield  {journal}
  {\bibinfo  {journal} {Acc. Chem. Res}\ }\textbf {\bibinfo {volume} {7}},\
  \bibinfo {pages} {129} (\bibinfo {year} {1974})}\BibitemShut {NoStop}%
\bibitem [{\citenamefont {{I. Tinoco Jr.}}(1974)}]{Itjr1974}%
  \BibitemOpen
  \bibfield  {author} {\bibinfo {author} {\bibnamefont {{I. Tinoco Jr.}}},\
  }\href {\doibase http://dx.doi.org/10.1063/1.430566} {\bibfield  {journal}
  {\bibinfo  {journal} {J. Chem. Phys.}\ }\textbf {\bibinfo {volume} {62}},\
  \bibinfo {pages} {1006} (\bibinfo {year} {1974})}\BibitemShut {NoStop}%
\bibitem [{\citenamefont {Keldysh}(1965)}]{Keldysh1965}%
  \BibitemOpen
  \bibfield  {author} {\bibinfo {author} {\bibfnamefont {L.~V.}\ \bibnamefont
  {Keldysh}},\ }\href@noop {} {\bibfield  {journal} {\bibinfo  {journal} {Sov.
  Phys. JEPT}\ }\textbf {\bibinfo {volume} {20}},\ \bibinfo {pages} {1307}
  (\bibinfo {year} {1965})}\BibitemShut {NoStop}%
\bibitem [{\citenamefont {Faisal}(1973)}]{Faisal1973}%
  \BibitemOpen
  \bibfield  {author} {\bibinfo {author} {\bibfnamefont {F.~H.~M.}\
  \bibnamefont {Faisal}},\ }\href {http://stacks.iop.org/0022-3700/6/i=4/a=011}
  {\bibfield  {journal} {\bibinfo  {journal} {J. Phys. B}\ }\textbf {\bibinfo
  {volume} {6}},\ \bibinfo {pages} {L89} (\bibinfo {year} {1973})}\BibitemShut
  {NoStop}%
\bibitem [{\citenamefont {Dreissigacker}\ and\ \citenamefont
  {Lein}(2014)}]{LeinPRA2014}%
  \BibitemOpen
  \bibfield  {author} {\bibinfo {author} {\bibfnamefont {I.}~\bibnamefont
  {Dreissigacker}}\ and\ \bibinfo {author} {\bibfnamefont {M.}~\bibnamefont
  {Lein}},\ }\href {\doibase 10.1103/PhysRevA.89.053406} {\bibfield  {journal}
  {\bibinfo  {journal} {Phys. Rev. A}\ }\textbf {\bibinfo {volume} {89}},\
  \bibinfo {pages} {053406} (\bibinfo {year} {2014})}\BibitemShut {NoStop}%
\bibitem [{\citenamefont {Amitay}\ \emph {et~al.}(2008)\citenamefont {Amitay},
  \citenamefont {Gandman}, \citenamefont {Chuntonov},\ and\ \citenamefont
  {Rybak}}]{AmitayPRL08}%
  \BibitemOpen
  \bibfield  {author} {\bibinfo {author} {\bibfnamefont {Z.}~\bibnamefont
  {Amitay}}, \bibinfo {author} {\bibfnamefont {A.}~\bibnamefont {Gandman}},
  \bibinfo {author} {\bibfnamefont {L.}~\bibnamefont {Chuntonov}}, \ and\
  \bibinfo {author} {\bibfnamefont {L.}~\bibnamefont {Rybak}},\ }\href@noop {}
  {\bibfield  {journal} {\bibinfo  {journal} {Phys. Rev. Lett.}\ }\textbf
  {\bibinfo {volume} {100}},\ \bibinfo {pages} {193002} (\bibinfo {year}
  {2008})}\BibitemShut {NoStop}%
\bibitem [{\citenamefont {Rybak}\ \emph {et~al.}(2011)\citenamefont {Rybak},
  \citenamefont {Amaran}, \citenamefont {Levin}, \citenamefont {Tomza},
  \citenamefont {Moszynski}, \citenamefont {Kosloff}, \citenamefont {Koch},\
  and\ \citenamefont {Amitay}}]{RybakPRL11}%
  \BibitemOpen
  \bibfield  {author} {\bibinfo {author} {\bibfnamefont {L.}~\bibnamefont
  {Rybak}}, \bibinfo {author} {\bibfnamefont {S.}~\bibnamefont {Amaran}},
  \bibinfo {author} {\bibfnamefont {L.}~\bibnamefont {Levin}}, \bibinfo
  {author} {\bibfnamefont {M.}~\bibnamefont {Tomza}}, \bibinfo {author}
  {\bibfnamefont {R.}~\bibnamefont {Moszynski}}, \bibinfo {author}
  {\bibfnamefont {R.}~\bibnamefont {Kosloff}}, \bibinfo {author} {\bibfnamefont
  {C.~P.}\ \bibnamefont {Koch}}, \ and\ \bibinfo {author} {\bibfnamefont
  {Z.}~\bibnamefont {Amitay}},\ }\href@noop {} {\bibfield  {journal} {\bibinfo
  {journal} {Phys. Rev. Lett.}\ }\textbf {\bibinfo {volume} {107}},\ \bibinfo
  {pages} {273001} (\bibinfo {year} {2011})}\BibitemShut {NoStop}%
\bibitem [{\citenamefont {Levin}\ \emph {et~al.}(2015)\citenamefont {Levin},
  \citenamefont {Skomorowski}, \citenamefont {Rybak}, \citenamefont {Kosloff},
  \citenamefont {Koch},\ and\ \citenamefont {Amitay}}]{LevinPRL15}%
  \BibitemOpen
  \bibfield  {author} {\bibinfo {author} {\bibfnamefont {L.}~\bibnamefont
  {Levin}}, \bibinfo {author} {\bibfnamefont {W.}~\bibnamefont {Skomorowski}},
  \bibinfo {author} {\bibfnamefont {L.}~\bibnamefont {Rybak}}, \bibinfo
  {author} {\bibfnamefont {R.}~\bibnamefont {Kosloff}}, \bibinfo {author}
  {\bibfnamefont {C.~P.}\ \bibnamefont {Koch}}, \ and\ \bibinfo {author}
  {\bibfnamefont {Z.}~\bibnamefont {Amitay}},\ }\href {\doibase
  10.1103/PhysRevLett.114.233003} {\bibfield  {journal} {\bibinfo  {journal}
  {Phys. Rev. Lett.}\ }\textbf {\bibinfo {volume} {114}},\ \bibinfo {pages}
  {233003} (\bibinfo {year} {2015})}\BibitemShut {NoStop}%
\bibitem [{\citenamefont {Ritchie}(1976{\natexlab{b}})}]{RitchiePRA1976}%
  \BibitemOpen
  \bibfield  {author} {\bibinfo {author} {\bibfnamefont {B.}~\bibnamefont
  {Ritchie}},\ }\href {\doibase 10.1103/PhysRevA.13.1411} {\bibfield  {journal}
  {\bibinfo  {journal} {Phys. Rev. A}\ }\textbf {\bibinfo {volume} {13}},\
  \bibinfo {pages} {1411} (\bibinfo {year} {1976}{\natexlab{b}})}\BibitemShut
  {NoStop}%
\bibitem [{\citenamefont {Reid.}(2003)}]{Reid2003}%
  \BibitemOpen
  \bibfield  {author} {\bibinfo {author} {\bibfnamefont {K.~L.}\ \bibnamefont
  {Reid.}},\ }\href {\doibase 10.1146/annurev.physchem.54.011002.103814}
  {\bibfield  {journal} {\bibinfo  {journal} {Annu. Rev. Phys. Chem}\ }\textbf
  {\bibinfo {volume} {54}},\ \bibinfo {pages} {397} (\bibinfo {year}
  {2003})}\BibitemShut {NoStop}%
\bibitem [{\citenamefont {Cooper}\ and\ \citenamefont {Zare}(1968)}]{cooper}%
  \BibitemOpen
  \bibfield  {author} {\bibinfo {author} {\bibfnamefont {J.}~\bibnamefont
  {Cooper}}\ and\ \bibinfo {author} {\bibfnamefont {R.~N.}\ \bibnamefont
  {Zare}},\ }in\ \href@noop {} {\emph {\bibinfo {booktitle} {Lectures In
  Theoretical Physics}}},\ Vol.~\bibinfo {volume} {9},\ \bibinfo {editor}
  {edited by\ \bibinfo {editor} {\bibnamefont {Gordon}}\ and\ \bibinfo {editor}
  {\bibnamefont {Breach}}}\ (\bibinfo  {publisher} {University of Colorado},\
  \bibinfo {address} {New York},\ \bibinfo {year} {1968})\ pp.\ \bibinfo
  {pages} {317--337}\BibitemShut {NoStop}%
\bibitem [{\citenamefont {Chandra}(1987)}]{ChandraJPhysB87}%
  \BibitemOpen
  \bibfield  {author} {\bibinfo {author} {\bibfnamefont {N.}~\bibnamefont
  {Chandra}},\ }\href@noop {} {\bibfield  {journal} {\bibinfo  {journal} {J.
  Phys. B}\ }\textbf {\bibinfo {volume} {20}},\ \bibinfo {pages} {3405}
  (\bibinfo {year} {1987})}\BibitemShut {NoStop}%
\bibitem [{\citenamefont {Bethe}\ and\ \citenamefont
  {Salpeter}(1957)}]{Hans1957}%
  \BibitemOpen
  \bibfield  {author} {\bibinfo {author} {\bibfnamefont {H.~A.}\ \bibnamefont
  {Bethe}}\ and\ \bibinfo {author} {\bibfnamefont {E.~E.}\ \bibnamefont
  {Salpeter}},\ }\href@noop {} {\emph {\bibinfo {title} {Quantum Mechanics of
  One and Two-Electron Atoms}}},\ \bibinfo {edition} {1st}\ ed.\ (\bibinfo
  {publisher} {Academic Press Inc.},\ \bibinfo {address} {111, Fifth Avenue,
  New York 3, New York/USA},\ \bibinfo {year} {1957})\BibitemShut {NoStop}%
\bibitem [{\citenamefont {Jin}\ \emph {et~al.}(2010{\natexlab{a}})\citenamefont
  {Jin}, \citenamefont {Le}, \citenamefont {Zhao}, \citenamefont {Lucchese},\
  and\ \citenamefont {Lin}}]{ChengPRA2010}%
  \BibitemOpen
  \bibfield  {author} {\bibinfo {author} {\bibfnamefont {C.}~\bibnamefont
  {Jin}}, \bibinfo {author} {\bibfnamefont {A.-T.}\ \bibnamefont {Le}},
  \bibinfo {author} {\bibfnamefont {S.-F.}\ \bibnamefont {Zhao}}, \bibinfo
  {author} {\bibfnamefont {R.~R.}\ \bibnamefont {Lucchese}}, \ and\ \bibinfo
  {author} {\bibfnamefont {C.~D.}\ \bibnamefont {Lin}},\ }\href {\doibase
  10.1103/PhysRevA.81.033421} {\bibfield  {journal} {\bibinfo  {journal} {Phys.
  Rev. A}\ }\textbf {\bibinfo {volume} {81}},\ \bibinfo {pages} {033421}
  (\bibinfo {year} {2010}{\natexlab{a}})}\BibitemShut {NoStop}%
\bibitem [{\citenamefont {Lucchese}\ \emph {et~al.}(1982)\citenamefont
  {Lucchese}, \citenamefont {Raseev},\ and\ \citenamefont
  {McKoy}}]{LuchessePRA1982}%
  \BibitemOpen
  \bibfield  {author} {\bibinfo {author} {\bibfnamefont {R.~R.}\ \bibnamefont
  {Lucchese}}, \bibinfo {author} {\bibfnamefont {G.}~\bibnamefont {Raseev}}, \
  and\ \bibinfo {author} {\bibfnamefont {V.}~\bibnamefont {McKoy}},\ }\href
  {\doibase 10.1103/PhysRevA.25.2572} {\bibfield  {journal} {\bibinfo
  {journal} {Phys. Rev. A}\ }\textbf {\bibinfo {volume} {25}},\ \bibinfo
  {pages} {2572} (\bibinfo {year} {1982})}\BibitemShut {NoStop}%
\bibitem [{\citenamefont {Bishop}(1967)}]{Bishop67}%
  \BibitemOpen
  \bibfield  {author} {\bibinfo {author} {\bibfnamefont {B.~M.}\ \bibnamefont
  {Bishop}},\ }\href@noop {} {\emph {\bibinfo {title} {Advances in Quantum
  Chemistry}}},\ \bibinfo {edition} {1st}\ ed.,\ Vol.~\bibinfo {volume} {3}\
  (\bibinfo  {publisher} {Academic Press Inc.},\ \bibinfo {address} {Berkeley
  Square House, London W.1},\ \bibinfo {year} {1967})\BibitemShut {NoStop}%
\bibitem [{\citenamefont {Jin}\ \emph {et~al.}(2010{\natexlab{b}})\citenamefont
  {Jin}, \citenamefont {Le}, \citenamefont {Zhao}, \citenamefont {Lucchese},\
  and\ \citenamefont {Lin}}]{ChengPRA10}%
  \BibitemOpen
  \bibfield  {author} {\bibinfo {author} {\bibfnamefont {C.}~\bibnamefont
  {Jin}}, \bibinfo {author} {\bibfnamefont {A.-T.}\ \bibnamefont {Le}},
  \bibinfo {author} {\bibfnamefont {S.-F.}\ \bibnamefont {Zhao}}, \bibinfo
  {author} {\bibfnamefont {R.~R.}\ \bibnamefont {Lucchese}}, \ and\ \bibinfo
  {author} {\bibfnamefont {C.~D.}\ \bibnamefont {Lin}},\ }\href {\doibase
  10.1103/PhysRevA.81.033421} {\bibfield  {journal} {\bibinfo  {journal} {Phys.
  Rev. A}\ }\textbf {\bibinfo {volume} {81}},\ \bibinfo {pages} {033421}
  (\bibinfo {year} {2010}{\natexlab{b}})}\BibitemShut {NoStop}%
\bibitem [{\citenamefont {Dill}(1976)}]{DillChemphys87}%
  \BibitemOpen
  \bibfield  {author} {\bibinfo {author} {\bibfnamefont {D.}~\bibnamefont
  {Dill}},\ }\href@noop {} {\bibfield  {journal} {\bibinfo  {journal} {J. Chem.
  Phys.}\ }\textbf {\bibinfo {volume} {65}},\ \bibinfo {pages} {1130} (\bibinfo
  {year} {1976})}\BibitemShut {NoStop}%
\bibitem [{\citenamefont {Oana}\ and\ \citenamefont
  {Krylov}(2009)}]{OanaJCP09}%
  \BibitemOpen
  \bibfield  {author} {\bibinfo {author} {\bibfnamefont {C.~M.}\ \bibnamefont
  {Oana}}\ and\ \bibinfo {author} {\bibfnamefont {A.~I.}\ \bibnamefont
  {Krylov}},\ }\href {\doibase http://dx.doi.org/10.1063/1.3231143} {\bibfield
  {journal} {\bibinfo  {journal} {J. Chem. Phys.}\ }\textbf {\bibinfo {volume}
  {131}},\ \bibinfo {pages} {124114} (\bibinfo {year} {2009})}\BibitemShut
  {NoStop}%
\bibitem [{\citenamefont {Edmonds}(1996)}]{edmonds}%
  \BibitemOpen
  \bibfield  {author} {\bibinfo {author} {\bibfnamefont {A.}~\bibnamefont
  {Edmonds}},\ }\href@noop {} {\emph {\bibinfo {title} {Angular Momentum in
  Quantum Mechanics}}},\ \bibinfo {edition} {4th}\ ed.\ (\bibinfo  {publisher}
  {Princeton University Press},\ \bibinfo {address} {Princeton, New Jersey},\
  \bibinfo {year} {1996})\BibitemShut {NoStop}%
\bibitem [{\citenamefont {Silver}(1976)}]{irreducible}%
  \BibitemOpen
  \bibfield  {author} {\bibinfo {author} {\bibfnamefont {B.~L.}\ \bibnamefont
  {Silver}},\ }\href@noop {} {\emph {\bibinfo {title} {Irreducible Tensor
  Methods: An Introduction or Chemists}}},\ \bibinfo {edition} {1st}\ ed.,\
  Vol.~\bibinfo {volume} {36}\ (\bibinfo  {publisher} {Academic Press,
  Inc.(London) LTD},\ \bibinfo {address} {24/28 Oval Road, London NW1},\
  \bibinfo {year} {1976})\ \bibinfo {note} {an optional note}\BibitemShut
  {NoStop}%
\bibitem [{\citenamefont {Rose}(1967)}]{Rose1967}%
  \BibitemOpen
  \bibfield  {author} {\bibinfo {author} {\bibfnamefont {M.}~\bibnamefont
  {Rose}},\ }\href@noop {} {\emph {\bibinfo {title} {Elementary Theory of
  Angular Momentum}}},\ \bibinfo {edition} {5th}\ ed.\ (\bibinfo  {publisher}
  {John Wiley \& Sons, Inc.},\ \bibinfo {address} {New York},\ \bibinfo {year}
  {1967})\BibitemShut {NoStop}%
\bibitem [{\citenamefont {Varshalovich}\ \emph {et~al.}(1988)\citenamefont
  {Varshalovich}, \citenamefont {Moskalev},\ and\ \citenamefont
  {Khersonskii}}]{Varshalovich1988}%
  \BibitemOpen
  \bibfield  {author} {\bibinfo {author} {\bibfnamefont {D.}~\bibnamefont
  {Varshalovich}}, \bibinfo {author} {\bibfnamefont {A.}~\bibnamefont
  {Moskalev}}, \ and\ \bibinfo {author} {\bibfnamefont {V.}~\bibnamefont
  {Khersonskii}},\ }\href@noop {} {\emph {\bibinfo {title} {Quantum Theory of
  Angular Momentum: Irreducible Tensors, Spherical Harmonics, Vector Coupling
  Coefficients, 3nj Symbols}}},\ \bibinfo {edition} {1st}\ ed.\ (\bibinfo
  {publisher} {Word Scientific Co. Pte. Ltd.},\ \bibinfo {address} {687
  Hartwell Street, Teaneck, NJ 07666},\ \bibinfo {year} {1988})\BibitemShut
  {NoStop}%
\bibitem [{\citenamefont {Peticolas}(1967)}]{Peticolas1967}%
  \BibitemOpen
  \bibfield  {author} {\bibinfo {author} {\bibfnamefont {W.~L.}\ \bibnamefont
  {Peticolas}},\ }\href {\doibase 10.1146/annurev.pc.18.100167.001313}
  {\bibfield  {journal} {\bibinfo  {journal} {Annu. Rev. Phys. Chem.}\ }\textbf
  {\bibinfo {volume} {18}},\ \bibinfo {pages} {233} (\bibinfo {year}
  {1967})}\BibitemShut {NoStop}%
\bibitem [{\citenamefont {McCLAIN}\ and\ \citenamefont
  {HARRIS}(1977)}]{McCLAIN19771}%
  \BibitemOpen
  \bibfield  {author} {\bibinfo {author} {\bibfnamefont {W.~M.}\ \bibnamefont
  {McCLAIN}}\ and\ \bibinfo {author} {\bibfnamefont {R.~A.}\ \bibnamefont
  {HARRIS}},\ }in\ \href {\doibase
  http://dx.doi.org/10.1016/B978-0-12-227203-5.50006-8} {\emph {\bibinfo
  {booktitle} {Excited States}}},\ \bibinfo {editor} {edited by\ \bibinfo
  {editor} {\bibfnamefont {E.~C.}\ \bibnamefont {LIM}}}\ (\bibinfo  {publisher}
  {Academic Press},\ \bibinfo {year} {1977})\ pp.\ \bibinfo {pages}
  {1--56}\BibitemShut {NoStop}%
\bibitem [{\citenamefont {Nascimento}(1983)}]{Marco1983}%
  \BibitemOpen
  \bibfield  {author} {\bibinfo {author} {\bibfnamefont {M.~A.~C.}\
  \bibnamefont {Nascimento}},\ }\href@noop {} {\bibfield  {journal} {\bibinfo
  {journal} {Chem. Phys.}\ }\textbf {\bibinfo {volume} {74}},\ \bibinfo {pages}
  {51} (\bibinfo {year} {1983})}\BibitemShut {NoStop}%
\bibitem [{\citenamefont {Yang}(1948)}]{Yang1948}%
  \BibitemOpen
  \bibfield  {author} {\bibinfo {author} {\bibfnamefont {C.~N.}\ \bibnamefont
  {Yang}},\ }\href {\doibase 10.1103/PhysRev.74.764} {\bibfield  {journal}
  {\bibinfo  {journal} {Phys. Rev.}\ }\textbf {\bibinfo {volume} {74}},\
  \bibinfo {pages} {764} (\bibinfo {year} {1948})}\BibitemShut {NoStop}%
\bibitem [{\citenamefont {Pollmann}\ \emph {et~al.}(1997)\citenamefont
  {Pollmann}, \citenamefont {Franke},\ and\ \citenamefont
  {Hormes}}]{PollmannSpectro97}%
  \BibitemOpen
  \bibfield  {author} {\bibinfo {author} {\bibfnamefont {J.}~\bibnamefont
  {Pollmann}}, \bibinfo {author} {\bibfnamefont {R.}~\bibnamefont {Franke}}, \
  and\ \bibinfo {author} {\bibfnamefont {J.}~\bibnamefont {Hormes}},\ }\href
  {\doibase http://dx.doi.org/10.1016/S1386-1425(96)01810-0} {\bibfield
  {journal} {\bibinfo  {journal} {Spectrochimica Acta Part A: Molecular and
  Biomolecular Spectroscopy}\ }\textbf {\bibinfo {volume} {53}},\ \bibinfo
  {pages} {491 } (\bibinfo {year} {1997})}\BibitemShut {NoStop}%
\bibitem [{\citenamefont {{University of Karlsruhe and Forschungszentrum
  Karlsruhe GmbH}}(2014)}]{turbomole6}%
  \BibitemOpen
  \bibfield  {author} {\bibinfo {author} {\bibnamefont {{University of
  Karlsruhe and Forschungszentrum Karlsruhe GmbH}}},\ }\href@noop {} {\enquote
  {\bibinfo {title} {{\textsc{TURBOMOLE 6.6 2014:}} program package for ab
  initio electronic structure calculations},}\ } (\bibinfo {year}
  {2014})\BibitemShut {NoStop}%
\bibitem [{Sup()}]{SuppMat}%
  \BibitemOpen
  \href@noop {} {}\bibinfo {note} {See supplemental material at [URL will be
  inserted by editor] for the expansion coefficients of the intermediate state
  wavefunctions obtained in the single center reexpansion and the Cartesian
  coordinates obtained in the geometry optimization}\BibitemShut {NoStop}%
\bibitem [{\citenamefont {Angeli}(2015)}]{dalton:2005}%
  \BibitemOpen
  \bibfield  {author} {\bibinfo {author} {\bibfnamefont {C.}~\bibnamefont
  {Angeli}},\ }\href@noop {} {\enquote {\bibinfo {title} {{\textsc{Dalton:}} a
  molecular electronic structure program, release 2.0},}\ } (\bibinfo {year}
  {2015})\BibitemShut {NoStop}%
\bibitem [{\citenamefont {Paterson}\ \emph {et~al.}(2006)\citenamefont
  {Paterson}, \citenamefont {Christiansen}, \citenamefont {Paw{\l}owski},
  \citenamefont {J{\o}rgensen}, \citenamefont {H{\"a}ttig}, \citenamefont
  {Helgaker},\ and\ \citenamefont {Sa{\l}ek}}]{paterson:2006}%
  \BibitemOpen
  \bibfield  {author} {\bibinfo {author} {\bibfnamefont {M.~J.}\ \bibnamefont
  {Paterson}}, \bibinfo {author} {\bibfnamefont {O.}~\bibnamefont
  {Christiansen}}, \bibinfo {author} {\bibfnamefont {F.}~\bibnamefont
  {Paw{\l}owski}}, \bibinfo {author} {\bibfnamefont {P.}~\bibnamefont
  {J{\o}rgensen}}, \bibinfo {author} {\bibfnamefont {C.}~\bibnamefont
  {H{\"a}ttig}}, \bibinfo {author} {\bibfnamefont {T.}~\bibnamefont
  {Helgaker}}, \ and\ \bibinfo {author} {\bibfnamefont {P.}~\bibnamefont
  {Sa{\l}ek}},\ }\href {\doibase http://dx.doi.org/10.1063/1.2163874}
  {\bibfield  {journal} {\bibinfo  {journal} {J. Chem. Phys.}\ }\textbf
  {\bibinfo {volume} {124}},\ \bibinfo {eid} {054322} (\bibinfo {year}
  {2006})}\BibitemShut {NoStop}%
\bibitem [{\citenamefont {H\"attig}\ and\ \citenamefont
  {J{\o}rgensen}(1998)}]{hattig:1998}%
  \BibitemOpen
  \bibfield  {author} {\bibinfo {author} {\bibfnamefont {C.}~\bibnamefont
  {H\"attig}}\ and\ \bibinfo {author} {\bibfnamefont {P.}~\bibnamefont
  {J{\o}rgensen}},\ }\href {\doibase http://dx.doi.org/10.1063/1.477581}
  {\bibfield  {journal} {\bibinfo  {journal} {J. Chem. Phys.}\ }\textbf
  {\bibinfo {volume} {109}},\ \bibinfo {pages} {9219} (\bibinfo {year}
  {1998})}\BibitemShut {NoStop}%
\bibitem [{\citenamefont {Werner}\ \emph {et~al.}(2012)\citenamefont {Werner},
  \citenamefont {Knowles}, \citenamefont {Knizia}, \citenamefont {Manby},
  \citenamefont {{Sch\"{u}tz}} \emph {et~al.}}]{molpro}%
  \BibitemOpen
  \bibfield  {author} {\bibinfo {author} {\bibfnamefont {H.-J.}\ \bibnamefont
  {Werner}}, \bibinfo {author} {\bibfnamefont {P.~J.}\ \bibnamefont {Knowles}},
  \bibinfo {author} {\bibfnamefont {G.}~\bibnamefont {Knizia}}, \bibinfo
  {author} {\bibfnamefont {F.~R.}\ \bibnamefont {Manby}}, \bibinfo {author}
  {\bibfnamefont {M.}~\bibnamefont {{Sch\"{u}tz}}},  \emph {et~al.},\
  }\href@noop {} {\enquote {\bibinfo {title} {Molpro, version 2012.1, a package
  of ab initio programs},}\ } (\bibinfo {year} {2012})\BibitemShut {NoStop}%
\bibitem [{\citenamefont {Pulm}\ \emph {et~al.}(1997)\citenamefont {Pulm},
  \citenamefont {Schramm}, \citenamefont {Hormes}, \citenamefont {Grimme},\
  and\ \citenamefont {Peyerimhoff}}]{pulm:1997}%
  \BibitemOpen
  \bibfield  {author} {\bibinfo {author} {\bibfnamefont {F.}~\bibnamefont
  {Pulm}}, \bibinfo {author} {\bibfnamefont {J.}~\bibnamefont {Schramm}},
  \bibinfo {author} {\bibfnamefont {J.}~\bibnamefont {Hormes}}, \bibinfo
  {author} {\bibfnamefont {S.}~\bibnamefont {Grimme}}, \ and\ \bibinfo {author}
  {\bibfnamefont {S.~D.}\ \bibnamefont {Peyerimhoff}},\ }\href {\doibase
  http://dx.doi.org/10.1016/S0301-0104(97)00258-9} {\bibfield  {journal}
  {\bibinfo  {journal} {Chem. Phys.}\ }\textbf {\bibinfo {volume} {224}},\
  \bibinfo {pages} {143 } (\bibinfo {year} {1997})}\BibitemShut {NoStop}%
\bibitem [{\citenamefont {Woon}\ and\ \citenamefont {Dunning}(1994)}]{DEW1994}%
  \BibitemOpen
  \bibfield  {author} {\bibinfo {author} {\bibfnamefont {D.~E.}\ \bibnamefont
  {Woon}}\ and\ \bibinfo {author} {\bibfnamefont {T.~H.}\ \bibnamefont
  {Dunning}},\ }\href {\doibase http://dx.doi.org/10.1063/1.466439} {\bibfield
  {journal} {\bibinfo  {journal} {J. Chem. Phys.}\ }\textbf {\bibinfo {volume}
  {100}},\ \bibinfo {pages} {2975} (\bibinfo {year} {1994})}\BibitemShut
  {NoStop}%
\bibitem [{\citenamefont {Casida}\ \emph {et~al.}(1998)\citenamefont {Casida},
  \citenamefont {Jamorski}, \citenamefont {Casida},\ and\ \citenamefont
  {Salahub}}]{Cme1998}%
  \BibitemOpen
  \bibfield  {author} {\bibinfo {author} {\bibfnamefont {M.~E.}\ \bibnamefont
  {Casida}}, \bibinfo {author} {\bibfnamefont {C.}~\bibnamefont {Jamorski}},
  \bibinfo {author} {\bibfnamefont {K.~C.}\ \bibnamefont {Casida}}, \ and\
  \bibinfo {author} {\bibfnamefont {D.~R.}\ \bibnamefont {Salahub}},\
  }\href@noop {} {\bibfield  {journal} {\bibinfo  {journal} {J. Chem. Phys.}\
  }\textbf {\bibinfo {volume} {108}},\ \bibinfo {pages} {4439} (\bibinfo {year}
  {1998})}\BibitemShut {NoStop}%
\bibitem [{\citenamefont {Falden}\ \emph {et~al.}(2009)\citenamefont {Falden},
  \citenamefont {Falster-Hansen}, \citenamefont {Bak}, \citenamefont
  {Rettrup},\ and\ \citenamefont {Sauer}}]{heid2009}%
  \BibitemOpen
  \bibfield  {author} {\bibinfo {author} {\bibfnamefont {H.~H.}\ \bibnamefont
  {Falden}}, \bibinfo {author} {\bibfnamefont {K.~R.}\ \bibnamefont
  {Falster-Hansen}}, \bibinfo {author} {\bibfnamefont {K.~L.}\ \bibnamefont
  {Bak}}, \bibinfo {author} {\bibfnamefont {S.}~\bibnamefont {Rettrup}}, \ and\
  \bibinfo {author} {\bibfnamefont {S.~P.~A.}\ \bibnamefont {Sauer}},\
  }\href@noop {} {\bibfield  {journal} {\bibinfo  {journal} {J. Phys. Chem. A}\
  }\textbf {\bibinfo {volume} {113}},\ \bibinfo {pages} {11995} (\bibinfo
  {year} {2009})}\BibitemShut {NoStop}%
\bibitem [{\citenamefont {Diedrich}\ and\ \citenamefont
  {Grimme}(2003)}]{Diedrich:2003}%
  \BibitemOpen
  \bibfield  {author} {\bibinfo {author} {\bibfnamefont {C.}~\bibnamefont
  {Diedrich}}\ and\ \bibinfo {author} {\bibfnamefont {S.}~\bibnamefont
  {Grimme}},\ }\href {\doibase 10.1021/jp0275802} {\bibfield  {journal}
  {\bibinfo  {journal} {J. Phys. Chem. A}\ }\textbf {\bibinfo {volume} {107}},\
  \bibinfo {pages} {2524} (\bibinfo {year} {2003})}\BibitemShut {NoStop}%
\bibitem [{\citenamefont {MATLAB}(2014)}]{simulinkR2014a}%
  \BibitemOpen
  \bibfield  {author} {\bibinfo {author} {\bibnamefont {MATLAB}},\ }\href@noop
  {} {\emph {\bibinfo {title} {version 7.10.0 (R2014a)}}}\ (\bibinfo
  {publisher} {The MathWorks Inc.},\ \bibinfo {address} {Natick,
  Massachusetts},\ \bibinfo {year} {2014})\BibitemShut {NoStop}%
\bibitem [{\citenamefont {Harding}\ \emph {et~al.}(2005)\citenamefont
  {Harding}, \citenamefont {Mikajlo}, \citenamefont {Powis}, \citenamefont
  {Barth}, \citenamefont {Joshi}, \citenamefont {Ulrich},\ and\ \citenamefont
  {Hergenhahn}}]{HardingChemPhys2005}%
  \BibitemOpen
  \bibfield  {author} {\bibinfo {author} {\bibfnamefont {C.~J.}\ \bibnamefont
  {Harding}}, \bibinfo {author} {\bibfnamefont {E.}~\bibnamefont {Mikajlo}},
  \bibinfo {author} {\bibfnamefont {I.}~\bibnamefont {Powis}}, \bibinfo
  {author} {\bibfnamefont {S.}~\bibnamefont {Barth}}, \bibinfo {author}
  {\bibfnamefont {S.}~\bibnamefont {Joshi}}, \bibinfo {author} {\bibfnamefont
  {V.}~\bibnamefont {Ulrich}}, \ and\ \bibinfo {author} {\bibfnamefont
  {U.}~\bibnamefont {Hergenhahn}},\ }\href {\doibase
  http://dx.doi.org/10.1063/1.2136150} {\bibfield  {journal} {\bibinfo
  {journal} {J. Chem. Phys.}\ }\textbf {\bibinfo {volume} {123}},\ \bibinfo
  {eid} {234310} (\bibinfo {year} {2005})}\BibitemShut {NoStop}%
\bibitem [{\citenamefont {Seideman}(2001)}]{SeidemanPRA2001}%
  \BibitemOpen
  \bibfield  {author} {\bibinfo {author} {\bibfnamefont {T.}~\bibnamefont
  {Seideman}},\ }\href {\doibase 10.1103/PhysRevA.64.042504} {\bibfield
  {journal} {\bibinfo  {journal} {Phys. Rev. A}\ }\textbf {\bibinfo {volume}
  {64}},\ \bibinfo {pages} {042504} (\bibinfo {year} {2001})}\BibitemShut
  {NoStop}%
\bibitem [{\citenamefont {Oana}\ and\ \citenamefont
  {Krylov}(2007)}]{OanaJCP07}%
  \BibitemOpen
  \bibfield  {author} {\bibinfo {author} {\bibfnamefont {C.~M.}\ \bibnamefont
  {Oana}}\ and\ \bibinfo {author} {\bibfnamefont {A.~I.}\ \bibnamefont
  {Krylov}},\ }\href {\doibase http://dx.doi.org/10.1063/1.2805393} {\bibfield
  {journal} {\bibinfo  {journal} {J. Chem. Phys.}\ }\textbf {\bibinfo {volume}
  {127}},\ \bibinfo {pages} {234106} (\bibinfo {year} {2007})}\BibitemShut
  {NoStop}%
\bibitem [{\citenamefont {Humeniuk}\ \emph {et~al.}(2013)\citenamefont
  {Humeniuk}, \citenamefont {Wohlgemuth}, \citenamefont {Suzuki},\ and\
  \citenamefont {Mitri\'c}}]{HumeniukJCP13}%
  \BibitemOpen
  \bibfield  {author} {\bibinfo {author} {\bibfnamefont {A.}~\bibnamefont
  {Humeniuk}}, \bibinfo {author} {\bibfnamefont {M.}~\bibnamefont
  {Wohlgemuth}}, \bibinfo {author} {\bibfnamefont {T.}~\bibnamefont {Suzuki}},
  \ and\ \bibinfo {author} {\bibfnamefont {R.}~\bibnamefont {Mitri\'c}},\
  }\href {\doibase http://dx.doi.org/10.1063/1.4820238} {\bibfield  {journal}
  {\bibinfo  {journal} {J. Chem. Phys.}\ }\textbf {\bibinfo {volume} {139}},\
  \bibinfo {pages} {134104} (\bibinfo {year} {2013})}\BibitemShut {NoStop}%
\bibitem [{\citenamefont {Boesl~von Grafenstein}\ and\ \citenamefont
  {Bornschlegl}(2006)}]{BoeslCPC06}%
  \BibitemOpen
  \bibfield  {author} {\bibinfo {author} {\bibfnamefont {U.}~\bibnamefont
  {Boesl~von Grafenstein}}\ and\ \bibinfo {author} {\bibfnamefont
  {A.}~\bibnamefont {Bornschlegl}},\ }\href {\doibase 10.1002/cphc.200600376}
  {\bibfield  {journal} {\bibinfo  {journal} {ChemPhysChem}\ }\textbf {\bibinfo
  {volume} {7}},\ \bibinfo {pages} {2085} (\bibinfo {year} {2006})}\BibitemShut
  {NoStop}%
\bibitem [{\citenamefont {Li}\ \emph {et~al.}(2006)\citenamefont {Li},
  \citenamefont {Sullivan}, \citenamefont {Al-Basheer}, \citenamefont {Pagni},\
  and\ \citenamefont {Compton}}]{LiJCP06}%
  \BibitemOpen
  \bibfield  {author} {\bibinfo {author} {\bibfnamefont {R.}~\bibnamefont
  {Li}}, \bibinfo {author} {\bibfnamefont {R.}~\bibnamefont {Sullivan}},
  \bibinfo {author} {\bibfnamefont {W.}~\bibnamefont {Al-Basheer}}, \bibinfo
  {author} {\bibfnamefont {R.~M.}\ \bibnamefont {Pagni}}, \ and\ \bibinfo
  {author} {\bibfnamefont {R.~N.}\ \bibnamefont {Compton}},\ }\href@noop {}
  {\bibfield  {journal} {\bibinfo  {journal} {J. Chem. Phys.}\ }\textbf
  {\bibinfo {volume} {125}},\ \bibinfo {eid} {144304} (\bibinfo {year}
  {2006})}\BibitemShut {NoStop}%
\bibitem [{\citenamefont {Breunig}\ \emph {et~al.}(2009)\citenamefont
  {Breunig}, \citenamefont {Urbasch}, \citenamefont {Horsch}, \citenamefont
  {Cordes}, \citenamefont {Koert},\ and\ \citenamefont
  {Weitzel}}]{BreunigCPC09}%
  \BibitemOpen
  \bibfield  {author} {\bibinfo {author} {\bibfnamefont {H.~G.}\ \bibnamefont
  {Breunig}}, \bibinfo {author} {\bibfnamefont {G.}~\bibnamefont {Urbasch}},
  \bibinfo {author} {\bibfnamefont {P.}~\bibnamefont {Horsch}}, \bibinfo
  {author} {\bibfnamefont {J.}~\bibnamefont {Cordes}}, \bibinfo {author}
  {\bibfnamefont {U.}~\bibnamefont {Koert}}, \ and\ \bibinfo {author}
  {\bibfnamefont {K.-M.}\ \bibnamefont {Weitzel}},\ }\href {\doibase
  10.1002/cphc.200900103} {\bibfield  {journal} {\bibinfo  {journal}
  {ChemPhysChem}\ }\textbf {\bibinfo {volume} {10}},\ \bibinfo {pages} {1199}
  (\bibinfo {year} {2009})}\BibitemShut {NoStop}%
\bibitem [{\citenamefont {Kr\"oner}(2015)}]{KroenerPCCP15}%
  \BibitemOpen
  \bibfield  {author} {\bibinfo {author} {\bibfnamefont {D.}~\bibnamefont
  {Kr\"oner}},\ }\href {\doibase 10.1039/C5CP02193F} {\bibfield  {journal}
  {\bibinfo  {journal} {Phys. Chem. Chem. Phys.}\ }\textbf {\bibinfo {volume}
  {17}},\ \bibinfo {pages} {19643} (\bibinfo {year} {2015})}\BibitemShut
  {NoStop}%
\bibitem [{\citenamefont {Aslangul}(2008)}]{Aslangul08}%
  \BibitemOpen
  \bibfield  {author} {\bibinfo {author} {\bibfnamefont {C.}~\bibnamefont
  {Aslangul}},\ }\href@noop {} {\emph {\bibinfo {title} {M\'ecanique Quantique
  2, D\'eveloppements et applications \`a basse \'energie}}},\ \bibinfo
  {edition} {1st}\ ed.\ (\bibinfo  {publisher} {De Boeck S.A},\ \bibinfo
  {address} {Rue des Minimes 39, B-1000, Bruxelles},\ \bibinfo {year}
  {2008})\BibitemShut {NoStop}%
\bibitem [{\citenamefont {Abramowitz}\ and\ \citenamefont
  {Stegun}(1972)}]{Abramowitz1972}%
  \BibitemOpen
  \bibfield  {author} {\bibinfo {author} {\bibfnamefont {M.}~\bibnamefont
  {Abramowitz}}\ and\ \bibinfo {author} {\bibfnamefont {I.~A.}\ \bibnamefont
  {Stegun}},\ }\href@noop {} {\emph {\bibinfo {title} {Handbook of Mathematical
  Functions with Formulas, Graphs and Mathematical Tables}}},\ \bibinfo
  {edition} {10th}\ ed.\ (\bibinfo  {publisher} {National Bureau of Standards,
  Applied Mathematics Series 55},\ \bibinfo {address} {20402, Washington
  D.C./USA},\ \bibinfo {year} {1972})\BibitemShut {NoStop}%
\bibitem [{\citenamefont {H\"{a}ttig}\ \emph
  {et~al.}(1998{\natexlab{a}})\citenamefont {H\"{a}ttig}, \citenamefont
  {Christiansen},\ and\ \citenamefont {J\o{}rgensen}}]{chattig:1998}%
  \BibitemOpen
  \bibfield  {author} {\bibinfo {author} {\bibfnamefont {C.}~\bibnamefont
  {H\"{a}ttig}}, \bibinfo {author} {\bibfnamefont {O.}~\bibnamefont
  {Christiansen}}, \ and\ \bibinfo {author} {\bibfnamefont {P.}~\bibnamefont
  {J\o{}rgensen}},\ }\href {\doibase http://dx.doi.org/10.1063/1.476262}
  {\bibfield  {journal} {\bibinfo  {journal} {J. Chem. Phys.}\ }\textbf
  {\bibinfo {volume} {96}},\ \bibinfo {pages} {8355} (\bibinfo {year}
  {1998}{\natexlab{a}})}\BibitemShut {NoStop}%
\bibitem [{\citenamefont {Monson}\ and\ \citenamefont
  {McClain}(1970{\natexlab{c}})}]{Monson:1970}%
  \BibitemOpen
  \bibfield  {author} {\bibinfo {author} {\bibfnamefont {P.~R.}\ \bibnamefont
  {Monson}}\ and\ \bibinfo {author} {\bibfnamefont {W.~M.}\ \bibnamefont
  {McClain}},\ }\href {\doibase http://dx.doi.org/10.1063/1.1673778} {\bibfield
   {journal} {\bibinfo  {journal} {J. Chem. Phys.}\ }\textbf {\bibinfo {volume}
  {53}},\ \bibinfo {pages} {29} (\bibinfo {year}
  {1970}{\natexlab{c}})}\BibitemShut {NoStop}%
\bibitem [{\citenamefont {Christiansen}\ \emph {et~al.}(1998)\citenamefont
  {Christiansen}, \citenamefont {J\o{}rgensen},\ and\ \citenamefont
  {H\"{a}ttig}}]{christiansen:1998}%
  \BibitemOpen
  \bibfield  {author} {\bibinfo {author} {\bibfnamefont {O.}~\bibnamefont
  {Christiansen}}, \bibinfo {author} {\bibfnamefont {P.}~\bibnamefont
  {J\o{}rgensen}}, \ and\ \bibinfo {author} {\bibfnamefont {C.}~\bibnamefont
  {H\"{a}ttig}},\ }\href {\doibase
  10.1002/(SICI)1097-461X(1998)68:1<1::AID-QUA1>3.0.CO;2-Z} {\bibfield
  {journal} {\bibinfo  {journal} {Inter. J. Quan. Chem.}\ }\textbf {\bibinfo
  {volume} {68}},\ \bibinfo {pages} {1} (\bibinfo {year} {1998})}\BibitemShut
  {NoStop}%
\bibitem [{\citenamefont {Sundholm}\ \emph {et~al.}(1994)\citenamefont
  {Sundholm}, \citenamefont {Rizzo},\ and\ \citenamefont
  {J\o{}rgensen}}]{sd1994}%
  \BibitemOpen
  \bibfield  {author} {\bibinfo {author} {\bibfnamefont {D.}~\bibnamefont
  {Sundholm}}, \bibinfo {author} {\bibfnamefont {A.}~\bibnamefont {Rizzo}}, \
  and\ \bibinfo {author} {\bibfnamefont {P.}~\bibnamefont {J\o{}rgensen}},\
  }\href@noop {} {\bibfield  {journal} {\bibinfo  {journal} {J. Chem. Phys.}\
  }\textbf {\bibinfo {volume} {101}},\ \bibinfo {pages} {4931} (\bibinfo {year}
  {1994})}\BibitemShut {NoStop}%
\bibitem [{\citenamefont {H\"{a}ttig}\ \emph
  {et~al.}(1998{\natexlab{b}})\citenamefont {H\"{a}ttig}, \citenamefont
  {Christiansen},\ and\ \citenamefont {J\o{}rgensen}}]{chh1998}%
  \BibitemOpen
  \bibfield  {author} {\bibinfo {author} {\bibfnamefont {C.}~\bibnamefont
  {H\"{a}ttig}}, \bibinfo {author} {\bibfnamefont {O.}~\bibnamefont
  {Christiansen}}, \ and\ \bibinfo {author} {\bibfnamefont {P.}~\bibnamefont
  {J\o{}rgensen}},\ }\href@noop {} {\bibfield  {journal} {\bibinfo  {journal}
  {J. Chem. Phys.}\ }\textbf {\bibinfo {volume} {108}},\ \bibinfo {pages}
  {8331} (\bibinfo {year} {1998}{\natexlab{b}})}\BibitemShut {NoStop}%
\end{thebibliography}%

\end{document}